\documentclass[twocolumn]{aastex701}
\usepackage{graphicx}	
\usepackage{amsmath}	
\usepackage{xspace}
\usepackage{xcolor}
\usepackage{tabularx}
\usepackage{array}
\newcolumntype{Y}{>{\raggedright\arraybackslash}X}
\makeatletter 
  \patchcmd{\NAT@citex}
    {\@citea\NAT@hyper@{%
      \NAT@nmfmt{\NAT@nm}%
      \hyper@natlinkbreak{\NAT@aysep\NAT@spacechar}{\@citeb\@extra@b@citeb}%
      \NAT@date}}
    {\@citea\NAT@nmfmt{\NAT@nm}%
    \NAT@aysep\NAT@spacechar\NAT@hyper@{\NAT@date}}{}{}

  \patchcmd{\NAT@citex}
    {\@citea\NAT@hyper@{%
      \NAT@nmfmt{\NAT@nm}%
      \hyper@natlinkbreak{\NAT@spacechar\NAT@@open\if*#1*\else#1\NAT@spacechar\fi}%
        {\@citeb\@extra@b@citeb}%
      \NAT@date}}
    {\@citea\NAT@nmfmt{\NAT@nm}%
    \NAT@spacechar\NAT@@open\if*#1*\else#1\NAT@spacechar\fi\NAT@hyper@{\NAT@date}}
    {}{}
\makeatother

\newcommand{\msun}{{\,\rm M_\odot}}

\newcommand{\cm}{\,{\rm cm}}

\newcommand{\K}{\,{\rm K}}
\newcommand{\Myr}{\,{\rm Myr}}
\newcommand{\yr}{\,{\rm yr}}

\newcommand{\mum}{\,\mu{\rm m}}

\makeatletter
\def\bhe@check@firstfloat\@elt#1#2\@nil{%
  \@if@notdblfloat{#1}{}{\check@deferlist@stuck\do@startpage}%
}
\patchcmd{\do@startpage}
  {\check@deferlist@stuck\do@startpage}
  {\@ifnum{\pagegrid@col>\@ne}{%
    \@ifx{\@deferlist\@empty}{}{%
      \expandafter\bhe@check@firstfloat\@deferlist\@nil
    }%
  }{\check@deferlist@stuck\do@startpage}}
  {}{\PackageError{bhe}{AASTeX float-check patch failed}{Check the AASTeX version.}}
\makeatother

\begin{document}

\title{How Little Red Dots Turn Blue: \\ Evolution of Kelvin-Helmholtz Contracting Black Hole Envelopes and Observational Diagnostics}

\author[orcid=0009-0007-4590-6550]{Zihao Wang}
\affiliation{Department of Astronomy, Peking University, Beijing 100871, China}
\affiliation{Kavli Institute for Astronomy and Astrophysics, Peking University, Beijing 100871, China}
\email[]{zihaowang25@stu.pku.edu.cn}  

\author[orcid=0000-0001-9840-4959]{Kohei Inayoshi}
\correspondingauthor{Kohei Inayoshi}
\email[]{inayoshi@pku.edu.cn}
\affiliation{Kavli Institute for Astronomy and Astrophysics, Peking University, Beijing 100871, China}
\email[]{inayoshi@pku.edu.cn}


\begin{abstract}
James Webb Space Telescope (JWST) observations have revealed compact, red-colored, broad-line emitters known as little red dots (LRDs), while a bluer counterpart population has been referred to as little blue dots (LBDs). 
Their physical connection remains unclear: they may represent different evolutionary stages of the same phenomenon, distinct populations, or different phases of the broader evolution of early galaxies and their black holes (BHs).
We propose a model in which a rapidly growing BH is embedded in a geometrically thick gaseous envelope accompanied by concurrent nuclear star formation. 
The model reproduces the observed spectral energy distributions (SEDs) of LRDs and LBDs, and naturally connects the two phases through Kelvin--Helmholtz contraction of the envelope and exposure of the nuclear accretion disk, as the gas supply declines due to external stellar activity or feedback.
We predict the corresponding evolutionary tracks in the UV--optical slope ($\beta_{\rm UV}$--$\beta_{\rm opt}$) plane which bridge the apparent gap between LRDs and LBDs. 
These evolutionary tracks start from the LRD regime and consistently exhibit a red excursion that extends toward $(\beta_{\rm UV},\beta_{\rm opt})\sim(2,-1)$ before turning toward the LBD regime.
This feature is largely insensitive to variations in the model parameters that govern the envelope structure, providing a robust signature of the transitional population. 
The transition lasts $\sim5$--$50\,\mathrm{Myr}$, depending primarily on the BH feedback efficiency and the gas-retention efficiency of the envelope. 
The abundance ratio of transitional systems to LRDs therefore provides a constraint on BH feedback and gas retention during the rapid growth of early supermassive BHs.
\end{abstract}


\section{Introduction}\label{sec:intro}
The James Webb Space Telescope (JWST) has revealed a large population of compact red sources at high redshifts, commonly referred to as ``little red dots" (LRDs). 
LRDs have been proposed as a potentially new class of active galactic nuclei (AGNs) and have become central to discussions of early black hole (BH) growth.
They are characterized by a distinctive V-shaped spectral energy distribution (SED) and broad Balmer emission lines~\citep[e.g.,][]{Greene2024,Kocevski2025,Hainline2024}, yet lack several canonical AGN signatures, such as strong X-ray emission, prominent near-to-mid-infrared excess, powerful radio jets or variability~\citep[e.g.,][]{Maiolino2025,Yue2024,Akins2024,Perezgonzalez2024,Gloudemans2025,Zhang2025a,Kokubo2025}. 

Subsequent observational and theoretical studies have provided the first clues to the physical nature of LRDs. Their spectra commonly exhibit broad Balmer emission lines together with prominent Balmer absorption features~\citep[e.g.,][]{Kocevski2023,Matthee2024,Maiolino2024,Lin2024,Taylor2025,Kocevski2025,Setton2025} and a spectral break near the Balmer limit wavelength~\citep[e.g.,][]{Greene2024,Furtak2024,Wang2024a}. The coexistence of these signatures, particularly in LRDs exhibiting exceptionally deep Balmer breaks~\citep[e.g.,][]{deGraaff2025,Naidu2025,Taylor2025a}, cannot be explained by evolved stellar populations alone, pointing instead to a central BH embedded within dense, high-covering-fraction gas~\citep{Inayoshi2025a,Ji2025,Yan2026}. In this picture, gas attenuation combined with thermal emission from an optically thick envelope with an effective temperature of $T_{\rm eff}\simeq5000\,\mathrm{K}$ naturally reproduces the red optical continuum and nearly flat infrared spectrum without requiring dust reddening~\citep[][]{Kido2025,Liu2025,Inayoshi2026,deGraaff2026,Naidu2025,Kokorev2026}. 
Continuous gas supply at super-Eddington accretion rates sustains
sufficiently high column densities in the gaseous envelope, which also suppresses X-ray emission and the propagation of radio jets~\citep{Madau2026,Inayoshi2025b,Kido2025}. 
This gas-enshrouded AGN scenario, often referred to as the black hole envelope (BHE) or BH-star model, therefore provides a compelling explanation for many observed properties of LRDs, 
although the detailed physical properties of the gaseous component remain under debate~(e.g., \citealt{Naidu2025,Naidu2026,Sneppen2026,Rusakov2026,deGraaff2026,Merida2026b,Merida2026c}; see \citealt{Inayoshi2025c} for a review).

At the same time, a new question has begun to emerge. Recent observations have identified a population of compact, blue broad-line AGNs at similar redshifts, sometimes referred to as ``little blue dots" (LBDs; \citealt{Asada2026,Brazzini2026,Geris2026}). Unlike LRDs, LBDs exhibit blue rest-frame UV--optical continua and follow the local quasar relation between UV and broad H$\alpha$ luminosities~\citep{Asada2026}. Nevertheless, they share many other observational properties with LRDs, including compact morphologies, weak (or moderate) hot-dust emission, X-ray nondetections, and extended wings in broad emission lines~\citep[e.g.,][]{Hainline2024,Juodzbalis2026,Rusakov2026,Brazzini2026,Scholtz2026}.
This phenomenological overlap naturally raises the question of whether LRDs and LBDs represent two distinct populations or different stages of the same underlying phenomenon. One class of models interprets them primarily as different viewing-angle manifestations of dust-obscured AGNs~\citep[e.g.,][]{Madau2026,Madau2026a}. Alternatively, the observed redshift evolution of the LRD number density has motivated an evolutionary picture in which LRDs gradually transform into LBDs~\citep[e.g.,][]{Inayoshi2025First,Asada2026,Merida2026}. A key distinction between these scenarios is therefore the existence, abundance, and lifetime of transitional systems connecting the two populations.

\begin{figure*}
    \centering
    \includegraphics[width=1\linewidth]{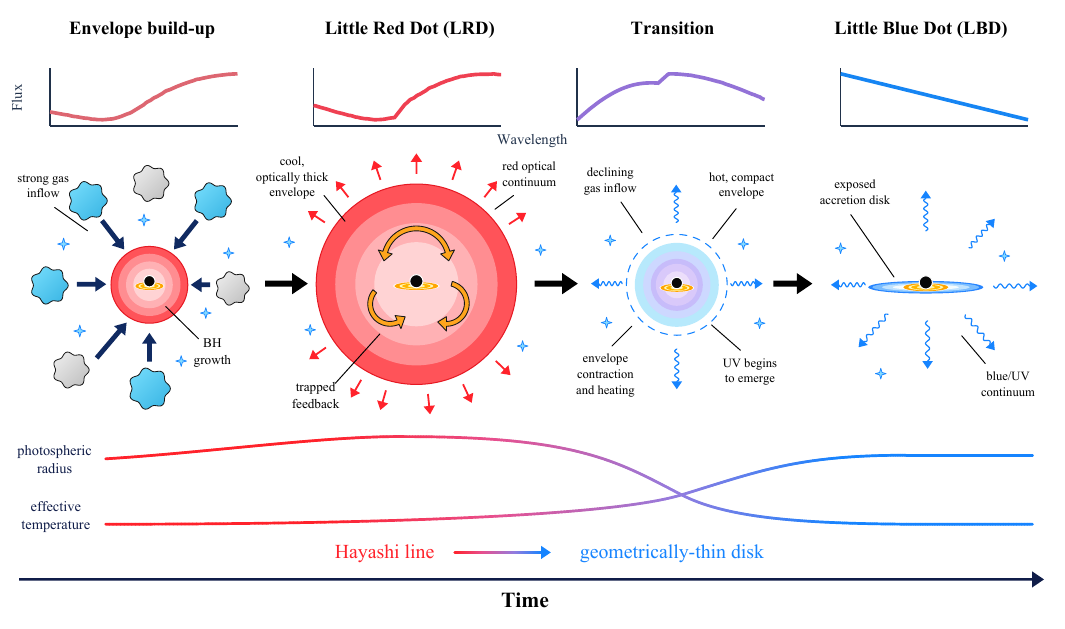}
    \caption{
    Schematic evolutionary pathway from an LRD to an LBD. Strong gas inflow initially drives BH growth and builds an extended envelope around the central BH and stellar component. During the LRD phase, the cool, optically thick BHE reprocesses the central emission into a red optical continuum and traps feedback. As the gas supply declines, the envelope contracts and heats, gradually evolving toward a compact LBD phase from which UV radiation emerges. The upper sequence schematically illustrates the corresponding spectral evolution, while the lower curves show the contraction of the photosphere and the increase in effective temperature, tracing the evolution from a Hayashi-line-regulated envelope toward a geometrically thin accretion disk.
    }
    \label{fig:cartoon}
    \vspace{5mm}
\end{figure*}

The BHE configuration predicts such an evolutionary sequence. Similar envelope systems have been studied independently in the contexts of seed BH formation~\citep[e.g.,][]{Volonteri2005a,Begelman2006,Shang2010,Inayoshi2016,Chon2018,Hu2022,Shi2024} and stellar tidal disruption events (TDEs; e.g., \citealt{Loeb1997,Strubbe2009,Coughlin2014,Metzger2016,Metzger2022,Ryu2021,Roth2016,Steinberg2023}), where rapidly accreting BHs are embedded within dense, optically thick inflows fueled by protogalactic gas collapse or stellar debris. As the envelope cools radiatively and undergoes Kelvin--Helmholtz contraction, its effective temperature rises, providing a plausible channel through which an LRD-like system may evolve toward an LBD-like state. Inspired by this analogy, we adopt the framework of \citet{Inayoshi2026}, in which LRDs arise from the coevolution of rapidly growing seed BHs and nascent nuclear starbursts, to investigate the evolutionary connection between LRDs and LBDs. We develop a time-dependent model that follows the evolution of the central BH coupled with its surrounding envelope under radiative cooling, gas inflows, BH accretion, and feedback (see Figure~\ref{fig:cartoon} for an illustration). The model predicts their evolutionary tracks, characteristic transition timescales, and observable properties of the intermediate phase.

This paper is organized as follows. In Section~\ref{sec:method}, we describe our BHE model. In Section~\ref{sec:BHE}, we present the envelope evolution under different physical conditions. In Section~\ref{sec:SED}, we examine the spectral evolution and its observational signatures. Finally, in Sections~\ref{sec:disc} and \ref{sec:conc}, we discuss the implications of our results and summarize our conclusions.

\section{Method}\label{sec:method}
\begin{table}
    \centering
    \caption{Model parameters adopted for the initial conditions (ICs) and evolution of the BHE.}
    \addtolength{\tabcolsep}{0.2pt}
    \def\arraystretch{1.2}
    \begin{tabular}{lcc}
        \hline
        Parameter & Description & Fiducial value \\
        \hline
        \hline
        \textbf{ICs} & & \\
        $M_{\rm BH,0}$ & Initial BH mass & $10^5\msun$ \\
        $M_{\rm env,0}$ & Initial envelope mass & $10^5\msun$ \\
        $T_{\rm eff,0}$ & Effective temperature & $10^4\K$ \\
        $\xi$ & Envelope density power law & 1 \\
        \hline

        \textbf{Evolution} & & \\
        $\dot{M}_{\rm ISM,0}$ & ISM accretion rate & $5\msun/\yr$ \\
        $t_{\rm fb}$ & Gas retention timescale & $20\Myr$ \\
        $f_{\rm env}$ & Envelope retention efficiency & 0.5 (fiducial) \\
        $\eta_{\rm fbk}$ & BH feedback efficiency & 0.01 (fiducial) \\
        
        \hline
        \end{tabular}
    \label{table:pars}
\end{table}

\subsection{Mechanical equilibrium}
We consider a newly formed BH seed embedded within a dense gaseous envelope sustained by the balance between AGN feedback and continuous mass inflow. 
This configuration forms a bloated, cool atmosphere with an effective temperature of $T_{\rm eff}\simeq 3,000{-}10,000\,\mathrm{K}$~\citep[e.g.,][]{Begelman2008, Hosokawa2013, Inayoshi2016}.


To describe the envelope structure, we adopt the mechanical framework of a pressure-supported gaseous envelope developed by \citet{Metzger2022} in the context of TDEs. 
Although their model was originally constructed for accretion from hot TDE debris, the underlying dynamical and thermal processes are relevant to our scenario. 
In both cases, gas accumulates around a BH to form an optically thick envelope that stores mass and energy, while the instantaneous accretion rate onto the BH is regulated by the global envelope structure. 
The subsequent evolution is governed by the competition between radiative cooling, inward mass transport, and accretion-powered heating.
This balance determines whether the envelope contracts, expands, or is gradually depleted over time.

We assume that the envelope is in virial equilibrium, such that its total energy satisfies
\begin{equation}
    E_{\rm tot}=U+K=\frac{U}{2},
\end{equation}
where $U$ and $K$ are the gravitational and kinetic/thermal energies, respectively. 
Following the structural prescription adopted by \citet{Metzger2022}, we assume that the envelope follows a power-law radial density profile within the characteristic radius $R_{\rm v}$ and transitions to an exponential cutoff beyond this radius:
\begin{equation}
    \rho(r)=C_{\rho}\frac{M_{\rm env}}{4\pi R_{\rm v}^3}
 \begin{cases}
 \left(\dfrac{r}{R_{\rm v}}\right)^{-\xi}, & r<R_{\rm v},\\[6pt]
 \exp\!\left[1-r/R_{\rm v}\right], & r>R_{\rm v},
 \end{cases}
\end{equation}
where $\xi$ denotes the power-law slope of the density profile and is expected to lie in the range of $1/2<\xi<3$~\citep{Coughlin2014}, and $C_{\rho}=(3-\xi)/(16-5\xi)$ is the normalization factor determined by mass conservation.
Throughout this work, we adopt a fiducial value of $\xi=1$.

The gravitational potential energy includes both the interaction
between the central BH and the envelope and the self-gravity of
the envelope. Under the adopted energy closure,
\begin{equation}
\begin{aligned}
2|E_{\rm tot}|
&=
4\pi G\int_0^\infty
\left[
M_{\rm BH}+M_{\rm env}(<r)
\right]\rho(r)\,r\,{\rm d}r, \\
&=
\frac{G}{R_{\rm v}}
\left(
C_{\rm BH}M_{\rm BH}M_{\rm env}
+
C_{\rm self}M_{\rm env}^2
\right),
\end{aligned}
\label{eq:Ubind}
\end{equation}
where $M_{\rm BH}$ is the central BH mass and
$M_{\rm env}(<r)$ is the envelope mass enclosed within radius $r$.
The dimensionless structural factors are
\begin{equation}
C_{\rm BH}
=
C_\rho\left[\frac{1}{2-\xi}+2\right]
=
\frac{(3-\xi)(5-2\xi)}
{(16-5\xi)(2-\xi)},
\end{equation}
and
\begin{equation}
C_{\rm self}
=
C_\rho^2
\left[
\frac{1}{(3-\xi)(5-2\xi)}
+
\frac{2}{3-\xi}
+
\frac{29}{8}
\right].
\end{equation}
The first coefficient describes the BH--envelope interaction,
while the second expresses the envelope self-gravity.
For our fiducial slope $\xi=1$, these coefficients become
\begin{equation}
C_\rho=\frac{2}{11},
\qquad
C_{\rm BH}=\frac{6}{11},
\qquad
C_{\rm self}=\frac{115}{726}.
\end{equation}
Therefore, the characteristic radius is given as
\begin{equation}
R_{\rm v}
=
\frac{G}{2|E_{\rm tot}|}
\left(
C_{\rm BH}M_{\rm BH}M_{\rm env}
+
C_{\rm self}M_{\rm env}^2
\right).
\label{eq:Rv}
\end{equation}
Note that this expression is valid for $\xi<2$. For a steeper density profile ($\xi\ge2$), the gravitational binding energy formally diverges at small radii in the power-law form and the solution becomes sensitive to the inner structure of the envelope.

It is worth noting that envelope self-gravity is not always negligible but rather dominant during the early gas-rich phase. 
A fully self-consistent treatment of self-gravity, however, depends on the detailed internal structure of the envelope and ultimately requires a hydrodynamic solution for the BHE (see the hydrostatic-equilibrium modeling in \citealt{Takasao2026}). Its implementation here should therefore be regarded as phenomenological, based on the adopted density profile. To assess the associated uncertainty, we additionally consider a limiting case in which envelope self-gravity is neglected (i.e., $C_{\rm self}=0$) in Appendix~\ref{apdx:xi}. We find that including self-gravity tends to delay the transition. We have verified that the envelope's overall properties and evolution are insensitive to the precise choice of $\xi$ over the range $1/2<\xi<2$.

\subsection{Photosphere determination}

The photospheric radius is defined by the optical-depth condition
\begin{equation}
\label{eq:phot_cond}
    \tau(R_{\rm ph})=\int_{R_{\rm ph}}^{\infty}\rho \kappa \,{\rm d}r=1.
\end{equation}
To evaluate the opacity, we adopt the simplified prescription from \citet{Begelman2008},
\begin{equation}
\label{eq:opacity}
    \kappa(T)=\frac{\kappa_{\rm es}}{1+(T/T_0)^{-s_0}},
\end{equation}
where $T_0=8000\,\mathrm{K}$ and $s_0=13$. 
This functional form approximately captures the sharp decline in opacity below $T\sim10^4\,\mathrm{K}$ associated with hydrogen recombination and the resulting H$^-$ bound-free and free-free opacity in partially ionized gas.
In the high-temperature limit, the opacity asymptotically approaches the electron-scattering value, $\kappa_{\rm es}=0.35 \cm^2 \rm g^{-1}$.

This simplified prescription has two important limitations. 
First, it neglects the dependence of opacity on gas density. 
Second, the actual opacity exhibits a more complex temperature dependence than represented by Equation~(\ref{eq:opacity}).
At high densities ($\rho \gg 10^{-11}~{\rm g~cm^{-3}}$), the adopted values of $T_0$ and $s_0$ provide a reasonable approximation to the overall opacity transition, but the true opacity shows a pronounced bump associated with hydrogen ionization near $T\simeq10^4\,\mathrm{K}$ and follows a tail of Kramers' opacity law ($\kappa\propto T^{-7/2}$) over the range $10^4 \lesssim T/\K \lesssim 2\times 10^4$.
At lower densities, the same functional form remains qualitatively valid, but the transition temperature shifts to lower values and the opacity decline becomes steeper.
For a more realistic treatment based on the opacity tables of \citet{Mayer2005}, we refer the reader to \citet{Takasao2026}.
Despite these limitations, we find that the evolution of the envelope photosphere is primarily controlled by the existence of the opacity cliff near $T\sim10^4~\K$ rather than by the detailed shape of the opacity curve. 
The analytic prescription in Equation~(\ref{eq:opacity}) therefore provides a useful approximation while allowing efficient exploration of a broad parameter space and long-term evolution of the BHE system.

The photospheric radius ($R_{\rm ph}$) and effective temperature ($T_{\rm eff}$) are determined by solving Equation~(\ref{eq:phot_cond}) together with the boundary condition
\begin{equation}
\label{eq:phot}
L_{\rm Edd}=4\pi \sigma_{\rm SB} T_{\rm eff}^4 R_{\rm ph}^2,
\end{equation}
with $T(R_{\rm ph})=T_{\rm eff}$. Here, $\sigma_{\rm SB}$ is the Stefan–Boltzmann constant.
Outside the photosphere, where the optical depth is below unity, radiation is allowed to escape efficiently.
The local radiation temperature is thus expected to be close to the effective temperature, and we approximate the temperature entering Equation (\ref{eq:phot_cond}) near the photosphere as $T\simeq T_{\rm eff}$. 
In practice, we solve iteratively for the photospheric location while imposing the boundary condition in Equation~(\ref{eq:phot}).

At low temperatures ($T_{\rm eff}<10^4\K$), the effective temperature is self-regulated by the strong temperature dependence of the opacity. As the envelope expands and $T_{\rm eff}$ decreases, the opacity drops sharply. This reduces the optical depth outside the photosphere and causes $R_{\rm ph}$ to move inward. According to Equation (\ref{eq:phot}), the resulting decrease in photospheric radius acts to increase $T_{\rm eff}$. 
As a result, the system exhibits a thermostat mechanism that stabilizes the photospheric temperature near a few thousand kelvin and prevents the envelope from cooling to arbitrarily low temperatures, analogous to the forbidden region for stellar atmospheres~\citep{Hayashi1961}.

In the high-temperature regime, where electron scattering dominates the opacity, the self-regulation mechanism described above no longer operates. 
As the envelope mass decreases, the photospheric radius contracts rapidly and the effective temperature inferred from Equation~(\ref{eq:phot}) rises. 
Formally, $T_{\rm eff}$ would diverge as $R_{\rm ph}\rightarrow 0$. 
In practice, however, such a solution is unphysical because it corresponds to a stage at which the bloated envelope has already contracted and transitioned into a geometrically thin accretion disk.
To characterize this transition, we impose an upper limit on the photospheric temperature based on the geometrically thin disk limit~\citep[e.g.,][]{Shakura1973}, in which viscous heating is balanced by radiative cooling,
\begin{equation}
T_{\rm disk}=\left(\frac{3GM_{\rm BH}\dot{M}_{\rm BH}}{8\pi\sigma_{\rm SB} R_{\rm in}^3}\right)^{1/4},
\label{eq:Tdisk}
\end{equation}
where $R_{\rm in}$ represents the disk inner radius and is set to the innermost stable circular orbit for a nonrotating BH, $R_{\rm in}=3R_{\rm Sch}$, and $R_{\rm Sch}=2GM_{\rm BH}/c^2$ is the Schwarzschild radius of the BH.
Once the photospheric temperature reaches this limit, we assume that the observable thermal emission follows the thin-disk solution.
We emphasize that this prescription affects only the post-processed photospheric properties and does not modify the underlying dynamical evolution of the envelope.

\subsection{Envelope evolution}

\begin{figure}
    \centering
    \includegraphics[width=0.95\linewidth]{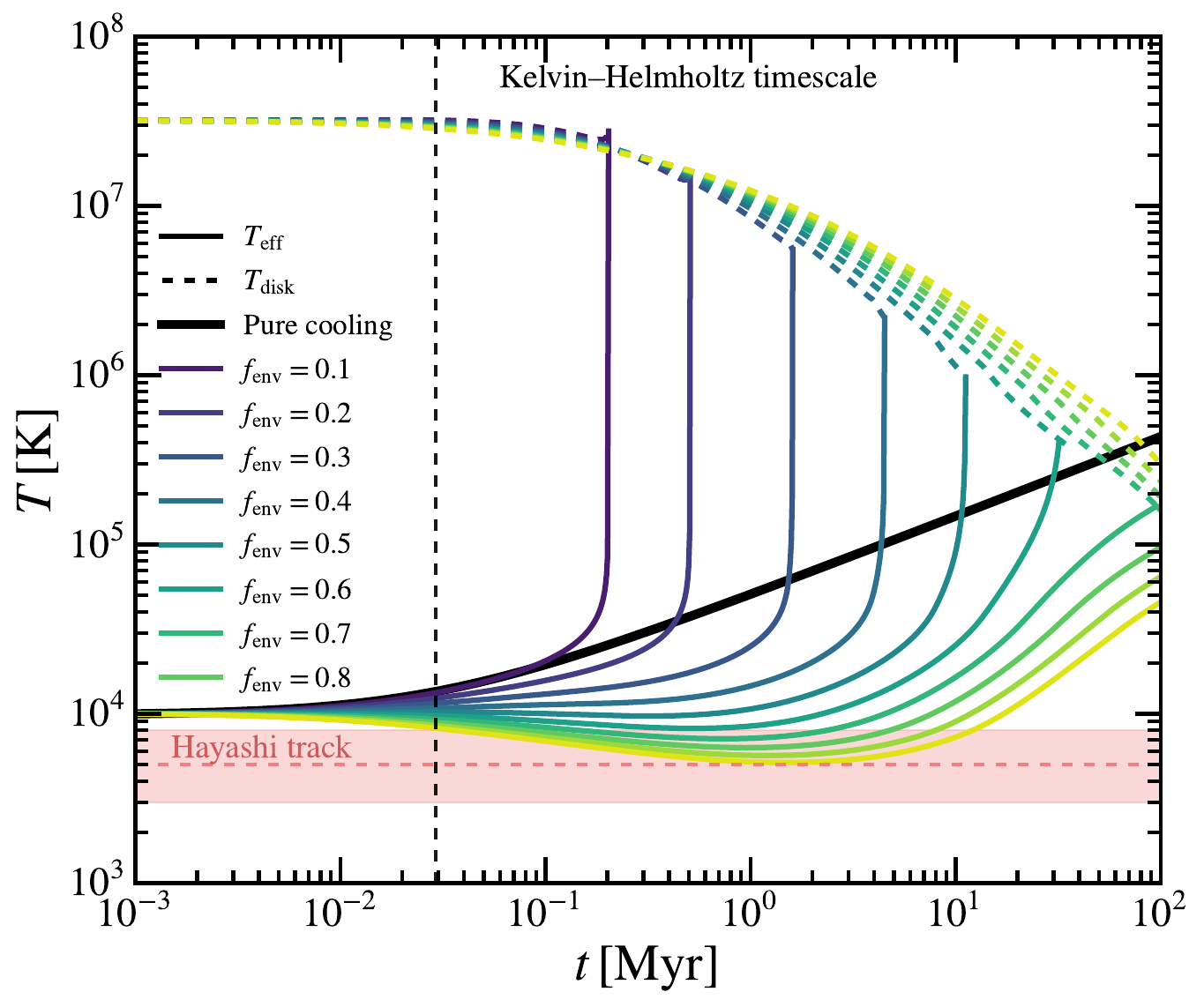}
    \caption{Evolution of the envelope effective temperature for the pure-cooling model (thick black curve) and accretion-only models with different gas-retention efficiencies $f_{\rm env}$ (colored curves). Solid curves show the envelope effective temperature $T_{\rm eff}$, while dashed curves show the capped disk surface temperature $T_{\rm disk}$ under the thin-disk assumption. The vertical dashed line marks the initial Kelvin--Helmholtz timescale. In the pure-cooling case, $T_{\rm eff}$ increases monotonically as the envelope contracts. In the accreting models, ISM inflow initially builds up the envelope and inflates $R_{\rm v}$ and $R_{\rm ph}$, driving $T_{\rm eff}$ downward and briefly bringing the system to the Hayashi track (shown as the red region, here defined approximately by $3,000\K<T<8,000\K$). As the envelope is later depleted by BH accretion, the temperature rises again toward the disk branch, with larger $f_{\rm env}$ delaying this turnover.}
    \label{fig:Teff_acc}
\end{figure}

We initialize the system with $M_{\rm BH,0}=M_{\rm env,0}=10^5\,\msun$ and $T_{\rm eff,0}=10^4\,\K$ to mimic a newly formed heavy BH seed~\citep[e.g.,][]{Pacucci2026,Baggen2026,Gu2026}. The exact choice of the initial mass is not critical, as subsequent rapid super-Eddington accretion quickly erases the memory of the initial conditions~\citep{Inayoshi2026}.
From the adopted temperature, we calculate the initial photospheric radius $R_{\rm ph,0}$ from Equation (\ref{eq:phot}), envelope virial radius $R_{\rm v,0}$ from Equation (\ref{eq:phot_cond}), and corresponding total energy $E_{\rm tot,\,0}$ by working backward.

In our model, the envelope evolution is governed by mass and energy conservation:
\begin{equation}
\label{eq:mass}
\frac{dM_{\rm env}}{dt}=\dot{M}_{\rm in}-\dot{M}_{\rm BH},
\end{equation}
\begin{equation}
\label{eq:energy}
\frac{dE_{\rm tot}}{dt}=-L_{\rm Edd}+\dot{E}_{\rm BH},
\end{equation}
where $\dot{M}_{\rm in}$ denotes the mass inflow rate retained by the envelope, while $\dot{M}_{\rm BH}$ and $\dot{E}_{\rm BH}$ denote the mass and energy fluxes associated with accretion onto the central BH, respectively. 
We neglect additional energy exchange associated with the external gas supply, including advected energy, dissipation during circularization and turbulent settling~\citep[e.g.,][]{Metzger2022,Sadowski2016}, and energy carried away by accretion-driven outflows~\citep[e.g.,][]{Tuna2026}. These contributions may be significant during the early phase of rapid gas supply but can act in competing directions in the envelope energy budget. Their importance, however, progressively diminishes as the inflow subsides.

We assume that the envelope radiates at the Eddington luminosity at all times, although the BH accretion rate itself does not need to remain at or above the Eddington accretion rate. The specific treatments of the three components are summarized below.\\

\noindent \textbf{Envelope accretion:} We assume that inflow from the interstellar medium (ISM) is regulated by feedback from core-collapse supernovae (SNe) in the nuclear region after the onset of the starburst. 
The gas inflow rate to the central region is approximated as
\begin{equation}
    \dot{M}_{\rm ISM}=\,\dot{M}_{\rm ISM,0}\,\exp(-t/t_{\rm fb}),
\end{equation}
where $\dot{M}_{\rm ISM,0}$ is the initial inflow rate to the galactic nuclei and $t_{\rm fb}$ is the gas retention timescale within which gas remains confined in the nuclear region against momentum injection from multiple SNe. In this study, we adopt $\dot{M}_{\rm ISM,0}\simeq5\msun\yr^{-1}$ and $t_{\rm fb}\simeq20\Myr$. Both parameters depend on redshift and environment through the halo accretion rate and the depth of the gravitational potential well; their detailed dependence is discussed in \citet{Inayoshi2026}. The values adopted here are broadly representative of massive dark matter halos with $M_{\rm h}\simeq10^{10-11}\msun$ at $z\simeq 5-10$ (virial temperature $T_{\rm vir}\simeq 10^6\K$).

We assume that only a fraction of the large-scale ISM inflow is ultimately retained by the envelope, while the rest is lost through outflows and other mass-loss processes:
\begin{equation}
    \dot{M}_{\rm in}=f_{\rm env} \dot{M}_{\rm ISM}.
\end{equation}
Here, $f_{\rm env}(\leq 1)$ is the gas-retention efficiency of the envelope, which accounts for the fact that not all inflowing gas can remain bound to the system.
\\

\noindent \textbf{BH accretion:} In such a dense gaseous environment with strong inflows, the BH may grow in the super-Eddington regime. Radiation-hydrodynamic simulations have shown that the BH accretion rate can be reduced relative to the large-scale inflow rate following a power-law profile as $\dot{M}_{\rm BH}\propto r^{p}$, with $p\sim0.5{-}0.8$~\citep{Hu2022}; similar behavior occurs more generally in radiatively inefficient flows~\citep{Yuan2014}. Here, we adopt the specific scaling
\begin{equation}
\dot{M}_{\rm BH}=\alpha \dot{M}_{\rm ISM},
\label{eq:bhdot}
\end{equation}
where
\begin{equation}
\alpha=\min\!\left[\left(\frac{10R_{\rm Sch}}{R_{\rm v}}\right)^{1/2},\,1\right].
\label{eq:alpha}
\end{equation}
This prescription corresponds to the case $p=0.5$ and ensures that the BH accretion rate approaches the inflow rate when the envelope becomes sufficiently compact while remaining suppressed when $R_{\rm v}\gg10R_{\rm Sch}$~\citep[e.g.,][]{Abramowicz2002}.
\\

\noindent \textbf{BH feedback:} Mass accretion onto the BH releases the gravitational energy that is carried outward through the envelope by convection and radiation. We parameterize the resulting feedback luminosity as
\begin{equation}
\dot{E}_{\rm BH}=\eta_{\rm fbk}\dot{M}_{\rm BH}c^2,
\end{equation}
where $\eta_{\rm fbk}\lesssim0.1$ is an effective feedback efficiency that encapsulates both intrinsic and extrinsic uncertainties in the coupling between the accreting BH and the envelope. The former includes uncertainties in the production and reprocessing of radiation, outflows, and jets from the inner accretion flow in the super-Eddington regime~\citep[e.g.,][]{Blandford1999,Metzger2022}, while the latter accounts for energy that may escape without coupling to the gas, for example, energy leakage through an envelope with incomplete covering~\citep[e.g.,][]{Sok2026,Ando2026,Deugenio2026,TangMT2026}. \\

Starting from the initial state, we evolve the envelope for 100 Myr by updating its mass and energy according to Equations~(\ref{eq:mass}) and~(\ref{eq:energy}). At each time step, we derive the corresponding envelope mass $M_{\rm env}$, envelope radius $R_{\rm v}$, photospheric radius $R_{\rm ph}$, and effective temperature $T_{\rm eff}$. The calculation is terminated once the envelope is either fully depleted ($M_{\rm env}\leq0$) or unbound ($E_{\rm tot}>0$). A summary of parameters adopted for the dynamical modeling is provided in Table~\ref{table:pars}. 

\section{Evolution of the envelope}\label{sec:BHE}

\subsection{Pure-cooling model}

As a sanity check, the thick black curve in Figure~\ref{fig:Teff_acc} shows the evolution of the envelope temperature in the absence of ISM inflow, BH accretion, and feedback; i.e., we set $\dot{M}_{\rm ISM}=\dot{M}_{\rm BH}=0$ and hold $M_{\rm env}=M_{\rm BH}=10^5\msun$ fixed throughout the calculation. The Kelvin--Helmholtz timescale,
\begin{equation}
    \tau_{\rm KH}=\frac{|E_{\rm tot,0}|}{L_{\rm Edd,0}},
\end{equation}
is of order $10^{4-5}$ yr, as marked by the vertical dashed line. After this point, the envelope undergoes rapid Kelvin--Helmholtz contraction, accompanied by a shrinking photospheric radius $R_{\rm ph}$. During this contraction phase, the effective temperature increases as
\begin{equation}
    T_{\rm eff}(t)=T_{\rm eff,0}\left(1+\frac{t}{\tau_{\rm KH}}\right)^{1/2}.
\end{equation}

\subsection{Accretion-only models}
\begin{figure*}
    \centering
    \includegraphics[width=0.95\linewidth]{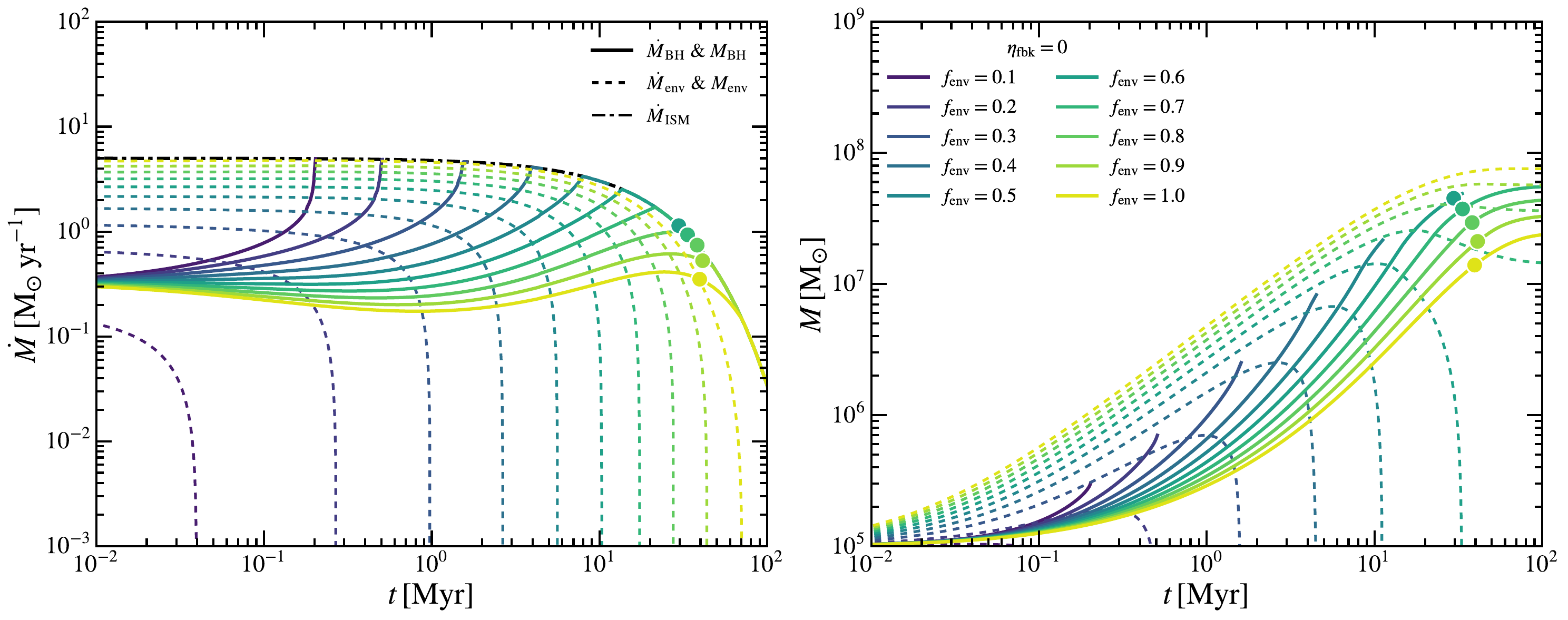}
    \caption{Mass-growth histories in the accretion-only models with $\eta_{\rm fbk}=0$ and different gas-retention efficiencies $f_{\rm env}$ (colored curves). In the left panel, solid, dashed, and dash-dotted curves show $\dot{M}_{\rm BH}$, $\dot{M}_{\rm env}$, and $\dot{M}_{\rm ISM}$, respectively. In the right panel, solid and dashed curves show the corresponding BH and envelope masses. Circles mark the time when the BH accretion rate drops below the super-Eddington regime. Lower $f_{\rm env}$ leads to a runaway depletion of the envelope, causing the accretion history to terminate rapidly, whereas larger $f_{\rm env}$ allows the envelope to grow and support BH accretion over much longer timescales.}
    \label{fig:EnvelopeEvo_acc}
\end{figure*}

We now consider mass exchange among the ISM, the BH, and the envelope while neglecting BH feedback (i.e., $\eta_{\rm fbk}=0$). We vary $f_{\rm env}$, which parameterizes how efficiently gas is retained in the envelope. Figure~\ref{fig:EnvelopeEvo_acc} shows the mass evolution of the envelope and the BH for different values of $f_{\rm env}$. The solid curves denote the BH mass evolution, while the dashed curves show the envelope mass evolution. For clarity, we plot only the phase during which the envelope mass remains positive; at the final stage, the envelope mass formally drops below zero as the envelope is fully depleted.

The mass growth of the BH and envelope is coupled through Equation~(\ref{eq:bhdot}), such that different envelope configurations imply different BH accretion rates, which in turn feed back on the envelope evolution in a nonlinear manner. When $f_{\rm env}$ is low, the envelope lifetime is short. In this regime, $R_{\rm v}$ shrinks because of rapid mass loss driven by BH accretion, which further accelerates the depletion of the envelope and eventually produces a diverging solution of $T_{\rm eff}$ at late times. By contrast, for larger $f_{\rm env}$, the system can survive for much longer, in some cases beyond the simulation time.

The resulting evolution of the effective temperature is overplotted in Figure~\ref{fig:Teff_acc}. Once mass exchange among the ISM, the BH, and the envelope is included, the evolution becomes qualitatively different from the pure-cooling case. At early times, the ISM inflow exceeds the BH accretion rate, so the envelope mass grows. The increasing envelope mass causes the envelope to expand, leading to larger $R_{\rm v}$ and $R_{\rm ph}$. This expansion lowers the effective temperature and drives the system toward the Hayashi track with $T_{\rm eff}\simeq3000$--$5000\K$. This low-temperature phase is only transient. As the evolution proceeds, BH accretion gradually catches up and begins to deplete the envelope. The temperature then turns upward and eventually approaches the disk-temperature branch once the envelope becomes sufficiently compact or nearly exhausted. A larger $f_{\rm env}$ delays this turnover by allowing the envelope-supported phase to persist longer.

\subsection{Feedback-involved models}\label{subsec:fbk}
\begin{figure*}
    \centering
    \includegraphics[width=0.95\linewidth]{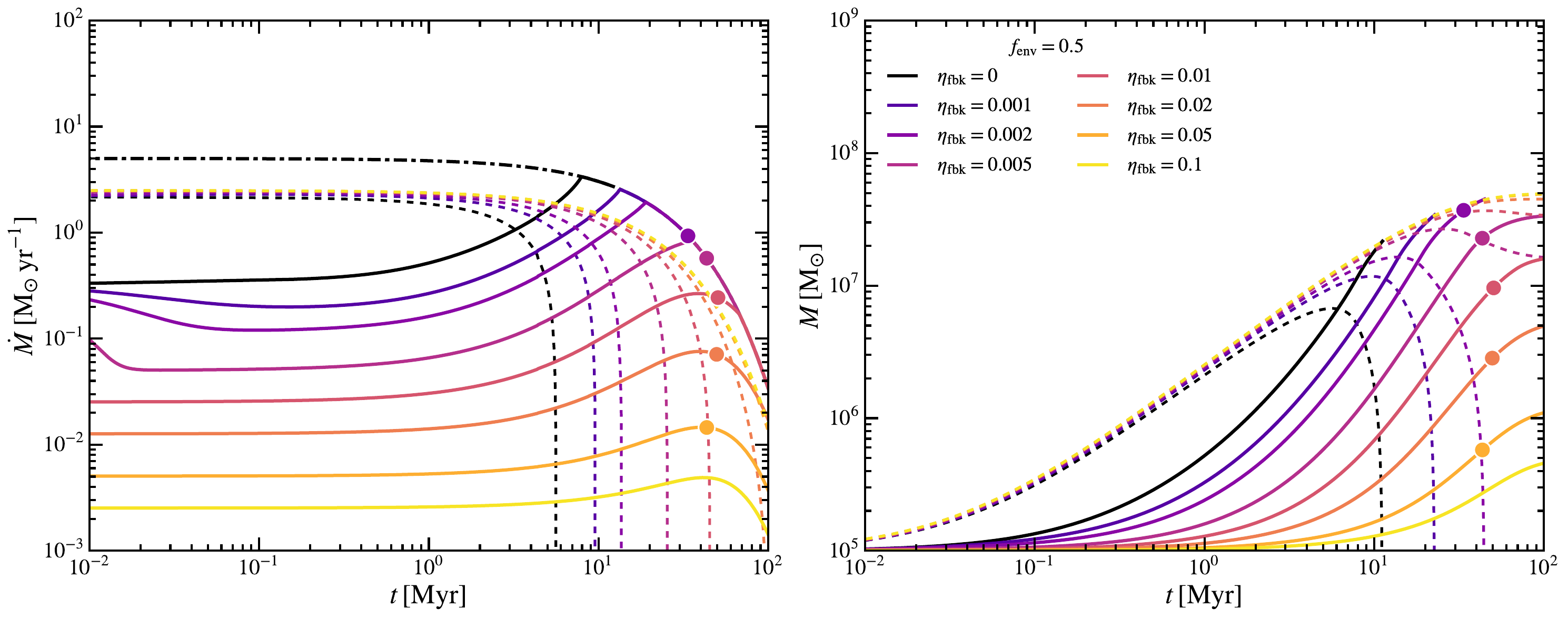}
    \caption{Similar to Figure~\ref{fig:EnvelopeEvo_acc} but shown for the feedback-involved models with fixed $f_{\rm env}=0.5$ and different feedback efficiencies $\eta_{\rm fbk}$. Increasing feedback inflates the envelope and suppresses BH accretion, thereby delaying envelope depletion and prolonging the system lifetime. At sufficiently large $\eta_{\rm fbk}$, the system enters a long-lived, feedback-regulated phase.}
    \label{fig:EnvelopeEvo_fbk}
\end{figure*}

We now consider the effect of BH accretion heating on the envelope evolution.
The feedback efficiency of BH accretion remains highly uncertain and is generally expected to be relatively low in the super-Eddington regime~\citep[e.g.,][]{Abramowicz2001,Watarai2001,Ohsuga2005,Jiang2019}. We therefore explore a range of $\eta_{\rm fbk}=0$--$0.1$. Figure~\ref{fig:EnvelopeEvo_fbk} shows the mass evolution of the envelope and the BH for different values of $\eta_{\rm fbk}$ at fixed $f_{\rm env}=0.5$. The corresponding solutions for other values of $f_{\rm env}$ are presented in Appendix~\ref{apdx:fenv}.

Once BH feedback is included, the injected heat increases the total energy of the envelope and causes it to expand. This produces a larger characteristic radius $R_{\rm v}$, which in turn reduces the BH accretion rate through $\dot{M}_{\rm BH}\propto R_{\rm v}^{-1/2}$ (see Equation~\ref{eq:bhdot}). The coupled heating and accretion response naturally gives rise to a self-regulated state. The feedback heating rate can be written as
\begin{equation}
    \dot{E}_{\rm BH}=\eta_{\rm fbk}\dot{M}_{\rm BH}c^2=
    \frac{\eta_{\rm fbk}}{\eta_{\rm rad}}\dot{m}_{\rm BH}L_{\rm Edd},
\label{eq:feedback_heating_rate}
\end{equation}
where $\dot{m}_{\rm BH}\equiv\dot{M}_{\rm BH}/\dot{M}_{\rm Edd}$, with $\dot{M}_{\rm Edd}\equiv L_{\rm Edd}/(\eta_{\rm rad}c^2)$ and $\eta_{\rm rad}=0.1$.
Balancing feedback heating against the radiative luminosity of the envelope then gives
\begin{equation}
    \dot{m}_{\rm BH,eq}=
    \frac{\eta_{\rm rad}}{\eta_{\rm fbk}}.
    \label{eq:mdot_fbk_equilibrium}
\end{equation}
If $\dot{m}_{\rm BH}>\dot{m}_{\rm BH,eq}$, feedback heating exceeds radiative losses, causing the envelope to expand and suppressing further accretion. Conversely, if $\dot{m}_{\rm BH}<\dot{m}_{\rm BH,eq}$, radiative cooling causes the envelope to contract, allowing the accretion rate to recover. This loop maintains the system near $\dot{m}_{\rm BH,eq}$ as long as the required accretion rate can be supplied by the ISM and the envelope can adjust sufficiently rapidly to the evolving equilibrium configuration. Eventually, the equilibrium breaks down either because the declining ISM inflow can no longer sustain $\dot{m}_{\rm BH,eq}$ or because the equilibrium radius contracts more rapidly than the envelope can respond on its Kelvin--Helmholtz timescale (see more details in Appendix~\ref{apdx:fenv}).

Feedback therefore slows the depletion of the envelope and allows the system to survive for a longer time. At the same time, the lower equilibrium accretion rate suppresses BH growth. Stronger feedback therefore produces a longer-lived envelope but a less massive BH at the final stage.

\begin{figure}
    \centering
    \includegraphics[width=0.95\linewidth]{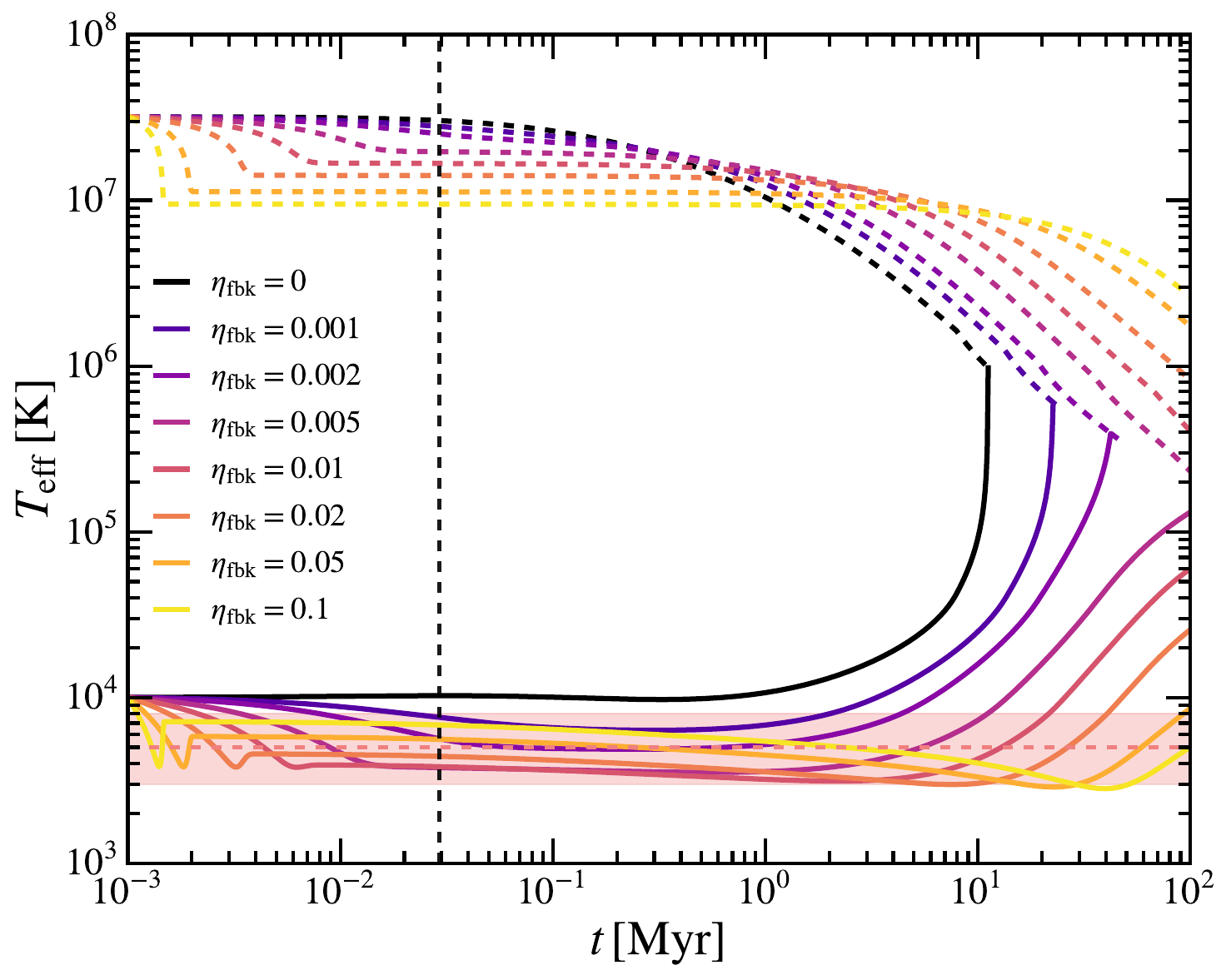}
    \caption{Similar to Figure~\ref{fig:Teff_acc}, but shown for the feedback-involved models with fixed $f_{\rm env}=0.5$ and different feedback efficiencies $\eta_{\rm fbk}$. Feedback rapidly inflates the envelope and drives $T_{\rm eff}$ toward the Hayashi track, largely erasing the memory of the initial conditions. Stronger feedback then prolongs this feedback-regulated, low-temperature phase and delays the eventual transition to the exposed-disk branch. }
    \label{fig:Teff_fbk}
\end{figure}

The resulting evolution of the effective temperature is shown in Figure~\ref{fig:Teff_fbk}. It is worth noting that once even a modest amount of feedback heating is included, the envelope expands and the memory of the initial conditions is rapidly erased. The effective temperature then drops toward the Hayashi track, where the steep temperature dependence of the opacity regulates the photosphere. Thereafter, the system remains near the Hayashi-track temperature for an extended period, until the envelope is finally close to depletion and $T_{\rm eff}$ rises again toward the exposed-disk branch $T_{\rm disk}$. As the BH continues to grow while the accretion rate declines, $T_{\rm disk}$ subsequently decreases, producing the downward evolution along the disk branch. A larger $\eta_{\rm fbk}$ delays this final upturn by prolonging the feedback-regulated envelope phase. 

\section{Evolution of the spectrum}\label{sec:SED}

\subsection{SED modeling}
We next convert the dynamical tracks into mock spectra to follow how an embedded LRD-like source evolves toward a bluer broad-line AGN phase. At each output time, the total SED is written as the sum of the BHE and surrounding stellar components,
\begin{equation}
    L_{\nu}^{\rm tot}(t)=L_{\nu}^{\rm BH}(t)+L_{\nu}^{\star}(t).
\end{equation}
The BH-related quantities entering the spectral calculation ($M_{\rm BH}$, $T_{\rm eff}$, $L_{\rm Edd}$, $\dot{M}_{\rm BH}$, and $\dot{M}_{\rm Edd}$) are taken directly from the evolutionary tracks described in Section~\ref{sec:method}. The additional parameters adopted for the SED modeling are summarized in Table~\ref{table:seds}.

\begin{table}
    \centering
    \caption{Model parameters in the SED modeling.}
    \addtolength{\tabcolsep}{0.2pt}
    \def\arraystretch{1.2}
    \begin{tabular}{lcc}
        \hline
        Parameter & Description & Fiducial value \\
        \hline
        \hline
        $N_{\rm H}$ & Hydrogen column density & $10^{23}\cm^{-2}$ \\
        $\Delta_{\rm sm}$ & Transition smoothness & $1\,\rm dex$ \\
        $\rm SFR_{0}$ & Nuclear star formation rate & $1\msun/\yr$ \\
        $A_{\rm V}$ & Dust attenuation & $0.1 \,\rm mag$ \\
        
        \hline
        \end{tabular}
    \label{table:seds}
\end{table}

\subsubsection{BHE component}
The BHE spectrum is modeled as thermal emission from the reprocessing photosphere. We adopt a single blackbody spectrum as a simplified description of the emergent continuum. This approximation is sufficient for our purpose of capturing the broad-band spectral evolution. Both theoretical radiative-transfer calculations~\citep[e.g.,][]{Liu2026} and empirical modeling~\citep[e.g.,][]{Naidu2025,Kokorev2026,Sun2026,deGraaff2026,Merida2026b,Merida2026c} have shown that a blackbody provides a good leading-order description of the overall continuum from optically thick black hole envelopes, while the spectra of some sources exhibit higher-order features, including deviations from the pure blackbody shape and individual emission and absorption features. These additional features are not modeled here; instead, we refer the reader to the synthetic spectral library of \citet{Liu2026}.

At each time step, we assume that the emergent luminosity is set by the Eddington luminosity of the instantaneous BH mass,
\begin{equation}
    L_{\rm src}(t)=L_{\rm Edd}[M_{\rm BH}(t)],
\end{equation}
and that the spectral shape is a blackbody with temperature $T_{\rm eff}(t)$. The blackbody spectrum is normalized to conserve the bolometric luminosity,
\begin{equation}
    L_{\nu}^{\rm BH}(t)
    =
    L_{\rm src}(t)
    \frac{\pi B_\nu[T_{\rm eff}(t)]}
   {\sigma_{\rm SB}T_{\rm eff}^4}.
\end{equation}

At late times, when the envelope temperature approaches that of an exposed accretion disk, the emergent spectrum is no longer expected to be well described by a single-temperature blackbody.
In a standard disk, where radiative cooling and viscous heating are balanced (e.g., $T_{\rm eff}\propto r^{-3/4}$), the spectrum is well described with a multicolor blackbody spectrum with $L_\nu\propto\nu^{1/3}$, or equivalently $L_\lambda\propto\lambda^{-7/3}$ over the characteristic intermediate-frequency regime~\citep[][]{Shakura1973}. Observationally, unobscured quasars exhibit even redder UV continua, with a typical slope of $\beta_{\rm UV}\simeq-1.5$~\citep[e.g.,][]{VandenBerk2001,Davis2007,Lusso2015}. The resulting UV--optical continuum during the LBD phase is therefore expected to be redder than the single-blackbody spectrum adopted here. However, given the substantial uncertainty in the intrinsic AGN spectrum, both the timing and extent of the transition from an envelope-dominated photosphere to a multicolor accretion-disk spectrum remain uncertain and cannot be determined within our simplified BHE model. We therefore do not explicitly incorporate this spectral transition, but caution that our predicted continua may be systematically bluer than those of fully exposed AGNs, once entering the LBD regime.

\subsubsection{Additional absorber}

As discussed by \citet{Inayoshi2025a}, dense gas surrounding an AGN can imprint a prominent Balmer break on the transmitted spectrum.
This is a key spectral feature for reproducing spectral slopes of some LRDs with a deep Balmer break. 
To capture the overall continuum shape of the gas-attenuated AGN spectrum, we adopt the following analytic transmission function to describe the continuum modification by dense gas absorption:
\begin{equation}
\mathcal{T}_{\rm Bal,0}(\lambda)=
\begin{cases}
\exp\!\left[-\left(\dfrac{\lambda}{\lambda_0}\right)^{s_1}
           -\left(\dfrac{\lambda}{\lambda_1}\right)^{s_2}\right],
& \lambda < \lambda_{\rm B,lim},\\[8pt]
s_3\,\tanh\!\left[s_4\left(\dfrac{\lambda}{\lambda_{\rm B,lim}}-1\right)\right],
& \lambda \ge \lambda_{\rm B,lim},
\end{cases}
\end{equation}
where $\lambda_{\rm B,lim}=3646$~{\AA} is the Balmer limit. The parameters $\lambda_0$ and $\lambda_1$ characterize the continuum attenuation redward of Ly$\alpha$ and blueward of the Balmer limit, respectively, while $s_4$ controls the sharpness of the Balmer break. The best-fit parameters for a hydrogen column density of
$N_{\rm H}=10^{23}\,\mathrm{cm^{-2}}$ and a covering fraction of
$50\%$ are
$\lambda_0=1293$~{\AA},
$\lambda_1=3965$~{\AA},
$s_1=-26.00$,
$s_2=2.337$,
$s_3=0.9905$, and
$s_4=34.70$
(see the Appendix of \citealt{Inayoshi2026a} for more details).

Strong Balmer-break absorption associated with such high-density gas is not required to reproduce the observed spectra of LBDs or normal broad-line AGNs,
possibly because their circumnuclear gas becomes less dense and/or has a lower covering fraction as the gas supply declines. Since the physical origin and evolution of the dense absorbing gas remain under debate, we adopt a simple phenomenological prescription in which the strength of the Balmer-break absorption is tied to the accretion state via a sigmoid function:
\begin{equation}
    w_{\rm Bal}(t)=\frac{1}{1+\exp(-x_{\rm Bal})},
\end{equation}
where
\begin{equation}
    x_{\rm Bal}=
    \frac{\log_{10}(\dot{m}_{\rm BH}/\dot{m}_{\rm crit})
}
    {\Delta_{\rm sm}},
\end{equation}
with $\dot{m}_{\rm crit}=1$ and $\Delta_{\rm sm}=1$, where the former sets the midpoint of the transition and the latter controls its smoothness. The resulting transmission curve is
\begin{equation}
    \mathcal{T}_{\rm Bal}(\lambda,t)
    =
    1-w_{\rm Bal}(t)
    \left[1-\mathcal{T}_{\rm Bal,0}(\lambda)\right].
\end{equation}
When the BH accretes at super-Eddington rates, $w_{\rm Bal}\simeq1$ and the full Balmer-break attenuation is applied. As the accretion rate declines below the Eddington limit, $w_{\rm Bal}$ smoothly approaches zero, mimicking the gradual disappearance of the dense absorbers. We note that this phenomenological treatment is introduced primarily to account for the uncertain nature of absorbing gas and to reproduce the continuum spectral shape of a subset of the observed LRD population. Nevertheless, as demonstrated in Appendix~\ref{apdx:BB}, our main conclusions are insensitive to the specific treatment of the absorber, including the transition smoothness and absorber column density.

Finally, for illustrative purposes, we apply complete absorption blueward of Ly$\alpha$ to mimic strong intergalactic medium (IGM) absorption at high redshifts,
\begin{equation}
    \mathcal{T}_{\rm Ly\alpha}(\lambda)=
    \begin{cases}
    0, & \lambda<\lambda_{\rm Ly\alpha},\\
    1, & \lambda\ge\lambda_{\rm Ly\alpha},
    \end{cases}
\end{equation}
where $\lambda_{\rm Ly\alpha}=1215.67${\AA}. We emphasize that this is included only for visualization and does not affect any of the subsequent calculations or conclusions.

\subsubsection{Stellar component}
For the stellar component, we adopt a simple nuclear star formation history tied to the same gas supply that feeds the envelope. We set the initial star formation rate to ${\rm SFR}_0=1\,\msun\,{\rm yr^{-1}}$, corresponding to the characteristic UV luminosity of LRDs, $M_{\rm UV}\simeq-19$~\citep[e.g.,][]{Kocevski2025}. This value can also be derived from the expected ISM inflow rates predicted by disk formation models~\citep[][]{Mo1998} driven by intense cold-stream accretion onto high-redshift galaxies~\citep[][]{Inayoshi2026}.

The star formation rate is then allowed to decay according to the same feedback-regulated gas retention prescription adopted in the dynamical model, as
\begin{equation}
    {\rm SFR}(t)={\rm SFR}_0 \exp(-t/t_{\rm fb}).
\end{equation}

We generate the stellar spectrum from the tabulated star formation history using low-metallicity stellar population templates with $Z=Z_\odot/50$, a Salpeter initial mass function~\citep{Salpeter1955} over $1$--$100\msun$, and modest dust attenuation of $A_{\rm V}=0.1$ following a Small Magellanic Cloud (SMC) attenuation curve~\citep[][]{Gordon2003}. We compute the nebular continuum and emission-line luminosities using precomputed {\tt CLOUDY} photoionization grids~\citep{Byler2017}, assuming that all stellar ionizing photons beyond the Lyman limit are absorbed and reprocessed into lines (i.e., an escape fraction of $f_{\rm esc}=0$). However, we do not model line emission reprocessed by the BHE; therefore, the strengths and profiles of the predicted emission lines should not be interpreted quantitatively.

\subsection{SED evolution}
\begin{figure*}
    \centering
    \includegraphics[width=1\linewidth]{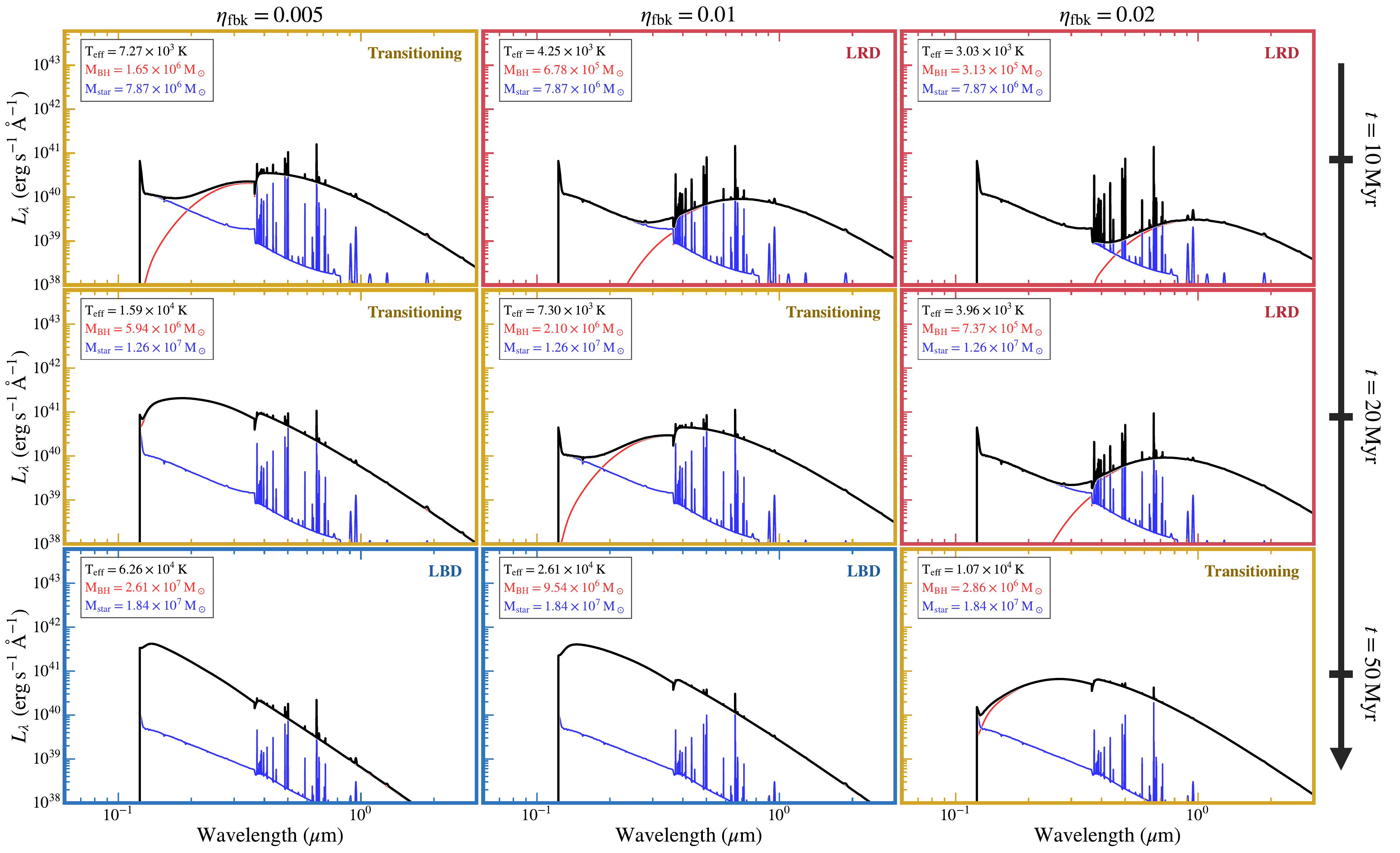}
    \caption{
    Rest-frame SEDs predicted by models with $f_{\rm env}=0.5$ for three feedback efficiencies, $\eta_{\rm fbk}=0.005$, $0.01$, and $0.02$ (left to right), at evolutionary times of $t=10$, 20, and 50 Myr (top to bottom). The spectra show the nuclear stellar component (blue), BHE emission (red), and their sum (black). Labels indicate the envelope effective temperature, BH mass, and stellar mass at each snapshot. The frame colors indicate the corresponding evolutionary stages, classified according to the continuum-slope criteria of \citet{Asada2026}. During the early phase, the cool photosphere with $T_{\rm eff}\simeq5000\K$, together with gas attenuation, produces an LRD-like spectrum. The weakest-feedback model has already left the LRD phase by the first snapshot because of its shorter evolutionary timescale. Stronger feedback keeps the envelope inflated for longer, delaying the fading of the red optical component and the transition toward a bluer, exposed broad-line AGN-like state.
    }
    \label{fig:SEDs}
\end{figure*}

\begin{figure*}
    \centering
    \includegraphics[width=1\linewidth]{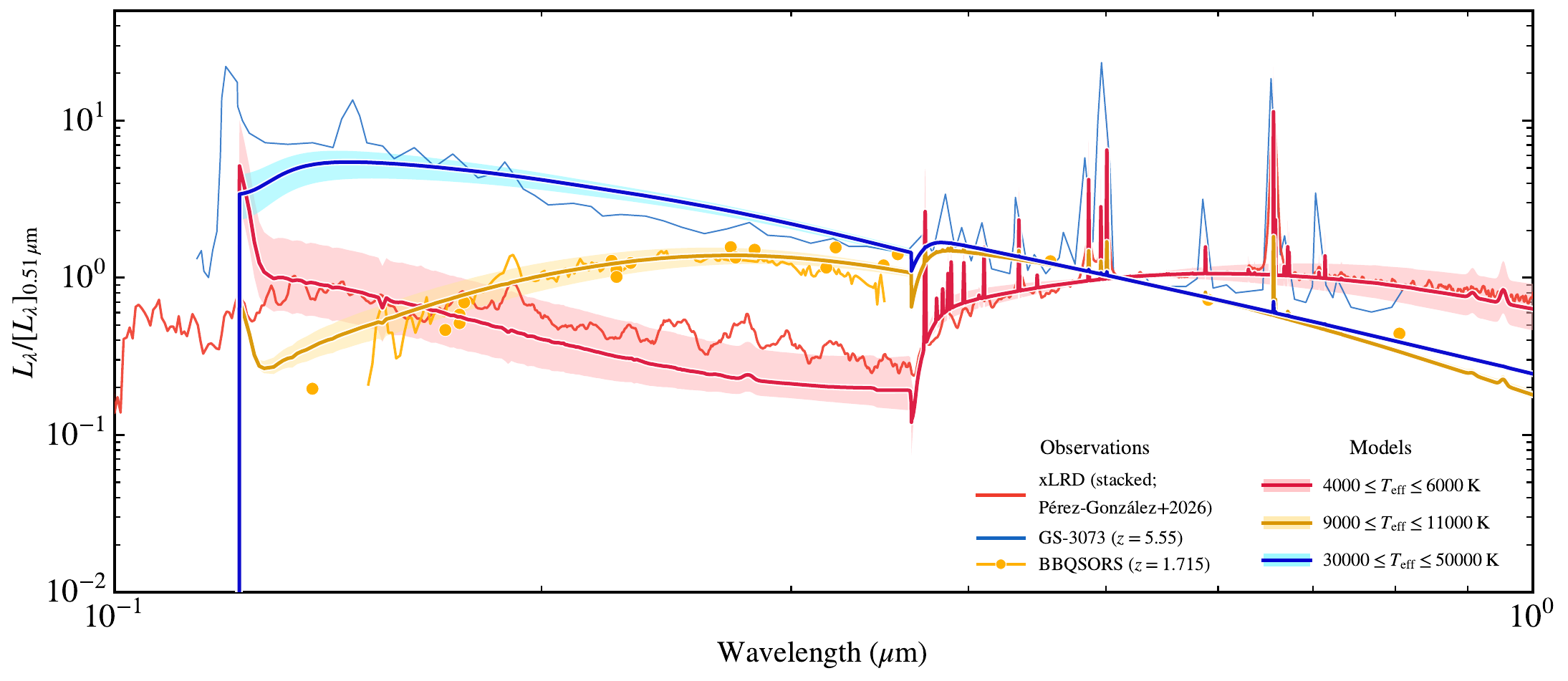}
    \caption{
    Rest-frame spectral energy distributions normalized at
    $0.51\,\mum$. The thick red, gold, and blue curves show the median model SEDs for $4000 \leq T_{\rm eff} \leq 6000\,{\rm K}$, $9000 \leq T_{\rm eff} \leq 11000\,{\rm K}$, and $30000 \leq T_{\rm eff} \leq 50000\,{\rm K}$, respectively, with the shaded regions enclosing the 16th--84th percentiles. The model adopts $f_{\rm env}=0.5$ and $\eta_{\rm fbk}=0.01$. 
    We compare these predictions with observations spanning the proposed LRD--LBD evolutionary sequence, including the stacked SED of extreme LRDs ($L_{5100\text{\AA}}/L_{2500\text{\AA}}>6.3$) from \citet{PG2026}, the Subaru/PFS spectrum and compiled UV--optical photometry of BBQSORS at $z=1.715$ as a candidate transitional object~\citep{Zhong2026}, and GS-3073 at $z=5.55$ as a representative LBD~\citep{Ubler2023,Brazzini2026}.
}
    \label{fig:observations}
\end{figure*}

Figure~\ref{fig:SEDs} shows the spectral evolution of the BHE model for three representative feedback efficiencies, $\eta_{\rm fbk}=0.005$, $0.01$, and $0.02$, at ages of 10, 20, and 50 Myr. 
The envelope retention efficiency is fixed at $f_{\rm env}=0.5$. Despite different feedback strengths, all models follow qualitatively similar evolutionary sequences where the feedback efficiency primarily regulates the evolutionary timescale.
The evolution is driven by gradual heating of the reprocessing envelope as it contracts. 
At early times, the BH remains embedded within a massive, optically thick envelope with an effective temperature of $T_{\rm eff} \simeq 5\times10^3~\K$. The young stellar population still contributes significantly to the UV continuum, whereas thermal emission from the cool envelope dominates the rest-frame optical continuum.

As the envelope contracts, the effective temperature increases to $\sim 10^4~\K$, shifting the flux peak toward shorter wavelengths ($\sim 3000~{\rm \AA}$). 
The envelope emission progressively outshines the stellar component, producing a spectrum that approaches a single blackbody while driving the UV continuum toward redder slopes, as the UV wavelengths sample the exponentially declining Wien tail of the blackbody spectrum.
Once the envelope becomes sufficiently depleted, its temperature approaches that of the accretion disk ($\sim 3\times 10^4-10^5~\K$). 
The flux peak then moves blueward of the Lyman break, leaving only the Rayleigh--Jeans tail at the UV--optical wavelengths. 
As a result, the continuum evolves from a thermal spectrum to approximately a single power-law shape.

As discussed in Section~\ref{subsec:fbk}, increasing the feedback efficiency $\eta_{\rm fbk}$ injects more energy into the envelope, inflating the photosphere and reducing the BH accretion rate through the larger virial radius. 
As a result, stronger feedback maintains a cooler and more extended envelope for a longer time, delaying the disappearance of the red optical continuum and the subsequent emergence of the hotter spectrum (see the middle row at $t=20~{\rm Myr}$ in Figure~\ref{fig:SEDs}).

Figure~\ref{fig:observations} compares modeled LRD spectra along the fiducial~\footnote{Here, ``fiducial'' refers only to representative values near the middle of the parameter ranges explored, rather than a preferred or best-fit model.} evolutionary track ($\eta_{\rm fbk}=0.01$ and $f_{\rm env}=0.5$) with observed spectra of several broad-line AGNs that span a sequence of AGN evolution. Here, all the spectra are normalized at rest-frame $0.51\,\mum$. The predicted continuum evolution provides a unified explanation for the observed diversity of compact AGN systems:
At the early stage ($t\simeq10$--$20\Myr$), the cool BHE maintains $T_{\rm eff}\simeq(4$--$6)\times10^3\,\K$, producing a continuum that, together with the compact stellar component, closely reproduces the characteristic SED of LRDs~\citep[e.g., the stacked LRD spectrum of][]{PG2026}.

At the intermediate stage ($t\simeq25-30\Myr$), the envelope heats to $T_{\rm eff}\sim10^4\K$, producing a nearly pure blackbody continuum spectrum. Remarkably, the predicted SED closely resembles a quasar at $z=1.715$ \citep[BBQSORS,][]{Zhong2026}, whose continuum spectrum is well fitted by a $\sim10^4\K$ blackbody and was proposed as a transitional object between LRDs and unobscured quasars. Our model reproduces both the Subaru/PFS spectrum and the UV--optical photometry, naturally explaining the higher blackbody temperature as expected from the contraction of the BHE. 
The agreement provides independent support for the presence of an intermediate evolutionary stage between LRDs and LBDs. The apparent rarity of such blackbody-dominated sources may reflect the relatively rapid passage through this particular stage of the transition, which we will discuss in Section~\ref{subsec:time}.

Eventually, the envelope becomes sufficiently depleted ($t\gtrsim50\Myr$) so that the continuum is dominated by the exposed accretion disk. Adopting the shallow $\beta=-7/3$ slope expected for a multicolor blackbody accretion disk at this late evolutionary stage, the predicted spectrum closely resembles that of GS-3073~\citep[][]{Ubler2023,Brazzini2026}, whose blue continuum is also characteristic of unobscured quasars.

The quantitative comparison suggests that the observed LRD, transitional, and LBD populations may represent different stages of a common BHE evolutionary sequence, providing a coherent physical picture for the observed diversity of broad-line emitters.

\begin{figure}
    \centering
    \includegraphics[width=0.95\linewidth]{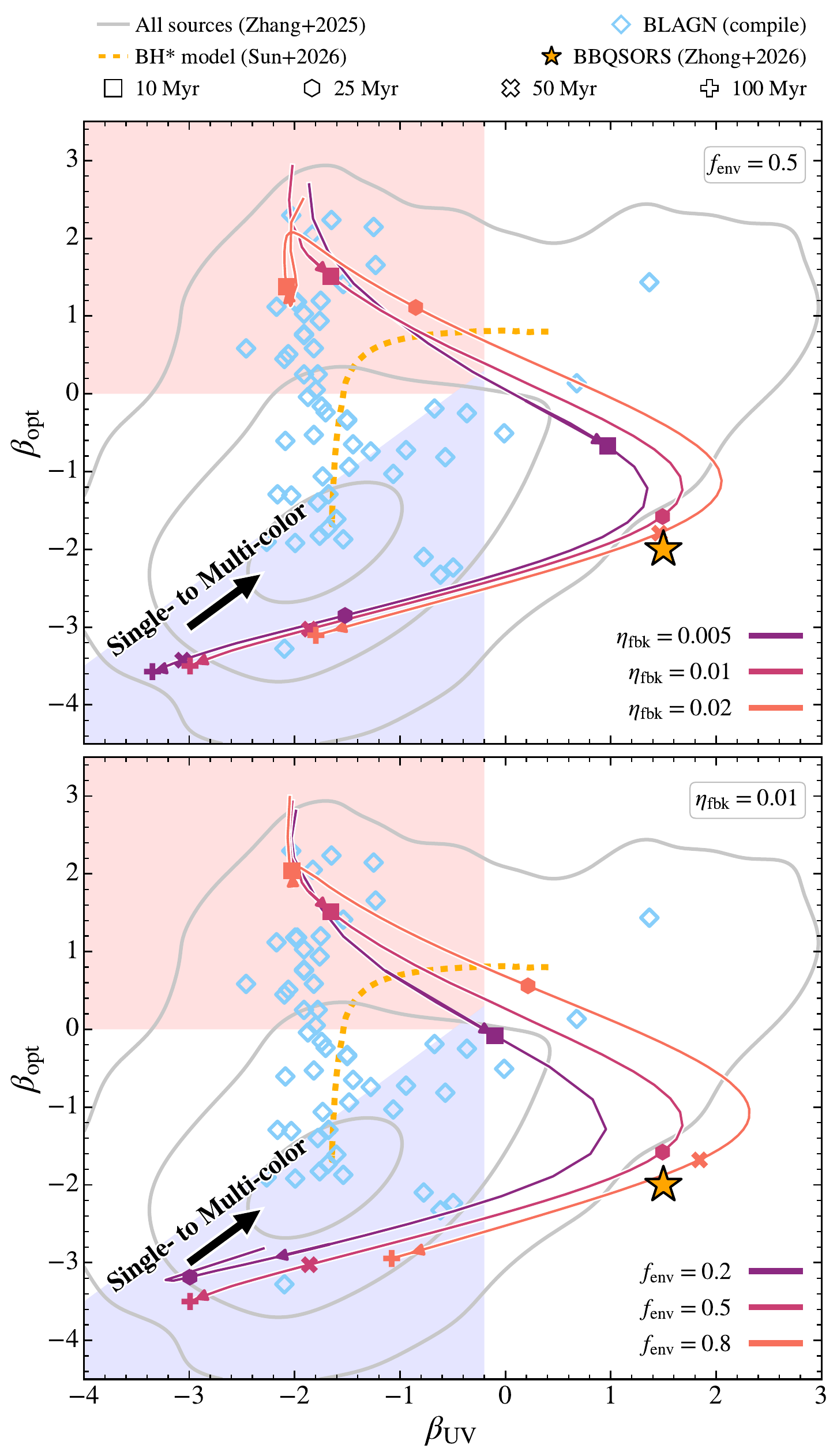}
    \caption{
    Evolutionary tracks in the $(\beta_{\rm UV},\beta_{\rm opt})$ plane, with symbols marking the labeled evolutionary times. The upper and lower panels vary the feedback efficiency $\eta_{\rm fbk}$ and envelope retention efficiency $f_{\rm env}$, respectively. The LRD and LBD regions are shaded in red and blue following \citet{Asada2026}, with arrows indicating uncertainties in the LBD spectral prescription. For comparison, we show $z>2$ JWST/NIRSpec PRISM sources from \citet{Zhang2025}, broad-line emitters from \citet{Hainline2024}, BBQSORS~\citep{Zhong2026}, and the BH* model of \citet{Sun2026} with varying BH-to-stellar luminosity ratios. The predicted tracks follow nearly identical trajectories, exhibiting a characteristic red excursion toward $(\beta_{\rm UV},\beta_{\rm opt})\sim(2,-1)$ before turning toward the LBD locus.
    }
    \label{fig:betas}
\end{figure}

\subsection{Evolution trajectory}

To quantify the spectral evolution, we measure the UV and optical continuum slopes by fitting $F_\lambda \propto \lambda^\beta$.
Figure~\ref{fig:betas} shows the corresponding evolutionary tracks in the $(\beta_{\rm UV},\beta_{\rm opt})$ plane for different model parameters.
Each symbol on the evolutionary track indicates the elapsed time since the initial state. 
For comparison, we overplot the distribution of all $z>2$ sources with JWST/NIRSpec PRISM spectra from \citet{Zhang2025}, together with compiled broad-line emitters from \citet{Hainline2024} and BBQSORS \citep{Zhong2026}.
Following \citet{Asada2026}, we define LRDs by
\begin{itemize}
    \item[1.] $\beta_{\rm UV}<-0.2$,
    \item[2.] $\beta_{\rm opt}>0$,
    \item[3.] $\beta_{\rm opt}-\beta_{\rm UV}>0.5$,
\end{itemize}
while LBDs satisfy
\begin{itemize}
    \item[1.] $\beta_{\rm UV}<-0.2$,
    \item[2.] $\beta_{\rm opt}-\beta_{\rm UV}<0.5$,
\end{itemize}
where $\beta_{\rm UV}$ and $\beta_{\rm opt}$ are the rest-frame UV and optical slopes over the wavelength ranges of rest-frame $1250$--$3600$\,{\AA} and $3600$--$7000$\,{\AA}, respectively.
The corresponding LRD and LBD regions are highlighted by the red and blue shaded areas, respectively.

Across the parameter space explored, all models follow nearly the same trajectory in the $(\beta_{\rm UV},\beta_{\rm opt})$ plane. The evolution begins in the LRD regime. As the envelope contracts and heats, the blackbody peak shifts from the optical toward the UV band, causing $\beta_{\rm opt}$ to decrease continuously. Before the flux peak moves blueward of the Lyman break, it passes through the observed UV window, producing a pronounced UV bump that drives $\beta_{\rm UV}$ toward increasingly positive values. During this phase, the evolutionary tracks naturally pass through the location of BBQSORS~\citep[][]{Zhong2026}, $(\beta_{\rm UV},\beta_{\rm opt})\sim(1.5,-2)$. As the envelope continues to heat, the peak of the blackbody spectrum (the Wien peak) shifts blueward of the Lyman break, leaving only the Rayleigh–Jeans tail in the observed UV–optical window. Consequently, $\beta_{\rm UV}$ turns blue again, and the tracks converge in the observed LBD region, marking the transition from an envelope-dominated system to an exposed accretion-disk-dominated AGN.

Although the evolutionary trajectory is largely insensitive to the adopted model parameters, the transition timescale depends strongly on both the feedback efficiency and the envelope retention efficiency. In our fiducial model ($\eta_{\rm fbk}=0.01$, $f_{\rm env}=0.5$), the system remains in the LRD regime for $\sim15$ Myr. It then spends $\sim20$ Myr in a transitional phase, defined as the period during which the system satisfies neither the LRD nor the LBD criteria, before finally entering the LBD regime.
Increasing either $\eta_{\rm fbk}$ or $f_{\rm env}$ prolongs both the LRD lifetime and the duration of the transitional phase by maintaining a more massive, cooler envelope, thereby delaying the emergence of the exposed accretion disk.

We note that a population of sources occupies the region between the LRD and LBD loci with $\beta_{\rm UV}\simeq-2$, which is not traversed by the fiducial evolutionary tracks presented here. However, this does not pose a challenge to the BHE framework. As demonstrated by \citet{Sun2026}, such systems can be reproduced naturally by varying the BH-to-stellar luminosity ratio, with a larger stellar contribution producing a bluer optical continuum (shown in the orange dashed lines).
In the present work, we fix the nuclear star formation history using a simple semianalytic prescription in which cold-stream accretion fuels a nuclear starburst with a constant star formation efficiency (i.e., the fraction of gas converted into stars during one local free-fall time). In reality, the star formation efficiency is likely to exhibit substantial diversity depending on the host halo mass, gas density, and other environmental or structural properties~\citep[e.g.,][]{Shen2026,Wang2025-SFE}. Allowing this additional degree of freedom would naturally broaden the predicted evolutionary tracks toward these populations without altering the underlying BHE evolutionary picture (see Appendix~\ref{apdx:BB}).

\section{Discussion}\label{sec:disc}

\begin{figure*}
    \centering
    \includegraphics[width=1\linewidth]{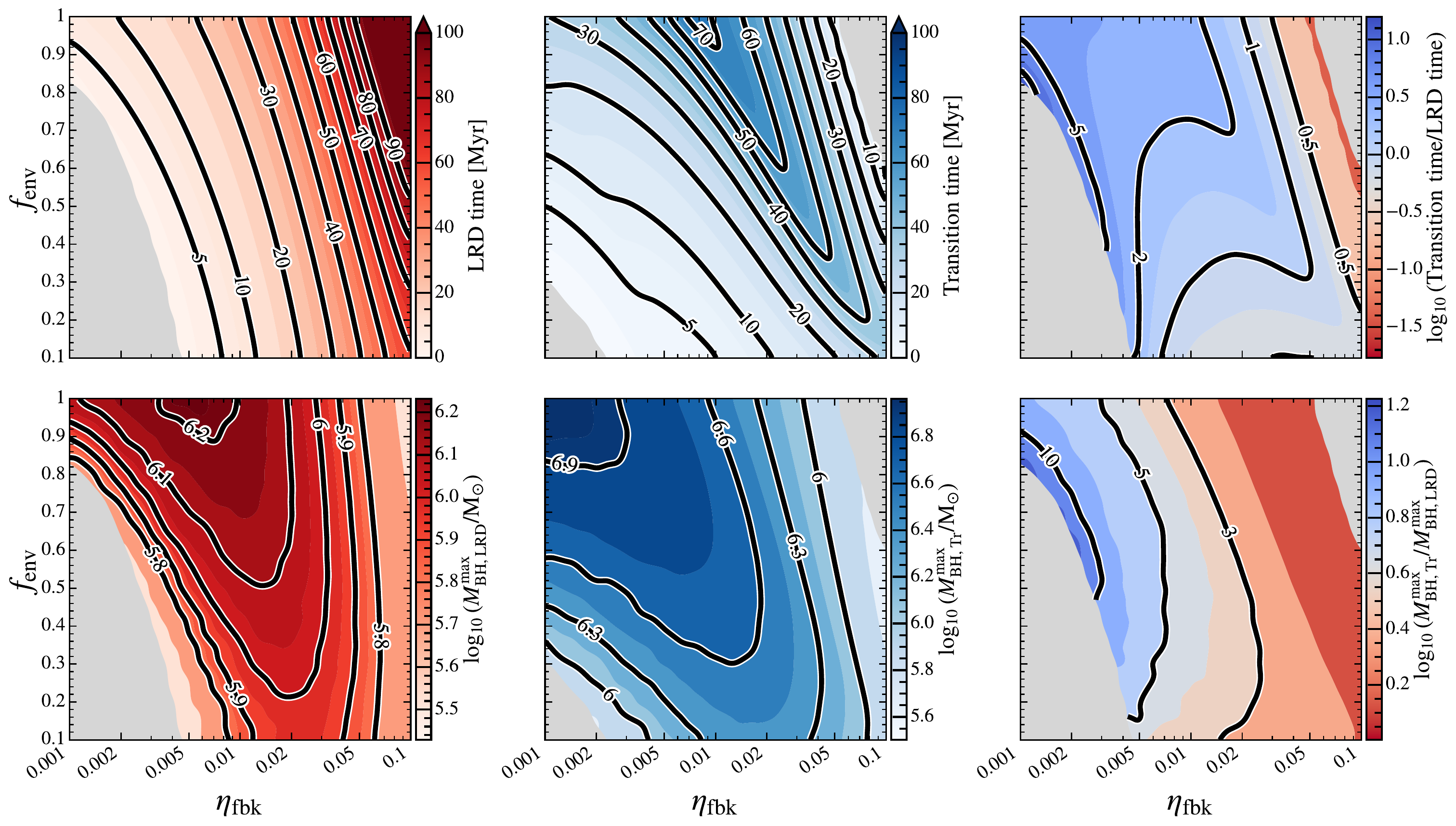}
    \caption{
    Effects of the model parameters on the evolutionary timescales and BH masses. \textit{Upper:} the time spent in the LRD phase, the time spent in the transitional phase, and the ratio of the transitional to LRD lifetimes, respectively. \textit{Lower:} the maximum BH mass during the LRD phase ($M^{\rm max}_{\rm BH,\,LRD}$), the maximum BH mass during the transitional phase ($M^{\rm max}_{\rm BH,\,Tr}$), and their ratio, respectively. Evolutionary stages are defined according to the locations of the models in the $(\beta_{\rm UV},\beta_{\rm opt})$ plane. Colored contours encode the measured quantity, with contour lines indicating the corresponding values. Gray regions denote models that do not pass through the corresponding evolutionary stage.
    }
    \label{fig:grids}
\end{figure*}

\subsection{Transition timescale and abundance}\label{subsec:time}
As shown above, the time spent in each evolutionary stage depends sensitively on the underlying BHE physics. This suggests that the relative cosmic abundance of LRDs, transitional systems, and LBDs provides a direct observational probe of the physical processes governing BH growth and envelope evolution.
To this end, we perform a broader parameter survey to quantify how the characteristic lifetimes of different evolutionary stages depend on the model parameters and discuss the resulting implications for the observed abundances of the three populations.

We sample $f_{\rm env}$ from 0.1 to 1 using 50 linearly spaced values and $\eta_{\rm fbk}$ from $10^{-3}$ to $0.1$ using 50 logarithmically spaced values, yielding a total of 2,500 model realizations.
In the upper panels of Figure~\ref{fig:grids}, we show, from left to right, the time spent in the LRD phase, the time spent in the transitional phase, and the ratio of the transitional to LRD lifetimes across the parameter space. Gray areas denote models that do not experience the corresponding phase.

The LRD lifetime depends on both the feedback efficiency and the envelope retention efficiency\footnote{In addition, the characteristic lifetime should depend on the external ISM supply, particularly through the gas-retention timescale $t_{\rm fb}$, which is inherited from the environment of the LRD host halo~\citep{Inayoshi2026}.}, increasing from $\sim5$ Myr to $\gtrsim50$ Myr toward higher $\eta_{\rm fbk}$ and $f_{\rm env}$. At $\eta_{\rm fbk}\gtrsim0.05$, feedback heating becomes strong enough to balance radiative cooling and maintain the system in the LRD phase throughout the entire simulation time of 100 Myr.
The LRD lifetimes shown in the upper-right corner should be regarded as lower limits.

The duration of the transitional phase shows a similar dependence, increasing from $\sim5$ Myr to $\gtrsim50$ Myr toward higher envelope retention efficiencies and intermediate feedback efficiencies. At sufficiently strong feedback, however, the prolonged LRD phase delays or prevents the system from reaching the transitional phase.

The present calculations follow only a single gas-accretion episode, terminating once the envelope is depleted and the system enters the LBD phase. The subsequent evolution is expected to depend on long-term gas replenishment and recurrent accretion episodes, which are beyond the scope of the current model. We therefore do not attempt to predict the lifetime of the LBD phase.

The evolutionary timescales can be directly translated into an upper limit on the cosmic abundance by introducing a duty cycle,
\begin{equation}
    f_{\rm duty}\equiv\frac{\Delta t}{t_{\rm H}(z)},
\end{equation}
defined as the fraction of the Hubble time $t_{\rm H}(z)$ that a system spends in a given evolutionary stage $\Delta t$. The corresponding number density is then
\begin{equation}
    \phi=\phi_{\rm h}(M_{\rm vir})\,f_{\rm duty}
    =\phi_{\rm h}(M_{\rm vir})\,\frac{\Delta t}{t_{\rm H}(z)},
\end{equation}
where $M_{\rm vir}$ is the halo virial mass and $\phi_{\rm h}(M_{\rm vir})$ denotes the number density of halos capable of hosting BHE systems.
The high ISM inflow rates required to sustain super-Eddington accretion, together with sufficiently long gas-retention timescales, favor host halos with virial temperatures of $T_{\rm vir}\simeq(1$--$2)\times10^6\K$~\citep[][]{Inayoshi2026}, corresponding to masses of $M_{\rm h}\simeq6\times10^{10}$--$3\times10^{11}\msun$ at $z\simeq4$--$6$, a range broadly consistent with current clustering constraints on LRD host halos~\citep[e.g.,][]{Lin2025,Zhang2026}. The corresponding halo abundance is
$\phi_{\rm h}\simeq4\times10^{-4}$--$5\times10^{-3}\,\mathrm{Mpc^{-3}}$.
An LRD lifetime of $t_{\rm LRD}\simeq20$ Myr corresponds to
$f_{\rm duty}\simeq0.01$, yielding
$\phi_{\rm LRD}\simeq4\times10^{-6}$--$5\times10^{-5}\,\mathrm{Mpc^{-3}}$.
Such a range is broadly consistent with the observed abundance,
$\phi_{\rm LRD}\simeq6\times10^{-5}\,\mathrm{Mpc^{-3}}$,
suggesting a feedback efficiency of $\eta_{\rm fbk}\sim\mathcal{O}(10^{-2})$.

Assuming a statistical steady state, in which BHE systems are continuously formed and evolve through the same sequence at an approximately constant birth rate, the relative abundance of different populations is proportional to the time spent in each evolutionary stage. The ratio of the transitional to LRD lifetimes therefore indicates the abundance of transitional systems relative to LRDs. The upper-right panel of Figure~\ref{fig:grids} shows this ratio. Across most of the parameter space, the two phases have comparable durations, implying that transitional systems should be nearly as abundant as LRDs.

However, unlike LRDs and LBDs, which occupy relatively concentrated loci in observable parameter space, transitional systems are expected to be distributed over a much broader region, as illustrated in the $(\beta_{\rm UV},\beta_{\rm opt})$ plane. Moreover, as shown in Appendix~\ref{apdx:BB}, a stronger stellar contribution at later times can shift the continuum toward the LBD selection region before the BHE transition is completed. 
Their lack of a distinctive selection locus may therefore explain why such systems have not yet been systematically identified despite their potentially large intrinsic abundance. We therefore expect a substantial population of transitional sources awaiting discovery in current and forthcoming spectroscopic surveys.

\subsection{BH masses in LRDs and LBDs}
Another important consequence of the model parameters is their impact on the BH accretion history and the final BH mass. The lower panels of Figure~\ref{fig:grids} illustrate this dependence.

The maximum BH mass during the LRD phase is determined by the competition between the accretion rate and the duration of the growth phase. At low feedback efficiencies, BH accretion proceeds efficiently, but the LRD phase is short-lived, allowing the BH to grow by less than $\sim1$ dex before the transition to an LBD. At the opposite extreme, strong feedback substantially prolongs the LRD lifetime but suppresses the accretion rate by inflating the envelope and reducing the gas inflow onto the BH, again resulting in only modest BH growth. Consequently, the largest BH masses are achieved at intermediate feedback efficiencies, $\eta_{\rm fbk}\sim\mathcal{O}(10^{-2})$, and high envelope retention efficiencies ($f_{\rm env}\gtrsim0.5$), producing maximum LRD-phase BH masses of $\sim10^{6}\msun$.

Such a mass range is consistent with BH masses inferred from the observed bolometric luminosities of LRDs within the BHE framework, assuming an Eddington ratio of $\lambda_{\rm Edd}=1$ ($\simeq10^6~\msun$;~\citealt{Umeda2026}), but is lower than those inferred from single-epoch virial estimators and dynamical measurements ($\gtrsim10^7~\msun$; e.g.,~\citealt[][]{Matthee2024,Kocevski2025,Juodzbalis2026b}). Such a discrepancy is often alleviated by invoking electron scattering to broaden the emission lines~\citep[e.g.,][]{Rusakov2026}. In the BHE scenario, however, part of the excess line width can arise naturally from the gravitational potential of the envelope itself. Under our assumption of a virialized BHE, the gas kinematics respond to the enclosed mass of both the BH and the envelope, such that interpreting the line width using the BH mass alone can lead to an overestimate of the virial BH mass.

\begin{figure}
    \centering
    \includegraphics[width=0.95\linewidth]{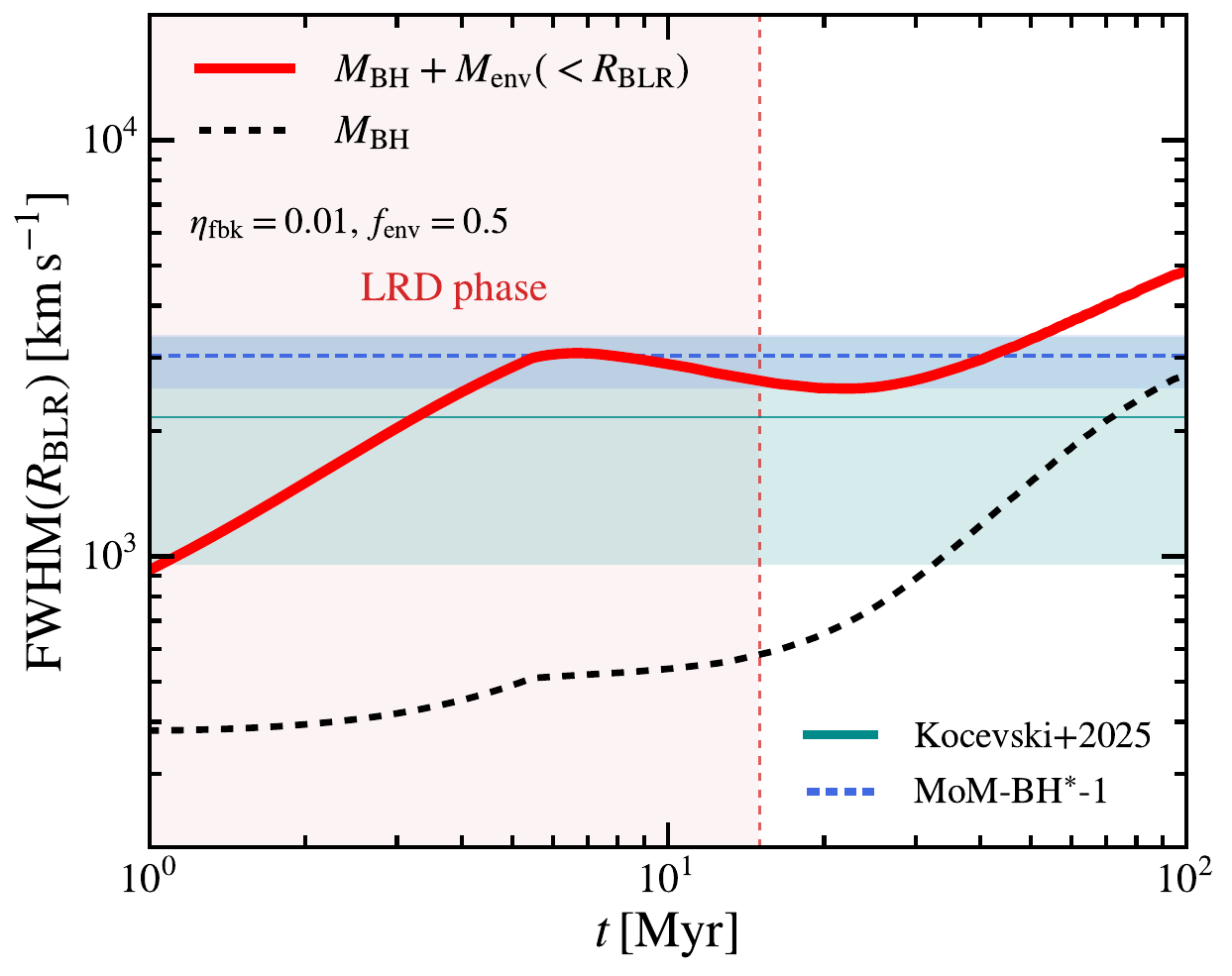}
    \caption{
    Evolution of the FWHM evaluated at $R_{\rm BLR}$ for the fiducial model with $f_{\rm env}=0.5$ and $\eta_{\rm fbk}=0.01$. The red solid and black dashed curves correspond to $M_{\rm dyn}=M_{\rm BH}+M_{\rm env}(<R_{\rm BLR})$ and $M_{\rm dyn}=M_{\rm BH}$, respectively. The solid and dashed horizontal bands indicate the observed broad-line FWHM distribution of the LRD sample from \citet{Kocevski2025} and the measurement of the individual source MoM-BH$^\ast$-1~\citep[][]{Naidu2025}, respectively. The red shaded region marks the LRD phase.
    }
    \label{fig:FWHM}
\end{figure}

To test this possibility, we adopt the empirical radius--luminosity relation obtained from reverberation mapping (RM) observations \citep{Wang2024} to estimate the characteristic location of the broad-line region (BLR),
\begin{equation}
R^{\rm RM}_{\rm BLR}
=2.04\times 10^{-2}\,{\rm pc}
\left(
\frac{L_{5100}}{10^{44}\,{\rm erg\,s^{-1}}}
\right)^{0.42}.
\end{equation}
Since broad-line emission cannot emerge from within the photosphere, we impose $R_{\rm ph}$ as a lower limit on the BLR radius,
\begin{equation}
R_{\rm BLR}
=
\max\left(R_{\rm ph},R^{\rm RM}_{\rm BLR}\right).
\end{equation}
Figure~\ref{fig:FWHM} shows the characteristic full width at half maximum (FWHM) evaluated at $R_{\rm BLR}$. Assuming a spherical distribution of clouds with $f_{\rm vir}=0.75$~\citep[][]{Netzer1990}, one obtains
\begin{equation}
    {\rm FWHM}=\sqrt{\frac{GM_{\rm dyn}}{2f_{\rm vir}R_{\rm BLR}}}.
\end{equation}
We consider two estimates of the dynamical mass: $M_{\rm dyn}=M_{\rm BH}$, corresponding to the conventional BH-dominated potential, and $M_{\rm dyn}=M_{\rm BH}+M_{\rm env}(<R_{\rm BLR})$, which additionally accounts for the gravitational contribution of the enclosed envelope mass. 
During the LRD phase, $M_{\rm env}(<R_{\rm BLR})$ typically accounts for $\sim25$--$100\%$ of the total envelope mass, making its contribution to the dynamical potential substantial.
This increases the predicted virial FWHM by a factor of $\sim2.5$--$6$ relative to the BH-only estimate, reaching $\sim2000$--$3000\,{\rm km\,s^{-1}}$, consistent with the observed broad-line width distribution of LRDs reported by \citet{Kocevski2025} and the measurement for the individual source MoM-BH$^\ast$-1 (${\rm FWHM}=3036^{+361}_{-506}\,{\rm km\,s^{-1}}$; \citealt{Naidu2025}). This agreement suggests that envelope gravity alone can explain the observed line widths without invoking additional electron-scattering broadening. At later times, as the BH grows and the envelope is depleted, the decreasing $M_{\rm env}(<R_{\rm BLR})/M_{\rm BH}$ causes the two predictions to gradually converge.

An important prediction of the BHE scenario is that any individual system evolving along this sequence must exhibit monotonically increasing BH mass, from the LRD phase, through the transitional phase, and finally into the LBD phase. The lower-middle panel of Figure~\ref{fig:grids} shows the maximum BH mass during the transitional phase, which also corresponds approximately to the initial BH mass of the LBD phase. During this stage, the BHE contracts rapidly as the virial radius decreases, leading to substantially more efficient BH accretion than during the LRD phase. Consequently, depending on the adopted model parameters, the BH can grow to $\sim10^{6.5}\msun$ before the envelope is completely dispersed.

This naturally leads to a statistical prediction for the BH mass functions of the two populations: the relative abundance of LBDs should increase toward higher BH masses, whereas LRDs should become progressively less common.
The lower-right panel of Figure~\ref{fig:grids} quantifies this effect by showing the ratio of the maximum BH mass during the transitional phase to that during the LRD phase. This mass offset is itself sensitive to the feedback strength, with weaker feedback producing a larger separation because of the more rapid BH growth during the transitional phase. Our fiducial model predicts a characteristic mass offset of a factor of $\sim3\text{--}5$ between the LRD and LBD phases. Interestingly, preliminary observational studies suggest that the fraction of LRDs decreases toward higher luminosities~\citep[e.g.,][]{Madau2026b}, qualitatively consistent with this prediction.

\subsection{The Hertzsprung--Russell diagram}
\begin{figure*}
    \centering
    \includegraphics[width=1\linewidth]{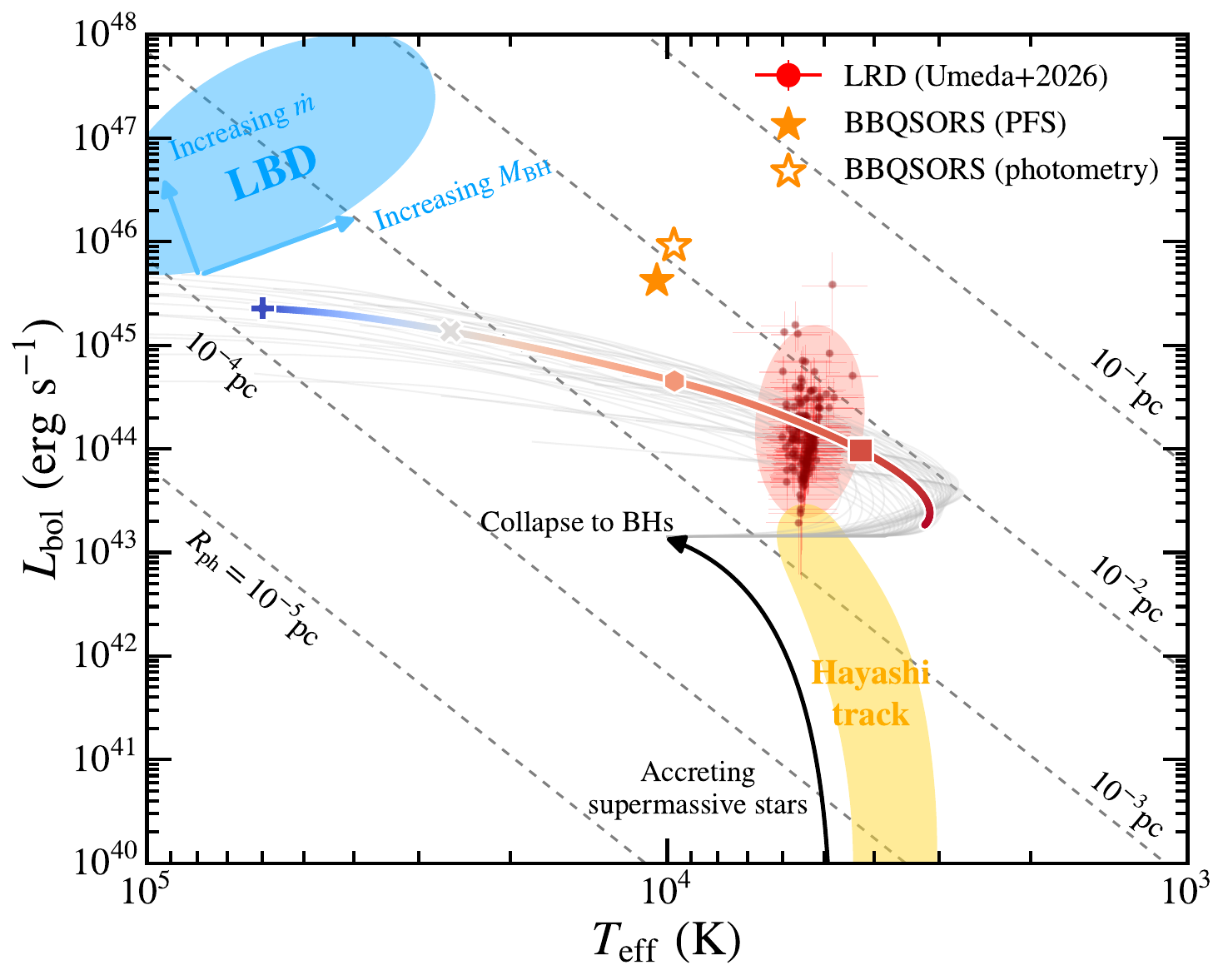}
    \caption{
    BH Hertzsprung--Russell (HR) diagram illustrating the proposed evolutionary sequence linking rapidly growing BH seeds, LRDs, transitional systems, and LBDs. The sequence begins with the formation of a BH seed following the collapse of a supermassive star \citep{Hosokawa2013}. The red shaded region denotes the LRD regime inferred by \citet{Umeda2026}, while the blue shaded region marks the expected location of LBDs, estimated from the accretion-disk temperature given by Equation~\ref{eq:Tdisk}. Arrows in the LBD region indicate the effects of increasing BH mass, $M_{\rm BH}$, and reduced accretion rate, $\dot{m}_{\rm BH}$.
    The colored curve shows the fiducial BHE evolutionary track, with the color indicating the evolutionary age and symbols marking selected epochs. Gray curves illustrate the evolutionary tracks of all BHE models that pass through the LRD phase. Gray dashed lines denote constant photospheric radii, following $L=4\pi\sigma_{\rm SB} R_{\rm ph}^2T_{\rm eff}^4$. The filled and open orange stars indicate the Subaru/PFS spectrum and broadband photometric measurements of BBQSORS, respectively.
    }
    \label{fig:HR}
\end{figure*}

The evolution of a rapidly growing black hole from an embedded LRD to an exposed broad-line AGN closely resembles the life cycle of a very massive star. Motivated by this analogy, we introduce a ``BH HR diagram'' and summarize the BHE evolutionary sequence in Figure~\ref{fig:HR}. The sequence begins with a rapidly accreting supermassive star that evolves near the Hayashi limit~\citep[][]{Hosokawa2013}. Once its core collapses, the newly formed BH seed remains embedded within the massive stellar envelope, producing a quasi-star-like configuration\footnote{Axisymmetric
numerical-relativity simulations found $\gtrsim 90\%$ of the mass of a rapidly rotating supermassive star collapses into a BH once it exceeds several~$\times 10^5~\msun$ \citep{Shibata2002, Shibata2016}. Therefore, a hypothetical gaseous envelope forming a quasi-star-like configuration must be replenished by sustained mass supply from the surrounding gas unless it can be supported by the stellar remnant.}~\citep[][]{Volonteri2005a,Begelman2006,Shang2010}. The first episode of BH activity then gives rise to an LRD phase~\citep{Inayoshi2025First}, during which the photospheric temperature remains close to the Hayashi temperature because of the strong temperature dependence of the envelope opacity (red region; \citealt{Umeda2026}).

The LRD phase terminates once the external gas supply is suppressed and can no longer replenish the envelope. The remaining envelope subsequently undergoes Kelvin--Helmholtz contraction, causing the photospheric radius to decrease and the effective temperature $T_{\rm eff}$ to rise. The gray curves show all available parameter combinations that enter the LRD regime, while the time-colored curve highlights our fiducial model. Despite differences in their detailed evolution, these tracks follow a common sequence: they depart from the LRD region, pass through the location occupied by BBQSORS, and eventually approach the LBD regime (blue region). In this exposed phase, the effective temperature is no longer set by the reprocessing envelope but asymptotically follows the accretion-disk temperature ($T_{\rm disk}$) given by Equation~(\ref{eq:Tdisk}). At a fixed Eddington-scaled accretion rate, increasing $M_{\rm BH}$ raises the bolometric luminosity while lowering $T_{\rm disk}$, moving the source toward the upper right of the diagram. At fixed $M_{\rm BH}$, increasing $\dot{m}_{\rm BH}$ raises both the luminosity and disk temperature, shifting the source toward the upper left, as indicated by the arrows.

\subsection{Model caveats and validation}

\begin{figure}
    \centering
    \includegraphics[width=1\linewidth]{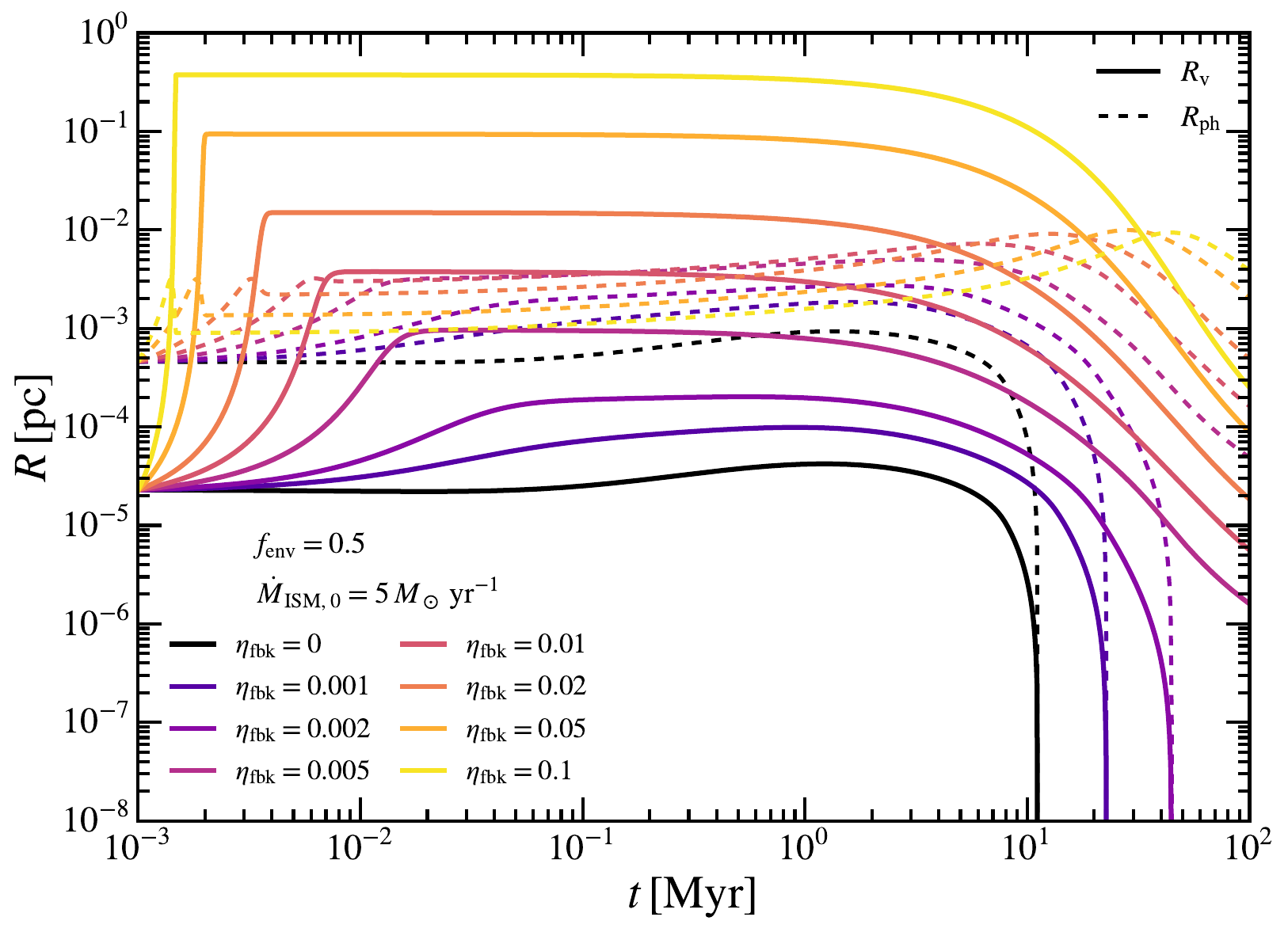}
    \caption{
    Evolution of $R_{\rm v}$ and $R_{\rm ph}$ in the feedback-involved models at fixed $f_{\rm env}=0.5$ and different feedback efficiencies $\eta_{\rm fbk}$. Stronger feedback drives a more rapid expansion of $R_{\rm v}$, while $R_{\rm ph}$ remains at smaller radii owing to the opacity regulation, leading to an increasing separation between the two radii.
    }
    \label{fig:R}
\end{figure}

Our model is designed primarily to capture the continuum spectral evolution driven by the Kelvin--Helmholtz contraction of a BHE. Despite the success of this simple framework in reproducing the observed diversity from LRDs to LBDs, several uncertainties and limitations remain.

First, our model follows the mechanical evolution of the envelope by tracking $R_{\rm v}$, while the photospheric radius $R_{\rm ph}$ and the corresponding effective temperature are calculated only in post-processing and do not affect the dynamical evolution. This treatment is self-consistent as long as $R_{\rm ph}\gtrsim R_{\rm v}$. Since the luminosity is assumed to emerge from $R_{\rm ph}$, material located substantially outside the photosphere should in principle no longer remain part of the bound, radiation-supported envelope.

We find that this condition can break down when the feedback is too strong or the BH accretion rate is too high. In such cases, $R_{\rm v}$ expands rapidly during the earliest stage ($\lesssim1$ Myr), while $R_{\rm ph}$ remains much smaller, leading to $R_{\rm ph}\ll R_{\rm v}$ and violating the underlying assumption of our treatment, as illustrated in Figure~\ref{fig:R}. This consistency requirement therefore places an upper limit on the ISM inflow rate for which a bound BHE can be maintained. By requiring $R_{\rm v}=R_{\rm ph}$ at the feedback-regulated equilibrium state $\dot m_{\rm BH}= \dot m_{\rm BH,eq}$ for the adopted initial conditions, we obtain the critical inflow rate
\begin{equation}
\dot M_{\rm ISM,crit}\simeq
0.45\,
\eta_{\rm fbk,0.1}^{-1}\,
M_{\rm BH,5}^{3/4}\,
q_{\rm env}^{1/17}\,
\msun\,{\rm yr^{-1}},
\end{equation}
where $\eta_{\rm fbk,0.1}\equiv\eta_{\rm fbk}/0.1$, $M_{\rm BH,5}\equiv M_{\rm BH}/10^5\msun$, and $q_{\rm env}\equiv M_{\rm env}/M_{\rm BH}$ is the envelope-to-BH mass ratio.
Above this rate, the equilibrium envelope would expand beyond its photosphere.

For the ISM inflow rate adopted in this work, our fiducial feedback efficiency of $\eta_{\rm fbk}\sim\mathcal{O}(10^{-2})$ remains within the self-consistent regime, whereas at substantially higher feedback efficiencies the assumption of a quasi-hydrostatic envelope may break down. In this regime, strong energy injection can render the envelope only weakly bound, and the system may instead develop inflows, outflows, or convective circulation rather than maintaining a single hydrostatic configuration~\citep[e.g.,][]{Begelman2008,Dotan2011,Coughlin2014,Fiacconi2016}. Indeed, radiation-pressure-supported envelopes near the Eddington limit are particularly susceptible to substantial mass loss once a hydrostatic solution can no longer be maintained.
At the same time, the precise boundary of this regime should not be over-interpreted. The rapid expansion of $R_{\rm v}$ relative to $R_{\rm ph}$ may partly reflect our simplified treatment of the unresolved envelope structure and opacity \citep[e.g.,][]{Kido2025,Takasao2026}. A self-consistent hydrodynamic and radiative-transfer calculation of the BHE structure could modify and potentially alleviate this behavior. Modeling this regime therefore requires a time-dependent hydrodynamic treatment of the coupled inflow–outflow structure and lies beyond the scope of our phenomenological model.

\section{Conclusion}\label{sec:conc}

We use a time-dependent model of an optically thick gaseous envelope surrounding a massive BH, in which radiative cooling, declining ISM inflow, BH accretion, and accretion-powered feedback are coupled, and translate the resulting evolution into mock SEDs including a coevolving nuclear stellar population. Our main conclusions are as follows.

\begin{itemize}
    \item \textbf{Kelvin--Helmholtz contraction of a gas-fed BHE can naturally transform an LRD into an LBD through an observable transitional phase.} Sustained inflow initially builds a cool, extended envelope whose photospheric temperature is regulated near the Hayashi track. As supernova feedback reduces the gas supply, the envelope can no longer be replenished and begins to contract. Its photosphere shrinks and heats until the accretion disk emerges, producing a continuous sequence from an LRD to a transitional source and finally to an LBD. The proposed AGN HR diagram provides a complementary view of this sequence, connecting a Hayashi-track-regulated BHE to an exposed accretion disk through envelope contraction.

    \item \textbf{The LRD--LBD transition follows a robust path in the $(\beta_{\rm UV},\beta_{\rm opt})$ plane.} As the blackbody peak moves through the UV, the tracks make a characteristic red excursion near $(\beta_{\rm UV},\beta_{\rm opt})\sim(2,-1)$ before turning blue and entering the LBD region. The same trajectory passes near BBQSORS, whose $\sim10^4\K$ blackbody-like continuum is consistent with an envelope caught during contraction. Although the track shape varies little across the feedback and envelope-retention parameters explored, these parameters strongly regulate the evolutionary timing.

    \item \textbf{The abundance of transitional systems relative to LRDs can constrain BH physics.} The LRD and transitional lifetimes span approximately $\lesssim5$--$50$ Myr across the model grid. In our fiducial case, the system remains an LRD for $\sim15$ Myr and then spends $\sim20$ Myr in the transitional phase, with both timescales increasing toward higher feedback and gas-retention efficiencies. Under a statistical steady-state assumption, the relative abundance of transitional systems and LRDs directly traces their relative lifetimes, providing an observational constraint on the feedback efficiency and gas retention that govern BHE evolution.

    \item \textbf{BH growth during the transition produces a systematic mass offset between LRDs and LBDs.} Intermediate feedback efficiencies produce maximum LRD-phase BH masses of $\sim10^6\msun$, while the BH can reach $\sim10^{6.5}\msun$ by the end of the transitional phase. This corresponds to a characteristic increase in BH mass by a factor of $\sim3$--$5$ between the two stages, implying that the relative abundance of LBDs should increase toward higher BH masses.
\end{itemize}

In conclusion, by combining a minimal set of physical ingredients, our model reproduces the observed diversity of broad-line emitters within a coherent evolutionary framework. This picture makes clear and testable observational predictions. Measurements of the number densities, continuum slopes, photospheric temperatures, and BH masses of transitional sources can directly test the proposed evolutionary sequence and constrain the efficiency of feedback during rapid BH growth.

Our model also provides a simple and readily implementable subgrid prescription for semi-analytic and cosmological simulations, as it requires only the ISM inflow history onto the nuclear region once the model parameters are specified. Coupling this prescription to a more realistic, time-dependent gas supply from galactic scales would allow the duration and possible recurrence of the subsequent LBD phase to be followed naturally, including the effects of long-term gas replenishment and repeated accretion episodes. Such an implementation would also follow the coevolution of the host galaxy, providing an additional channel for the disappearance of the LRD signature~\citep[e.g.,][]{Zhang2026,Billand2026} that is not captured in the present model.

A further important step will be to extend the current framework by incorporating a self-consistent hydrodynamic treatment of the BHE structure, or ultimately to follow the BHE evolution directly with dedicated radiation-hydrodynamic simulations. Such calculations would enable a more realistic treatment of line formation and departures from spherical symmetry.

\begin{acknowledgments}
We greatly thank Yoshihisa Asada and Rosa M\'erida for jointly shaping the ideas presented in this work. We also thank Anna de Graaff, Dale Kocevski, Jorryt Matthee, Rohan Naidu, and Zijian Zhang for insightful discussions. We also thank Hiroya Umeda for providing the LRD data used in this work. KI was supported by the National Natural Science Foundation of China (12573015, W2532003, 1251101148, 12233001, 12473037), the Beijing Natural Science Foundation (IS25003), and the China Manned Space Program (CMS-CSST-2025-A09).
We sincerely thank the organizers of the conferences, {\it Unraveling Little Red Dots: Linking JWST Discoveries with Simulations to Understand Early Galaxies} and {\it Charting Cosmic Dawn in Copenhagen 2026} for their hospitality and for stimulating discussions that were valuable to this work.
\end{acknowledgments}




\appendix

\section{Impact of the envelope self-gravity}\label{apdx:xi}

\begin{figure}
    \centering
    \includegraphics[width=0.9\linewidth]{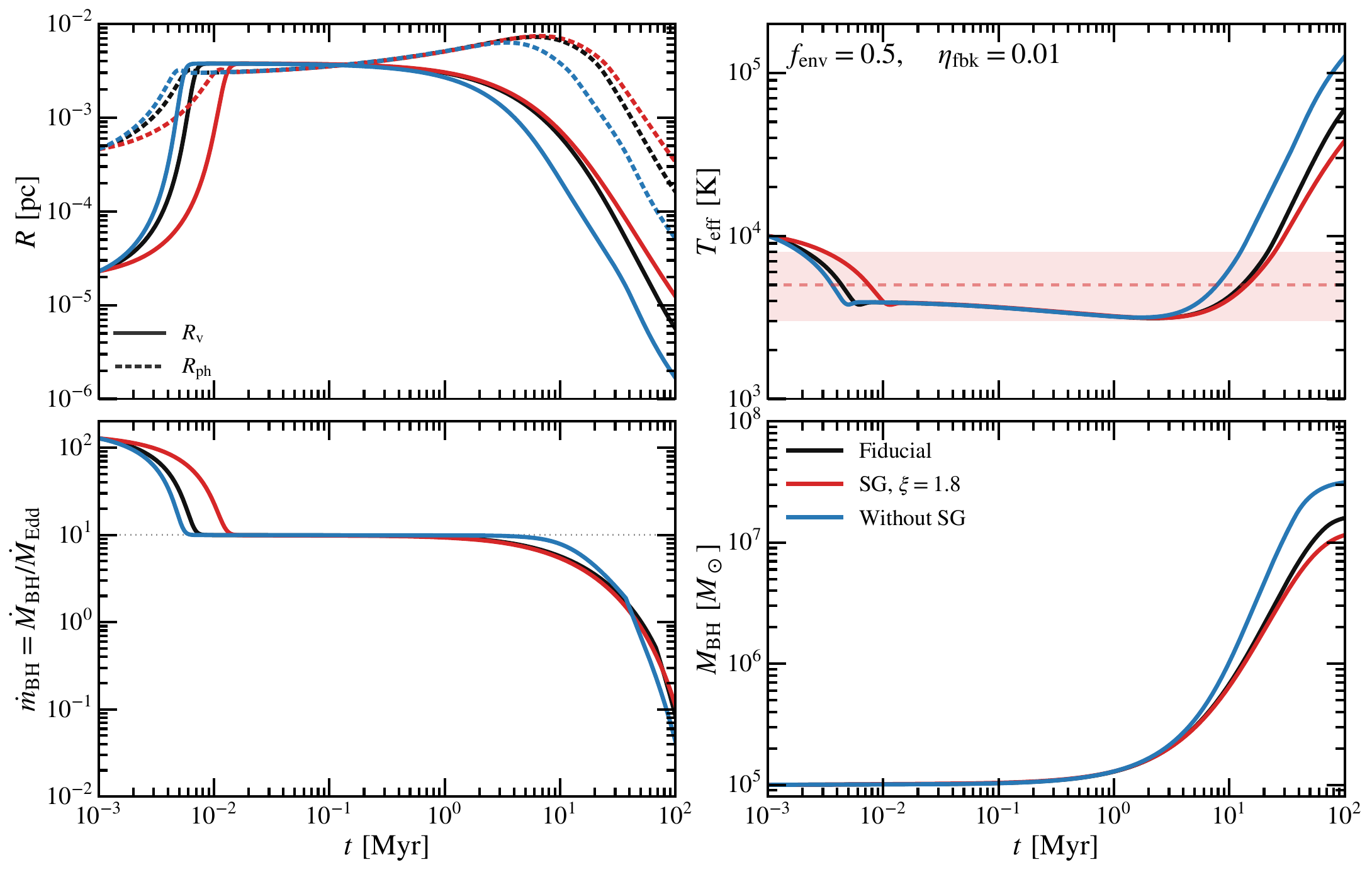}
    \caption{
    Comparison of BHE evolution for different assumptions about the envelope structure at fixed $f_{\rm env}=0.5$ and $\eta_{\rm fbk}=0.01$. Black, blue, and red curves represent the fiducial model, the model with a steeper density profile ($\xi=1.8$), and the limiting case without envelope self-gravity ($C_{\rm self}=0$), respectively. The panels show the evolution of the envelope and photospheric radii ($R_{\rm v}$ and $R_{\rm ph}$; top left), effective temperature ($T_{\rm eff}$; top right), Eddington-scaled BH accretion rate ($\dot{m}_{\rm BH}=\dot{M}_{\rm BH}/\dot{M}_{\rm Edd}$; bottom left), and BH mass ($M_{\rm BH}$; bottom right).
    }
    \label{fig:model_variants}
\end{figure}

\begin{figure}
   \centering
    \includegraphics[width=0.95\linewidth]{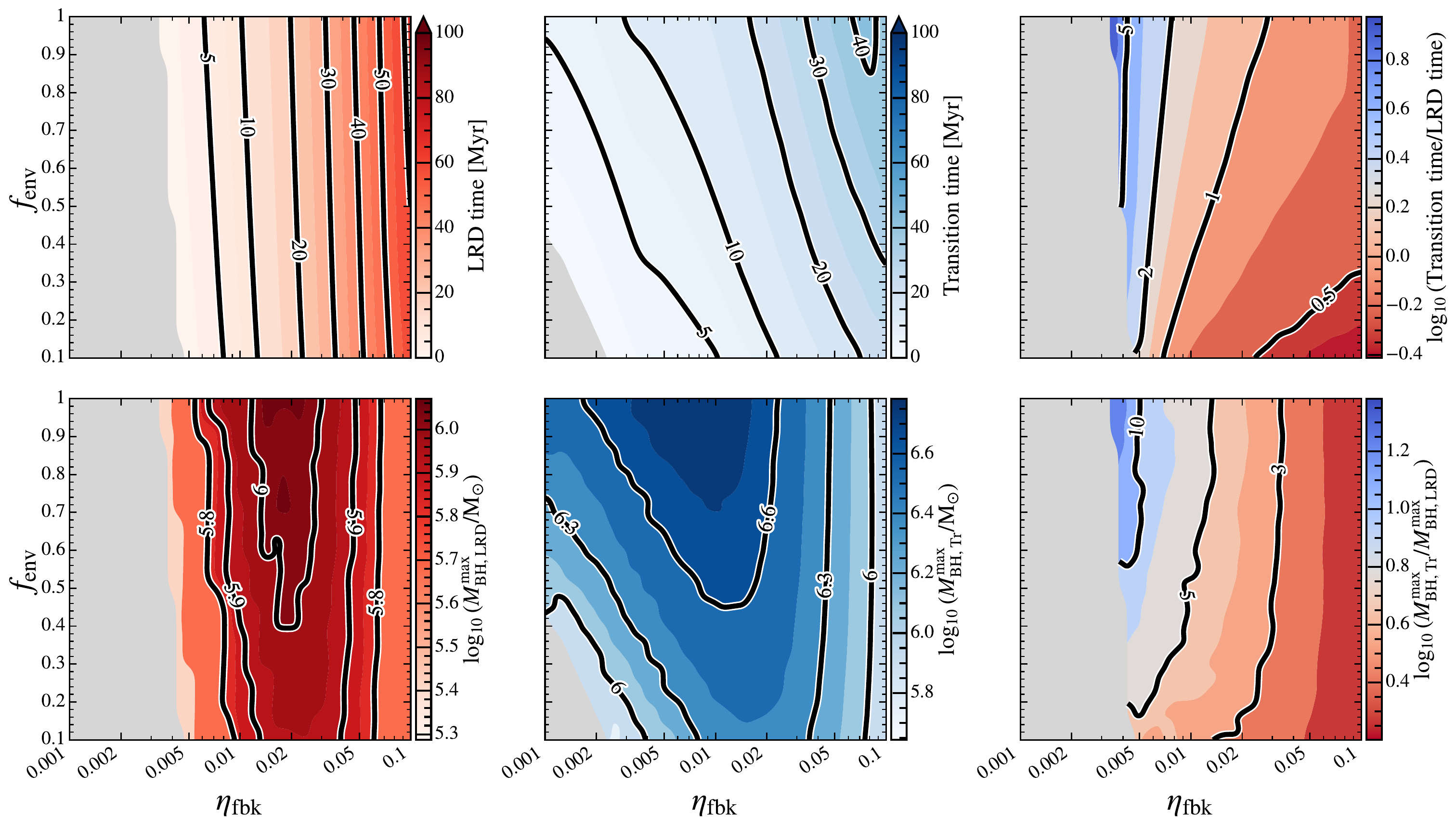}
    \caption{
    Same as Figure~\ref{fig:grids}, but for models without envelope self-gravity (SG). Compared with the fiducial models, the evolutionary timescales show a weaker dependence on $f_{\rm env}$, while both the LRD and transitional phases are generally shorter.
    }
    \label{fig:grid_noSG}
\end{figure}

In the main text, we evaluate the total binding energy of the system in Equation~(\ref{eq:Ubind}), including both the gravitational potential of the central BH and the self-gravity of the envelope. The unresolved internal structure of the envelope, however, introduces uncertainties into both terms. To assess its impact, we consider two model variants. In the first, we adopt a steeper density profile with $\xi=1.8$. In the second, we neglect the envelope self-gravity by setting $C_{\rm self}=0$, providing a limiting case for the uncertainty associated with the self-gravity term.

In Figure~\ref{fig:model_variants}, we compare the effects of these structural assumptions at fixed physical parameters. Steepening the envelope density profile to $\xi=1.8$ changes both the BH--envelope and envelope self-gravity contributions to the binding energy. In our calculations, the envelope remains modestly more extended and BH accretion proceeds slightly more slowly, delaying the evolution out of the LRD regime. Applying the same mock-observation procedure yields color tracks and parameter-space trends that remain broadly similar to those of the fiducial model.

In contrast, neglecting envelope self-gravity ($C_{\rm self}=0$)
accelerates the evolution out of the LRD regime, particularly at large
$f_{\rm env}$. At fixed total energy and masses, removing the
$M_{\rm env}^{2}$ contribution reduces the inferred envelope radius,
increasing the BH accretion rate and accelerating the transfer of gas
from the envelope to the BH. It also removes the additional structural
response to gas retention provided by the quadratic self-gravity term.
Although $f_{\rm env}$ still affects the gas reservoir and the
BH--envelope binding energy, its effect is no longer amplified by
envelope self-gravity. Consequently, the LRD residence time and the
maximum BH mass reached during the LRD phase depend much less strongly
on $f_{\rm env}$, as shown in Figure~\ref{fig:grid_noSG}.

\section{Dependence of BHE evolution on the physical parameters}
\label{apdx:fenv}

\begin{figure}
    \centering
    \includegraphics[width=1\linewidth]{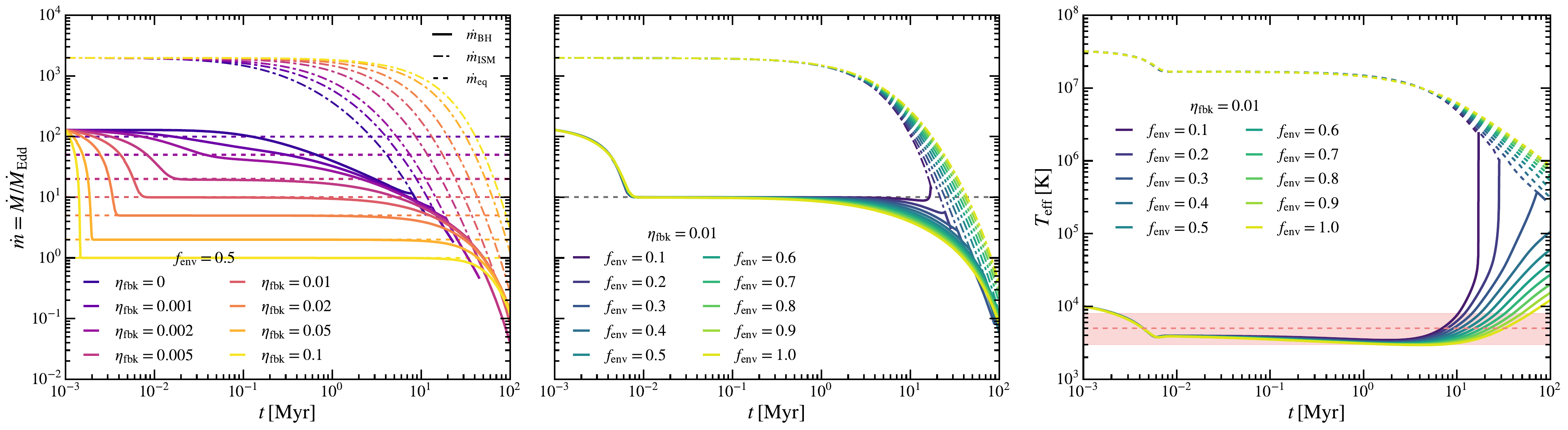}
    \caption{
    Dependence of BHE evolution on the physical parameters.
    \textit{Left:} Dimensionless BH accretion rate at fixed $f_{\rm env}=0.5$ for different feedback efficiencies, $\eta_{\rm fbk}$. Dashed lines denote the feedback-regulated equilibrium accretion rates, while dash-dotted lines show the ISM inflow rates. 
    \textit{Middle:} Same as the left panel, but at fixed $\eta_{\rm fbk}$ while varying $f_{\rm env}$. The envelope retention efficiency has little effect on the BH accretion history, but slightly shifts the equilibrium timescale.
    \textit{Right:} Evolution of the effective temperature for different $f_{\rm env}$. Because the photospheric radius depends on the envelope mass, $T_{\rm eff}$ becomes increasingly sensitive to $f_{\rm env}$ at late times as the envelope approaches depletion, with larger $f_{\rm env}$ producing a more extended photosphere and hence a lower effective temperature.
    }
    \label{fig:dotm}
\end{figure}

In the main text, we presented the absolute BH accretion rates for different feedback efficiencies. Feedback regulates BH accretion through the response of the envelope and yields a quasi-steady plateau in the dimensionless accretion rate.
The left panel of Figure~\ref{fig:dotm} shows the evolution of $\dot{m}_{\rm BH}$ for different $\eta_{\rm fbk}$ at fixed $f_{\rm env}=0.5$, together with the corresponding equilibrium rate and the dimensionless ISM supply rate.

The accretion history can be divided into three stages. Initially, the Kelvin--Helmholtz response of the envelope is sufficiently rapid for $R_{\rm v}$ to follow its evolving equilibrium value. Feedback heating and radiative cooling are therefore approximately balanced, giving
\begin{equation}
    \dot{m}_{\rm BH,eq}
    =\frac{\eta_{\rm rad}}{\eta_{\rm fbk}}.
\end{equation}
This balance produces the plateau seen in Figure~\ref{fig:dotm}. As the BH grows and the ISM supply declines, the equilibrium radius moves progressively inward. Eventually, the increasing thermal inertia of the envelope prevents it from contracting rapidly enough to follow this moving equilibrium. The envelope then remains more extended than $R_{\rm v,eq}$, causing $\dot{M}_{\rm BH}\propto R_{\rm v}^{-1/2}$ to fall progressively below $\dot{M}_{\rm BH,eq}$. At still later times, the declining ISM inflow becomes the dominant limitation, and the BH accretion rate approaches the available supply.

The middle panel shows the corresponding evolution at fixed
$\eta_{\rm fbk}=0.01$ while varying $f_{\rm env}$. The equilibrium accretion rate is independent of $f_{\rm env}$, such that the BH accretion rate is largely unchanged. A larger retention efficiency builds a more massive envelope and increases its thermal inertia. Consequently, the envelope becomes unable to follow the contracting equilibrium radius at an earlier time, even though the larger gas reservoir allows the subsequent embedded phase to persist for longer.

The effective temperature, however, depends sensitively on the envelope mass. Increasing $f_{\rm env}$ produces a more massive and optically thicker envelope, pushing the photosphere to larger radii. Because $L\simeq L_{\rm Edd}$ remains nearly constant, the larger photospheric radius directly implies a lower effective temperature.

\section{Spectral dependence of the absorption and stellar treatment}\label{apdx:BB}
In the main text, we adopt a simple prescription in which the Balmer absorption is smoothly weakened as the system evolves. Here, we test the sensitivity of our results to this phenomenological treatment. We consider two main uncertainties: the smoothness of the transition, characterized by $\Delta_{\rm sm}$, and the functional form of the transmission curve, characterized by the hydrogen
column density $N_{\rm H}$. Throughout this comparison, we adopt the fiducial BH parameters, $\eta_{\rm fbk}=0.01$ and $f_{\rm env}=0.5$.

\begin{figure}
    \centering
    \includegraphics[width=1\linewidth]{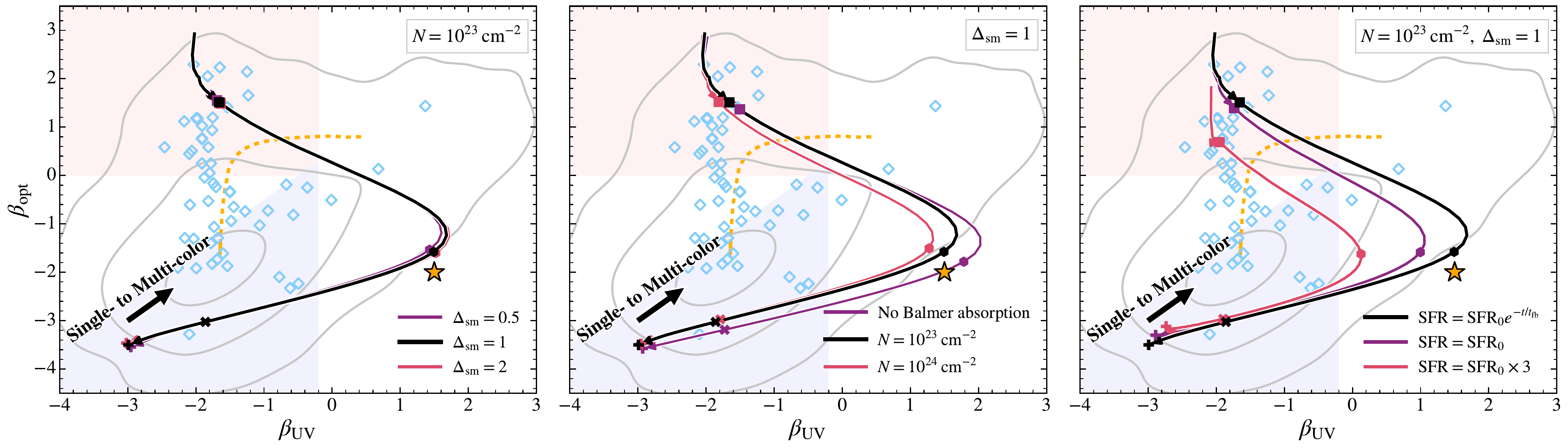}
    \caption{
    Same as Figure~\ref{fig:betas}, but showing the dependence of the evolutionary tracks on the spectral treatment. The left panel varies the Balmer-absorption transition smoothness, $\Delta_{\rm sm}=0.5$, $1$, and $2$, at fixed $N_{\rm H}=10^{23}\,\mathrm{cm^{-2}}$. The middle panel compares different absorption strengths, including no Balmer absorption, $N_{\rm H}=10^{23}\,\mathrm{cm^{-2}}$, and $N_{\rm H}=10^{24}\,\mathrm{cm^{-2}}$, at fixed $\Delta_{\rm sm}=1$. The right panel compares the fiducial evolving star formation history with constant ${\rm SFR}={\rm SFR}_0$ and ${\rm SFR}=3{\rm SFR}_0$. A stronger stellar contribution maintains a bluer UV continuum and reduces the extent of the red excursion. In all panels, the underlying BHE evolution adopts the fiducial parameters $\eta_{\rm fbk}=0.01$ and $f_{\rm env}=0.5$, and the fiducial model adopted in the main text is shown by a thick black curve.
    }
    \label{fig:BBtreatment}
\end{figure}

The left panel of Figure~\ref{fig:BBtreatment} shows the evolutionary tracks in the $(\beta_{\rm UV},\beta_{\rm opt})$ plane for the fiducial transmission function adopted in the main text, varying only $\Delta_{\rm sm}$. The observational data and time markers are the same as in Figure~\ref{fig:betas}. The tracks are nearly indistinguishable. A more gradual weakening of the absorption produces only a marginally larger maximum $\beta_{\rm UV}$ during the red excursion. Since this prescription modifies only the emergent spectrum and does not affect the underlying BHE dynamics, neither the evolutionary timescale nor the overall location of the track in the $(\beta_{\rm UV},\beta_{\rm opt})$ plane changes appreciably.

These tests demonstrate that the main conclusions presented in the main text are robust against uncertainties in both the strength and transition of the Balmer absorption. The stellar contribution, however, has a larger impact on the evolutionary tracks. In the main text, we assume that nuclear star formation follows the same gas supply as the BHE and is therefore suppressed as feedback reduces the inflow. Star formation on larger host-galaxy scales, however, may not respond to this nuclear feedback in the same manner. To bracket this uncertainty, we additionally consider models with a constant ${\rm SFR}={\rm SFR}_0$ and ${\rm SFR}=3{\rm SFR}_0$ throughout the evolution.

A stronger stellar component contributes increasingly to, and can eventually dominate, the UV continuum, while the optical continuum remains primarily dominated by the BHE. Consequently, $\beta_{\rm UV}$ remains bluer at a given $\beta_{\rm opt}$, reducing the extent of the red excursion and shifting the evolutionary track inward in the $(\beta_{\rm UV},\beta_{\rm opt})$ plane.

\bibliography{reference}{}

@ARTICLE{Metzger2022,
       author = {{Metzger}, Brian D.},
        title = "{Cooling Envelope Model for Tidal Disruption Events}",
      journal = {\apjl},
         year = 2022,
        month = sep,
       volume = {937},
       number = {1},
          eid = {L12},
        pages = {L12},
          doi = {10.3847/2041-8213/ac90ba},
archivePrefix = {arXiv},
       eprint = {2207.07136},
 primaryClass = {astro-ph.HE},
       adsurl = {https://ui.adsabs.harvard.edu/abs/2022ApJ...937L..12M}
}

@ARTICLE{Inayoshi2026,
       author = {{Inayoshi}, Kohei and {Murase}, Kohta and {Kashiyama}, Kazumi},
        title = "{Spectral Uniformity of Little Red Dots: A Natural Outcome of Coevolving Seed Black Holes and Nascent Starbursts}",
      journal = {\apj},
         year = 2026,
        month = mar,
       volume = {1000},
       number = {1},
          eid = {90},
        pages = {90},
          doi = {10.3847/1538-4357/ae42ce},
archivePrefix = {arXiv},
       eprint = {2509.19422},
 primaryClass = {astro-ph.GA},
       adsurl = {https://ui.adsabs.harvard.edu/abs/2026ApJ..1000...90I}
}

@ARTICLE{Hu2022,
       author = {{Hu}, Haojie and {Inayoshi}, Kohei and {Haiman}, Zolt{\'a}n and {Quataert}, Eliot and {Kuiper}, Rolf},
        title = "{Long-term Evolution of Supercritical Black Hole Accretion with Outflows: A Subgrid Feedback Model for Cosmological Simulations}",
      journal = {\apj},
         year = 2022,
        month = aug,
       volume = {934},
       number = {2},
          eid = {132},
        pages = {132},
          doi = {10.3847/1538-4357/ac75d8},
archivePrefix = {arXiv},
       eprint = {2203.14994},
 primaryClass = {astro-ph.HE},
       adsurl = {https://ui.adsabs.harvard.edu/abs/2022ApJ...934..132H}
}

@ARTICLE{Blandford1999,
       author = {{Blandford}, Roger D. and {Begelman}, Mitchell C.},
        title = "{On the fate of gas accreting at a low rate on to a black hole}",
      journal = {\mnras},
         year = 1999,
        month = feb,
       volume = {303},
       number = {1},
        pages = {L1-L5},
          doi = {10.1046/j.1365-8711.1999.02358.x},
archivePrefix = {arXiv},
       eprint = {astro-ph/9809083},
 primaryClass = {astro-ph},
       adsurl = {https://ui.adsabs.harvard.edu/abs/1999MNRAS.303L...1B}
}

@ARTICLE{Begelman2008,
       author = {{Begelman}, Mitchell C. and {Rossi}, Elena M. and {Armitage}, Philip J.},
        title = "{Quasi-stars: accreting black holes inside massive envelopes}",
      journal = {\mnras},
         year = 2008,
        month = jul,
       volume = {387},
       number = {4},
        pages = {1649-1659},
          doi = {10.1111/j.1365-2966.2008.13344.x},
archivePrefix = {arXiv},
       eprint = {0711.4078},
 primaryClass = {astro-ph},
       adsurl = {https://ui.adsabs.harvard.edu/abs/2008MNRAS.387.1649B}
}

@ARTICLE{Byler2017,
       author = {{Byler}, Nell and {Dalcanton}, Julianne J. and {Conroy}, Charlie and {Johnson}, Benjamin D.},
        title = "{Nebular Continuum and Line Emission in Stellar Population Synthesis Models}",
      journal = {\apj},
         year = 2017,
        month = may,
       volume = {840},
       number = {1},
          eid = {44},
        pages = {44},
          doi = {10.3847/1538-4357/aa6c66},
archivePrefix = {arXiv},
       eprint = {1611.08305},
 primaryClass = {astro-ph.GA},
       adsurl = {https://ui.adsabs.harvard.edu/abs/2017ApJ...840...44B}
}

@article{Coughlin2014,
   title={HYPERACCRETION DURING TIDAL DISRUPTION EVENTS: WEAKLY BOUND DEBRIS ENVELOPES AND JETS},
   volume={781},
   ISSN={1538-4357},
   url={http://dx.doi.org/10.1088/0004-637X/781/2/82},
   DOI={10.1088/0004-637x/781/2/82},
   number={2},
   journal={The Astrophysical Journal},
   publisher={American Astronomical Society},
   author={Coughlin, Eric R. and Begelman, Mitchell C.},
   year={2014},
   month=Jan, pages={82} }

@ARTICLE{Hosokawa2013,
       author = {{Hosokawa}, Takashi and {Yorke}, Harold W. and {Inayoshi}, Kohei and {Omukai}, Kazuyuki and {Yoshida}, Naoki},
        title = "{Formation of Primordial Supermassive Stars by Rapid Mass Accretion}",
      journal = {\apj},
         year = 2013,
        month = dec,
       volume = {778},
       number = {2},
          eid = {178},
        pages = {178},
          doi = {10.1088/0004-637X/778/2/178},
archivePrefix = {arXiv},
       eprint = {1308.4457},
 primaryClass = {astro-ph.SR},
       adsurl = {https://ui.adsabs.harvard.edu/abs/2013ApJ...778..178H}
}

@ARTICLE{Inayoshi2016,
       author = {{Inayoshi}, Kohei and {Haiman}, Zolt{\'a}n and {Ostriker}, Jeremiah P.},
        title = "{Hyper-Eddington accretion flows on to massive black holes}",
      journal = {\mnras},
         year = 2016,
        month = jul,
       volume = {459},
       number = {4},
        pages = {3738-3755},
          doi = {10.1093/mnras/stw836},
archivePrefix = {arXiv},
       eprint = {1511.02116},
 primaryClass = {astro-ph.HE},
       adsurl = {https://ui.adsabs.harvard.edu/abs/2016MNRAS.459.3738I}
}

@ARTICLE{Mayer2005,
       author = {{Mayer}, Michael and {Duschl}, Wolfgang J.},
        title = "{Rosseland and Planck mean opacities for primordial matter}",
      journal = {\mnras},
         year = 2005,
        month = apr,
       volume = {358},
       number = {2},
        pages = {614-631},
          doi = {10.1111/j.1365-2966.2005.08826.x},
archivePrefix = {arXiv},
       eprint = {astro-ph/0411613},
 primaryClass = {astro-ph},
       adsurl = {https://ui.adsabs.harvard.edu/abs/2005MNRAS.358..614M}
}

@misc{Takasao2026,
      title={A Magnetized Black Hole Envelope Model for Little Red Dots}, 
      author={Shinsuke Takasao and Kohei Inayoshi},
      year={2026},
      eprint={2605.21589},
      archivePrefix={arXiv},
      primaryClass={astro-ph.GA},
      url={https://arxiv.org/abs/2605.21589}, 
}

@ARTICLE{Hayashi1961,
       author = {{Hayashi}, Chushiro},
        title = "{Stellar Evolution in Early Phases of Gravitational Contraction}",
      journal = {\pasj},
         year = 1961,
        month = dec,
       volume = {13},
       number = {4},
        pages = {450-452},
          doi = {10.1093/pasj/13.4.450},
       adsurl = {https://ui.adsabs.harvard.edu/abs/1961PASJ...13..450H}
}

@ARTICLE{Shakura1973,
       author = {{Shakura}, N.~I. and {Sunyaev}, R.~A.},
        title = "{Black holes in binary systems. Observational appearance.}",
      journal = {\aap},
         year = 1973,
        month = jan,
       volume = {24},
        pages = {337-355},
       adsurl = {https://ui.adsabs.harvard.edu/abs/1973A&A....24..337S}
}

@misc{Zhang2026,
      title={Why Little Red Dots Disappear at z < 3: Evolution of Number Density and Halo Mass}, 
      author={Chenxuan Zhang and Huanian Zhang and Qingwen Wu and Luis C. Ho and Jian-Min Wang},
      year={2026},
      eprint={2606.02773},
      archivePrefix={arXiv},
      primaryClass={astro-ph.GA},
      url={https://arxiv.org/abs/2606.02773}, 
}

@misc{Lin2025,
      title={The Large-scale Environments of Low-luminosity AGNs at $3.9 < z < 6$ and Implications for Their Host Dark Matter Halos from a Complete NIRCam Grism Redshift Survey}, 
      author={Xiaojing Lin and Xiaohui Fan and Fengwu Sun and Junyu Zhang and Eiichi Egami and Jakob M. Helton and Feige Wang and Haowen Zhang and Andrew J. Bunker and Zheng Cai and Zhiyuan Ji and Xiangyu Jin and Roberto Maiolino and Maria Anne Pudoka and Pierluigi Rinaldi and Brant Robertson and Sandro Tacchella and Wei Leong Tee and Yang Sun and Christopher N. A. Willmer and Chris Willott and Yongda Zhu},
      year={2025},
      eprint={2505.02896},
      archivePrefix={arXiv},
      primaryClass={astro-ph.GA},
      url={https://arxiv.org/abs/2505.02896}, 
}

@misc{Kocevski2025,
      title={The Rise of Faint, Red AGN at $z>4$: A Sample of Little Red Dots in the JWST Extragalactic Legacy Fields}, 
      author={Dale D. Kocevski and Steven L. Finkelstein and Guillermo Barro and Anthony J. Taylor and Antonello Calabrò and Brivael Laloux and Johannes Buchner and Jonathan R. Trump and Gene C. K. Leung and Guang Yang and Mark Dickinson and Pablo G. Pérez-González and Fabio Pacucci and Kohei Inayoshi and Rachel S. Somerville and Elizabeth J. McGrath and Hollis B. Akins and Micaela B. Bagley and Laura Bisigello and Rebecca A. A. Bowler and Adam Carnall and Caitlin M. Casey and Yingjie Cheng and Nikko J. Cleri and Luca Costantin and Fergus Cullen and Kelcey Davis and Callum T. Donnan and James S. Dunlop and Richard S. Ellis and Henry C. Ferguson and Seiji Fujimoto and Adriano Fontana and Mauro Giavalisco and Andrea Grazian and Norman A. Grogin and Nimish P. Hathi and Michaela Hirschmann and Marc Huertas-Company and Benne W. Holwerda and Garth Illingworth and Stéphanie Juneau and Jeyhan S. Kartaltepe and Anton M. Koekemoer and Wenxiu Li and Ray A. Lucas and Dan Magee and Charlotte Mason and Derek J. McLeod and Ross J. McLure and Lorenzo Napolitano and Casey Papovich and Nor Pirzkal and Giulia Rodighiero and Paola Santini and Stephen M. Wilkins and L. Y. Aaron Yung},
      year={2025},
      eprint={2404.03576},
      archivePrefix={arXiv},
      primaryClass={astro-ph.GA},
      url={https://arxiv.org/abs/2404.03576}, 
}

@article{Zhang2025,
   title={Unexpected clustering pattern in dwarf galaxies challenges formation models},
   volume={642},
   ISSN={1476-4687},
   url={http://dx.doi.org/10.1038/s41586-025-08965-5},
   DOI={10.1038/s41586-025-08965-5},
   number={8066},
   journal={Nature},
   publisher={Springer Science and Business Media LLC},
   author={Zhang, Ziwen and Chen, Yangyao and Rong, Yu and Wang, Huiyuan and Mo, Houjun and Luo, Xiong and Li, Hao},
   year={2025},
   month=may, pages={47–52} }

@article{Maiolino2024,
   title={JADES: The diverse population of infant black holes at 4 &lt; z &lt; 11: Merging, tiny, poor, but mighty},
   volume={691},
   ISSN={1432-0746},
   url={http://dx.doi.org/10.1051/0004-6361/202347640},
   DOI={10.1051/0004-6361/202347640},
   journal={Astronomy \&amp; Astrophysics},
   publisher={EDP Sciences},
   author={Maiolino, Roberto and Scholtz, Jan and Curtis-Lake, Emma and Carniani, Stefano and Baker, William and de Graaff, Anna and Tacchella, Sandro and Übler, Hannah and D’Eugenio, Francesco and Witstok, Joris and Curti, Mirko and Arribas, Santiago and Bunker, Andrew J. and Charlot, Stéphane and Chevallard, Jacopo and Eisenstein, Daniel J. and Egami, Eiichi and Ji, Zhiyuan and Jones, Gareth C. and Lyu, Jianwei and Rawle, Tim and Robertson, Brant and Rujopakarn, Wiphu and Perna, Michele and Sun, Fengwu and Venturi, Giacomo and Williams, Christina C. and Willott, Chris},
   year={2024},
   month=nov, pages={A145} }

@misc{Naidu2025,
      title={A "Black Hole Star" Reveals the Remarkable Gas-Enshrouded Hearts of the Little Red Dots}, 
      author={Rohan P. Naidu and Jorryt Matthee and Harley Katz and Anna de Graaff and Pascal Oesch and Aaron Smith and Jenny E. Greene and Gabriel Brammer and Andrea Weibel and Raphael Hviding and John Chisholm and Ivo Labbé and Robert A. Simcoe and Callum Witten and Hakim Atek and Josephine F. W. Baggen and Sirio Belli and Rachel Bezanson and Leindert A. Boogaard and Sownak Bose and Alba Covelo-Paz and Pratika Dayal and Yoshinobu Fudamoto and Lukas J. Furtak and Emma Giovinazzo and Andy Goulding and Max Gronke and Kasper E. Heintz and Michaela Hirschmann and Garth Illingworth and Akio K. Inoue and Benjamin D. Johnson and Joel Leja and Ecaterina Leonova and Ian McConachie and Michael V. Maseda and Priyamvada Natarajan and Erica Nelson and David J. Setton and Irene Shivaei and David Sobral and Mauro Stefanon and Sandro Tacchella and Sune Toft and Alberto Torralba and Pieter van Dokkum and Arjen van der Wel and Marta Volonteri and Fabian Walter and Bingjie Wang and Darach Watson},
      year={2025},
      eprint={2503.16596},
      archivePrefix={arXiv},
      primaryClass={astro-ph.GA},
      url={https://arxiv.org/abs/2503.16596}, 
}

@ARTICLE{Matthee2024,
       author = {{Matthee}, Jorryt and {Naidu}, Rohan P. and {Brammer}, Gabriel and {Chisholm}, John and {Eilers}, Anna-Christina and {Goulding}, Andy and {Greene}, Jenny and {Kashino}, Daichi and {Labbe}, Ivo and {Lilly}, Simon J. and {Mackenzie}, Ruari and {Oesch}, Pascal A. and {Weibel}, Andrea and {Wuyts}, Stijn and {Xiao}, Mengyuan and {Bordoloi}, Rongmon and {Bouwens}, Rychard and {van Dokkum}, Pieter and {Illingworth}, Garth and {Kramarenko}, Ivan and {Maseda}, Michael V. and {Mason}, Charlotte and {Meyer}, Romain A. and {Nelson}, Erica J. and {Reddy}, Naveen A. and {Shivaei}, Irene and {Simcoe}, Robert A. and {Yue}, Minghao},
        title = "{Little Red Dots: An Abundant Population of Faint Active Galactic Nuclei at z {\ensuremath{\sim}} 5 Revealed by the EIGER and FRESCO JWST Surveys}",
      journal = {\apj},
         year = 2024,
        month = mar,
       volume = {963},
       number = {2},
          eid = {129},
        pages = {129},
          doi = {10.3847/1538-4357/ad2345},
archivePrefix = {arXiv},
       eprint = {2306.05448},
 primaryClass = {astro-ph.GA},
       adsurl = {https://ui.adsabs.harvard.edu/abs/2024ApJ...963..129M}
}

@ARTICLE{Greene2024,
       author = {{Greene}, Jenny E. and {Labbe}, Ivo and {Goulding}, Andy D. and {Furtak}, Lukas J. and {Chemerynska}, Iryna and {Kokorev}, Vasily and {Dayal}, Pratika and {Volonteri}, Marta and {Williams}, Christina C. and {Wang}, Bingjie and {Setton}, David J. and {Burgasser}, Adam J. and {Bezanson}, Rachel and {Atek}, Hakim and {Brammer}, Gabriel and {Cutler}, Sam E. and {Feldmann}, Robert and {Fujimoto}, Seiji and {Glazebrook}, Karl and {de Graaff}, Anna and {Khullar}, Gourav and {Leja}, Joel and {Marchesini}, Danilo and {Maseda}, Michael V. and {Matthee}, Jorryt and {Miller}, Tim B. and {Naidu}, Rohan P. and {Nanayakkara}, Themiya and {Oesch}, Pascal A. and {Pan}, Richard and {Papovich}, Casey and {Price}, Sedona H. and {van Dokkum}, Pieter and {Weaver}, John R. and {Whitaker}, Katherine E. and {Zitrin}, Adi},
        title = "{UNCOVER Spectroscopy Confirms the Surprising Ubiquity of Active Galactic Nuclei in Red Sources at z > 5}",
      journal = {\apj},
         year = 2024,
        month = mar,
       volume = {964},
       number = {1},
          eid = {39},
        pages = {39},
          doi = {10.3847/1538-4357/ad1e5f},
archivePrefix = {arXiv},
       eprint = {2309.05714},
 primaryClass = {astro-ph.GA},
       adsurl = {https://ui.adsabs.harvard.edu/abs/2024ApJ...964...39G}
}

@misc{Furtak2024,
      title={A high black hole to host mass ratio in a lensed AGN in the early Universe}, 
      author={Lukas J. Furtak and Ivo Labbé and Adi Zitrin and Jenny E. Greene and Pratika Dayal and Iryna Chemerynska and Vasily Kokorev and Tim B. Miller and Andy D. Goulding and Anna de Graaff and Rachel Bezanson and Gabriel B. Brammer and Sam E. Cutler and Joel Leja and Richard Pan and Sedona H. Price and Bingjie Wang and John R. Weaver and Katherine E. Whitaker and Hakim Atek and Ákos Bogdán and Stéphane Charlot and Emma Curtis-Lake and Pieter van Dokkum and Ryan Endsley and Yoshinobu Fudamoto and Seiji Fujimoto and Karl Glazebrook and Stéphanie Juneau and Danilo Marchesini and Michael V. Maseda and Erica Nelson and Pascal A. Oesch and Adèle Plat and David J. Setton and Daniel P. Stark and Christina C. Williams},
      year={2024},
      eprint={2308.05735},
      archivePrefix={arXiv},
      primaryClass={astro-ph.GA},
      url={https://arxiv.org/abs/2308.05735}, 
}

@misc{Taylor2025,
      title={Broad-Line AGN at 3.5<z<6: The Black Hole Mass Function and a Connection with Little Red Dots}, 
      author={Anthony J. Taylor and Steven L. Finkelstein and Dale D. Kocevski and Junehyoung Jeon and Volker Bromm and Ricardo O. Amorin and Pablo Arrabal Haro and Bren E. Backhaus and Micaela B. Bagley and Eduardo Bañados and Rachana Bhatawdekar and Madisyn Brooks and Antonello Calabro and Oscar A. Chavez Ortiz and Yingjie Cheng and Nikko J. Cleri and Justin W. Cole and Kelcey Davis and Mark Dickinson and Callum Donnan and James S. Dunlop and Richard S. Ellis and Vital Fernandez and Adriano Fontana and Seiji Fujimoto and Mauro Giavalisco and Andrea Grazian and Jingsong Guo and Nimish P. Hathi and Benne W. Holwerda and Michaela Hirschmann and Kohei Inayoshi and Jeyhan S. Kartaltepe and Yana Khusanova and Anton M. Koekemoer and Vasily Kokorev and Rebecca L. Larson and Gene C. K. Leung and Ray A. Lucas and Derek J. McLeod and Lorenzo Napolitano and Masafusa Onoue and Fabio Pacucci and Casey Papovich and Pablo G. Pérez-González and Nor Pirzkal and Rachel S. Somerville and Jonathan R. Trump and Stephen M. Wilkins and L. Y. Aaron Yung and Haowen Zhang},
      year={2025},
      eprint={2409.06772},
      archivePrefix={arXiv},
      primaryClass={astro-ph.GA},
      url={https://arxiv.org/abs/2409.06772}, 
}

@article{Mo1998,
   title={The formation of galactic discs},
   volume={295},
   ISSN={1365-2966},
   url={http://dx.doi.org/10.1046/j.1365-8711.1998.01227.x},
   DOI={10.1046/j.1365-8711.1998.01227.x},
   number={2},
   journal={Monthly Notices of the Royal Astronomical Society},
   publisher={Oxford University Press (OUP)},
   author={Mo, H. J. and Mao, S. and White, S. D. M.},
   year={1998},
   month=apr, pages={319–336} }

@ARTICLE{Begelman2006,
       author = {{Begelman}, Mitchell C. and {Volonteri}, Marta and {Rees}, Martin J.},
        title = "{Formation of supermassive black holes by direct collapse in pre-galactic haloes}",
      journal = {\mnras},
         year = 2006,
        month = jul,
       volume = {370},
       number = {1},
        pages = {289-298},
          doi = {10.1111/j.1365-2966.2006.10467.x},
archivePrefix = {arXiv},
       eprint = {astro-ph/0602363},
 primaryClass = {astro-ph},
       adsurl = {https://ui.adsabs.harvard.edu/abs/2006MNRAS.370..289B}
}

@misc{Wang2024,
      title={Revisiting the H$\beta$ Size-Luminosity Relation Using a Uniform Reverberation-Mapping Analysis}, 
      author={Shu Wang and Jong-Hak Woo},
      year={2024},
      eprint={2408.15872},
      archivePrefix={arXiv},
      primaryClass={astro-ph.GA},
      url={https://arxiv.org/abs/2408.15872}, 
}

@misc{Kido2025,
      title={Black Hole Envelopes in Little Red Dots}, 
      author={Daisaburo Kido and Kunihito Ioka and Kenta Hotokezaka and Kohei Inayoshi and Christopher M. Irwin},
      year={2025},
      eprint={2505.06965},
      archivePrefix={arXiv},
      primaryClass={astro-ph.HE},
      url={https://arxiv.org/abs/2505.06965}, 
}

@misc{Inayoshi2025First,
      title={Little Red Dots as the Very First Activity of Black Hole Growth}, 
      author={Kohei Inayoshi},
      year={2025},
      eprint={2503.05537},
      archivePrefix={arXiv},
      primaryClass={astro-ph.GA},
      url={https://arxiv.org/abs/2503.05537}, 
}

@misc{Lin2024,
      title={A SPectroscopic survey of biased halos In the Reionization Era (ASPIRE): Broad-line AGN at $z=4-5$ revealed by JWST/NIRCam WFSS}, 
      author={Xiaojing Lin and Feige Wang and Xiaohui Fan and Zheng Cai and Jaclyn B. Champagne and Fengwu Sun and Marta Volonteri and Jinyi Yang and Joseph F. Hennawi and Eduardo Bañados and Aaron Barth and Anna-Christina Eilers and Emanuele Paolo Farina and Weizhe Liu and Xiangyu Jin and Hyunsung D. Jun and Alessandro Lupi and Koki Kakiichi and Chiara Mazzucchelli and Masafusa Onoue and Zhiwei Pan and Elia Pizzati and Sofía Rojas-Ruiz and Jan-Torge Schindler and Benny Trakhtenbrot and Yue Shen and Maxime Trebitsch and Ming-Yang Zhuang and Ryan Endsley and Romain A. Meyer and Zihao Li and Mingyu Li and Maria Pudoka and Wei Leong Tee and Yunjing Wu and Haowen Zhang},
      year={2024},
      eprint={2407.17570},
      archivePrefix={arXiv},
      primaryClass={astro-ph.GA},
      url={https://arxiv.org/abs/2407.17570}, 
}

@misc{Pacucci2026,
      title={The Little Red Dots Are Direct Collapse Black Holes}, 
      author={Fabio Pacucci and Andrea Ferrara and Dale D. Kocevski},
      year={2026},
      eprint={2601.14368},
      archivePrefix={arXiv},
      primaryClass={astro-ph.GA},
      url={https://arxiv.org/abs/2601.14368}, 
}

@ARTICLE{Baggen2026,
       author = {{Baggen}, Josephine F.~W. and {Scoggins}, Matthew T. and {van Dokkum}, Pieter and {Haiman}, Zolt{\'a}n and {Torralba}, Alberto and {Matthee}, Jorryt},
        title = "{Connecting the Dots: UV-Bright Companions of Little Red Dots as Lyman-Werner Sources Enabling Direct Collapse Black Hole Formation}",
      journal = {arXiv e-prints},
         year = 2026,
        month = feb,
          eid = {arXiv:2602.02702},
        pages = {arXiv:2602.02702},
          doi = {10.48550/arXiv.2602.02702},
archivePrefix = {arXiv},
       eprint = {2602.02702},
 primaryClass = {astro-ph.GA},
       adsurl = {https://ui.adsabs.harvard.edu/abs/2026arXiv260202702B}
}

@misc{Maiolino2025,
      title={JWST meets Chandra: a large population of Compton thick, feedback-free, and intrinsically X-ray weak AGN, with a sprinkle of SNe}, 
      author={Roberto Maiolino and Guido Risaliti and Matilde Signorini and Bartolomeo Trefoloni and Ignas Juodzbalis and Jan Scholtz and Hannah Uebler and Francesco D'Eugenio and Stefano Carniani and Andy Fabian and Xihan Ji and Giovanni Mazzolari and Elena Bertola and Marcella Brusa and Andrew J. Bunker and Stephane Charlot and Andrea Comastri and Giovanni Cresci and Christa Noel DeCoursey and Eiichi Egami and Fabrizio Fiore and Roberto Gilli and Michele Perna and Sandro Tacchella and Giacomo Venturi},
      year={2024},
      eprint={2405.00504},
      archivePrefix={arXiv},
      primaryClass={astro-ph.GA},
      url={https://arxiv.org/abs/2405.00504}, 
}

@misc{Yue2024,
      title={Stacking X-ray Observations of "Little Red Dots": Implications for their AGN Properties}, 
      author={Minghao Yue and Anna-Christina Eilers and Tonima Tasnim Ananna and Christos Panagiotou and Erin Kara and Takamitsu Miyaji},
      year={2024},
      eprint={2404.13290},
      archivePrefix={arXiv},
      primaryClass={astro-ph.GA},
      url={https://arxiv.org/abs/2404.13290}, 
}

@misc{Akins2024,
      title={COSMOS-Web: The over-abundance and physical nature of "little red dots"--Implications for early galaxy and SMBH assembly}, 
      author={Hollis B. Akins and Caitlin M. Casey and Erini Lambrides and Natalie Allen and Irham T. Andika and Malte Brinch and Jaclyn B. Champagne and Olivia Cooper and Xuheng Ding and Nicole E. Drakos and Andreas Faisst and Steven L. Finkelstein and Maximilien Franco and Seiji Fujimoto and Fabrizio Gentile and Steven Gillman and Ghassem Gozaliasl and Santosh Harish and Christopher C. Hayward and Michaela Hirschmann and Olivier Ilbert and Jeyhan S. Kartaltepe and Dale D. Kocevski and Anton M. Koekemoer and Vasily Kokorev and Daizhong Liu and Arianna S. Long and Henry Joy McCracken and Jed McKinney and Masafusa Onoue and Louise Paquereau and Alvio Renzini and Jason Rhodes and Brant E. Robertson and Marko Shuntov and John D. Silverman and Takumi S. Tanaka and Sune Toft and Benny Trakhtenbrot and Francesco Valentino and Jorge Zavala},
      year={2024},
      eprint={2406.10341},
      archivePrefix={arXiv},
      primaryClass={astro-ph.GA},
      url={https://arxiv.org/abs/2406.10341}, 
}

@article{Gloudemans2025,
   title={Another Piece to the Puzzle: Radio Detection of a JWST-detected Active Galactic Nucleus Candidate},
   volume={986},
   ISSN={1538-4357},
   url={http://dx.doi.org/10.3847/1538-4357/adddb9},
   DOI={10.3847/1538-4357/adddb9},
   number={2},
   journal={The Astrophysical Journal},
   publisher={American Astronomical Society},
   author={Gloudemans, Anniek J. and Duncan, Kenneth J. and Eilers, Anna-Christina and Farina, Emanuele Paolo and Harikane, Yuichi and Inayoshi, Kohei and Lambrides, Erini and Vardoulaki, Eleni},
   year={2025},
   month=jun, pages={130} }

@misc{Perezgonzalez2024,
      title={What is the nature of Little Red Dots and what is not, MIRI SMILES edition}, 
      author={Pablo G. Pérez-González and Guillermo Barro and George H. Rieke and Jianwei Lyu and Marcia Rieke and Stacey Alberts and Christina Williams and Kevin Hainline and Fengwu Sun and David Puskas and Marianna Annunziatella and William M. Baker and Andrew J. Bunker and Eiichi Egami and Zhiyuan Ji and Benjamin D. Johnson and Brant Robertson and Bruno Rodriguez Del Pino and Wiphu Rujopakarn and Irene Shivaei and Sandro Tacchella and Christopher N. A. Willmer and Chris Willott},
      year={2024},
      eprint={2401.08782},
      archivePrefix={arXiv},
      primaryClass={astro-ph.GA},
      url={https://arxiv.org/abs/2401.08782}, 
}

@misc{Hainline2024,
      title={An Investigation Into The Selection and Colors of Little Red Dots and Active Galactic Nuclei}, 
      author={Kevin N. Hainline and Roberto Maiolino and Ignas Juodzbalis and Jan Scholtz and Hannah Ubler and Francesco D'Eugenio and Jakob M. Helton and Yang Sun and Fengwu Sun and Brant Robertson and Sandro Tacchella and Andrew J. Bunker and Stefano Carniani and Stephane Charlot and Emma Curtis-Lake and Eiichi Egami and Benjamin D. Johnson and Xiaojing Lin and Jianwei Lyu and Pablo G. Perez-Gonzalez and Pierluigi Rinaldi and Maddie S. Silcock and Giacomo Venturi and Christina C. Williams and Christopher N. A. Willmer and Chris Willott and Junyu Zhang and Yongda Zhu},
      year={2024},
      eprint={2410.00100},
      archivePrefix={arXiv},
      primaryClass={astro-ph.GA},
      url={https://arxiv.org/abs/2410.00100}, 
}

@article{Shang2010,
   title={Supermassive black hole formation by direct collapse: keeping protogalactic gas H2free in dark matter haloes with virial temperaturesTvir&gt;rsim 104K},
   volume={402},
   ISSN={1365-2966},
   url={http://dx.doi.org/10.1111/j.1365-2966.2009.15960.x},
   DOI={10.1111/j.1365-2966.2009.15960.x},
   number={2},
   journal={Monthly Notices of the Royal Astronomical Society},
   publisher={Oxford University Press (OUP)},
   author={Shang, Cien and Bryan, Greg L. and Haiman, Z.},
   year={2010},
   month=feb, pages={1249–1262} }

@ARTICLE{Kocevski2023,
       author = {{Kocevski}, Dale D. and {Onoue}, Masafusa and {Inayoshi}, Kohei and {Trump}, Jonathan R. and {Arrabal Haro}, Pablo and {Grazian}, Andrea and {Dickinson}, Mark and {Finkelstein}, Steven L. and {Kartaltepe}, Jeyhan S. and {Hirschmann}, Michaela and {Aird}, James and {Holwerda}, Benne W. and {Fujimoto}, Seiji and {Juneau}, St{\'e}phanie and {Amor{\'\i}n}, Ricardo O. and {Backhaus}, Bren E. and {Bagley}, Micaela B. and {Barro}, Guillermo and {Bell}, Eric F. and {Bisigello}, Laura and {Calabr{\`o}}, Antonello and {Cleri}, Nikko J. and {Cooper}, M.~C. and {Ding}, Xuheng and {Grogin}, Norman A. and {Ho}, Luis C. and {Hutchison}, Taylor A. and {Inoue}, Akio K. and {Jiang}, Linhua and {Jones}, Brenda and {Koekemoer}, Anton M. and {Li}, Wenxiu and {Li}, Zhengrong and {McGrath}, Elizabeth J. and {Molina}, Juan and {Papovich}, Casey and {P{\'e}rez-Gonz{\'a}lez}, Pablo G. and {Pirzkal}, Nor and {Wilkins}, Stephen M. and {Yang}, Guang and {Yung}, L.~Y. Aaron},
        title = "{Hidden Little Monsters: Spectroscopic Identification of Low-mass, Broad-line AGNs at z > 5 with CEERS}",
      journal = {\apjl},
         year = 2023,
        month = sep,
       volume = {954},
       number = {1},
          eid = {L4},
        pages = {L4},
          doi = {10.3847/2041-8213/ace5a0},
archivePrefix = {arXiv},
       eprint = {2302.00012},
 primaryClass = {astro-ph.GA},
       adsurl = {https://ui.adsabs.harvard.edu/abs/2023ApJ...954L...4K}
}

@ARTICLE{Wang2024a,
       author = {{Wang}, Bingjie and {Leja}, Joel and {de Graaff}, Anna and {Brammer}, Gabriel B. and {Weibel}, Andrea and {van Dokkum}, Pieter and {Baggen}, Josephine F.~W. and {Suess}, Katherine A. and {Greene}, Jenny E. and {Bezanson}, Rachel and {Cleri}, Nikko J. and {Hirschmann}, Michaela and {Labb{\'e}}, Ivo and {Matthee}, Jorryt and {McConachie}, Ian and {Naidu}, Rohan P. and {Nelson}, Erica and {Oesch}, Pascal A. and {Setton}, David J. and {Williams}, Christina C.},
        title = "{RUBIES: Evolved Stellar Populations with Extended Formation Histories at z {\ensuremath{\sim}} 7─8 in Candidate Massive Galaxies Identified with JWST/NIRSpec}",
      journal = {\apjl},
         year = 2024,
        month = jul,
       volume = {969},
       number = {1},
          eid = {L13},
        pages = {L13},
          doi = {10.3847/2041-8213/ad55f7},
archivePrefix = {arXiv},
       eprint = {2405.01473},
 primaryClass = {astro-ph.GA},
       adsurl = {https://ui.adsabs.harvard.edu/abs/2024ApJ...969L..13W}
}

@ARTICLE{deGraaff2025,
       author = {{de Graaff}, Anna and {Rix}, Hans-Walter and {Naidu}, Rohan P. and {Labb{\'e}}, Ivo and {Wang}, Bingjie and {Leja}, Joel and {Matthee}, Jorryt and {Katz}, Harley and {Greene}, Jenny E. and {Hviding}, Raphael E. and {Baggen}, Josephine and {Bezanson}, Rachel and {Boogaard}, Leindert A. and {Brammer}, Gabriel and {Dayal}, Pratika and {van Dokkum}, Pieter and {Goulding}, Andy D. and {Hirschmann}, Michaela and {Maseda}, Michael V. and {McConachie}, Ian and {Miller}, Tim B. and {Nelson}, Erica and {Oesch}, Pascal A. and {Setton}, David J. and {Shivaei}, Irene and {Weibel}, Andrea and {Whitaker}, Katherine E. and {Williams}, Christina C.},
        title = "{A remarkable ruby: Absorption in dense gas, rather than evolved stars, drives the extreme Balmer break of a little red dot at z = 3.5}",
      journal = {\aap},
         year = 2025,
        month = sep,
       volume = {701},
          eid = {A168},
        pages = {A168},
          doi = {10.1051/0004-6361/202554681},
archivePrefix = {arXiv},
       eprint = {2503.16600},
 primaryClass = {astro-ph.GA},
       adsurl = {https://ui.adsabs.harvard.edu/abs/2025A&A...701A.168D}
}

@ARTICLE{Taylor2025a,
       author = {{Taylor}, Anthony J. and {Kokorev}, Vasily and {Kocevski}, Dale D. and {Akins}, Hollis B. and {Cullen}, Fergus and {Dickinson}, Mark and {Finkelstein}, Steven L. and {Arrabal Haro}, Pablo and {Bromm}, Volker and {Giavalisco}, Mauro and {Inayoshi}, Kohei and {Juneau}, St{\'e}phanie and {Leung}, Gene C.~K. and {P{\'e}rez-Gonz{\'a}lez}, Pablo G. and {Somerville}, Rachel S. and {Trump}, Jonathan R. and {Amor{\'\i}n}, Ricardo O. and {Barro}, Guillermo and {Burgarella}, Denis and {Brooks}, Madisyn and {Carnall}, Adam C. and {Casey}, Caitlin M. and {Cheng}, Yingjie and {Chisholm}, John and {Chworowsky}, Katherine and {Davis}, Kelcey and {Donnan}, Callum T. and {Dunlop}, James S. and {Ellis}, Richard S. and {Fern{\'a}ndez}, Vital and {Fujimoto}, Seiji and {Grogin}, Norman A. and {Gupta}, Ansh R. and {Hathi}, Nimish P. and {Jung}, Intae and {Hirschmann}, Michaela and {Kartaltepe}, Jeyhan S. and {Koekemoer}, Anton M. and {Larson}, Rebecca L. and {Leung}, Ho-Hin and {Llerena}, Mario and {Lucas}, Ray A. and {McLeod}, Derek J. and {McLure}, Ross and {Napolitano}, Lorenzo and {Papovich}, Casey and {Stanton}, Thomas M. and {Tripodi}, Roberta and {Wang}, Xin and {Wilkins}, Stephen M. and {Yung}, L.~Y. Aaron and {Zavala}, Jorge A.},
        title = "{CAPERS-LRD-z9: A Gas-enshrouded Little Red Dot Hosting a Broad-line Active Galactic Nucleus at z = 9.288}",
      journal = {\apjl},
         year = 2025,
        month = aug,
       volume = {989},
       number = {1},
          eid = {L7},
        pages = {L7},
          doi = {10.3847/2041-8213/ade789},
archivePrefix = {arXiv},
       eprint = {2505.04609},
 primaryClass = {astro-ph.GA},
       adsurl = {https://ui.adsabs.harvard.edu/abs/2025ApJ...989L...7T}
}

@ARTICLE{Inayoshi2025a,
       author = {{Inayoshi}, Kohei and {Maiolino}, Roberto},
        title = "{Extremely Dense Gas around Little Red Dots and High-redshift Active Galactic Nuclei: A Nonstellar Origin of the Balmer Break and Absorption Features}",
      journal = {\apjl},
         year = 2025,
        month = feb,
       volume = {980},
       number = {2},
          eid = {L27},
        pages = {L27},
          doi = {10.3847/2041-8213/adaebd},
archivePrefix = {arXiv},
       eprint = {2409.07805},
 primaryClass = {astro-ph.GA},
       adsurl = {https://ui.adsabs.harvard.edu/abs/2025ApJ...980L..27I}
}

@ARTICLE{Liu2025,
       author = {{Liu}, Hanpu and {Jiang}, Yan-Fei and {Quataert}, Eliot and {Greene}, Jenny E. and {Ma}, Yilun},
        title = "{The Balmer Break and Optical Continuum of Little Red Dots from Super-Eddington Accretion}",
      journal = {\apj},
         year = 2025,
        month = nov,
       volume = {994},
       number = {1},
          eid = {113},
        pages = {113},
          doi = {10.3847/1538-4357/ae0c19},
archivePrefix = {arXiv},
       eprint = {2507.07190},
 primaryClass = {astro-ph.GA},
       adsurl = {https://ui.adsabs.harvard.edu/abs/2025ApJ...994..113L}
}

@ARTICLE{Madau2026,
       author = {{Madau}, Piero},
        title = "{Chasing the light: Shadowing, collimation, and the super-Eddington growth of infant black holes in JWST broad-line AGNs}",
      journal = {\aap},
         year = 2026,
        month = apr,
       volume = {708},
          eid = {A116},
        pages = {A116},
          doi = {10.1051/0004-6361/202659244},
archivePrefix = {arXiv},
       eprint = {2501.09854},
 primaryClass = {astro-ph.HE},
       adsurl = {https://ui.adsabs.harvard.edu/abs/2026A&A...708A.116M}
}

@ARTICLE{Madau2026a,
       author = {{Madau}, Piero and {Maiolino}, Roberto},
        title = "{Little Red Dots as Obscured Little Blue Dots: A Super-Eddington Unification Model}",
      journal = {arXiv e-prints},
         year = 2026,
        month = feb,
          eid = {arXiv:2602.22386},
        pages = {arXiv:2602.22386},
          doi = {10.48550/arXiv.2602.22386},
archivePrefix = {arXiv},
       eprint = {2602.22386},
 primaryClass = {astro-ph.GA},
       adsurl = {https://ui.adsabs.harvard.edu/abs/2026arXiv260222386M}
}

@ARTICLE{Yan2026,
       author = {{Yan}, Zu and {Inayoshi}, Kohei and {Chen}, Kejian and {Guo}, Jingsong},
        title = "{Balmer Transition Signatures from Gas-enshrouded, Dust-poor Active Galactic Nuclei}",
      journal = {\apj},
         year = 2026,
        month = may,
       volume = {1002},
       number = {2},
          eid = {159},
        pages = {159},
          doi = {10.3847/1538-4357/ae5dbd},
archivePrefix = {arXiv},
       eprint = {2512.11050},
 primaryClass = {astro-ph.GA},
       adsurl = {https://ui.adsabs.harvard.edu/abs/2026ApJ..1002..159Y}
}

@ARTICLE{Ji2025,
       author = {{Ji}, Xihan and {Maiolino}, Roberto and {{\"U}bler}, Hannah and {Scholtz}, Jan and {D'Eugenio}, Francesco and {Sun}, Fengwu and {Perna}, Michele and {Turner}, Hannah and {Carniani}, Stefano and {Arribas}, Santiago and {Bennett}, Jake S. and {Bunker}, Andrew and {Charlot}, St{\'e}phane and {Cresci}, Giovanni and {Curti}, Mirko and {Egami}, Eiichi and {Fabian}, Andy and {Inayoshi}, Kohei and {Isobe}, Yuki and {Jones}, Gareth and {Juod{\v{z}}balis}, Ignas and {Kumari}, Nimisha and {Lyu}, Jianwei and {Mazzolari}, Giovanni and {Parlanti}, Eleonora and {Robertson}, Brant and {Rodr{\'\i}guez Del Pino}, Bruno and {Schneider}, Raffaella and {Sijacki}, Debora and {Tacchella}, Sandro and {Trinca}, Alessandro and {Valiante}, Rosa and {Venturi}, Giacomo and {Volonteri}, Marta and {Willott}, Chris and {Witten}, Callum and {Witstok}, Joris},
        title = "{BlackTHUNDER ─ A non-stellar Balmer break in a black hole-dominated little red dot at z = 7.04}",
      journal = {\mnras},
         year = 2025,
        month = dec,
       volume = {544},
       number = {4},
        pages = {3900-3935},
          doi = {10.1093/mnras/staf1867},
archivePrefix = {arXiv},
       eprint = {2501.13082},
 primaryClass = {astro-ph.GA},
       adsurl = {https://ui.adsabs.harvard.edu/abs/2025MNRAS.544.3900J}
}

@ARTICLE{Inayoshi2025b,
       author = {{Inayoshi}, Kohei and {Kimura}, Shigeo S. and {Noda}, Hirofumi},
        title = "{Weakness of X-rays and variability in high-redshift active galactic nuclei with super-Eddington accretion}",
      journal = {\pasj},
         year = 2025,
        month = aug,
       volume = {77},
       number = {4},
        pages = {811-822},
          doi = {10.1093/pasj/psaf050},
archivePrefix = {arXiv},
       eprint = {2412.03653},
 primaryClass = {astro-ph.HE},
       adsurl = {https://ui.adsabs.harvard.edu/abs/2025PASJ...77..811I}
}

@misc{Brazzini2026,
      title={The Little Blue and Red Dots Rosetta Stones: Non-Gaussian broad lines, hot dust, and X-ray weakness}, 
      author={M. Brazzini and F. D'Eugenio and R. Maiolino and J. Lyu and C. DeCoursey and H. Übler and X. Ji and I. Juodžbalis and J. Scholtz and G. C. Jones and K. Hainline and E. Dalla Bontà and P. G. P érez-González and S. Geris and A. Harshan and C. Feruglio and M. Bischetti and G. Mazzolari and G. Rieke and S. Alberts and B. Trefoloni and S. Carniani and E. Parlanti and A. Marconi and G. Risaliti and C. Ramos Almeida and P. Rinaldi and M. Perna and S. Zamora and I. Lamperti and G. Venturi and G. Cresci and Andrew J. Bunker and L. R. Ivey},
      year={2026},
      eprint={2601.22214},
      archivePrefix={arXiv},
      primaryClass={astro-ph.GA},
      url={https://arxiv.org/abs/2601.22214}, 
}

@ARTICLE{Rusakov2026,
       author = {{Rusakov}, V. and {Watson}, D. and {Nikopoulos}, G.~P. and {Brammer}, G. and {Gottumukkala}, R. and {Harvey}, T. and {Heintz}, K.~E. and {Damgaard}, R. and {Sim}, S.~A. and {Sneppen}, A. and {Vijayan}, A.~P. and {Adams}, N. and {Austin}, D. and {Conselice}, C.~J. and {Goolsby}, C.~M. and {Toft}, S. and {Witstok}, J.},
        title = "{Little red dots as young supermassive black holes in dense ionized cocoons}",
      journal = {\nat},
         year = 2026,
        month = jan,
       volume = {649},
       number = {8097},
        pages = {574-579},
          doi = {10.1038/s41586-025-09900-4},
archivePrefix = {arXiv},
       eprint = {2503.16595},
 primaryClass = {astro-ph.GA},
       adsurl = {https://ui.adsabs.harvard.edu/abs/2026Natur.649..574R}
}

@ARTICLE{Scholtz2026,
       author = {{Scholtz}, J. and {D'Eugenio}, F. and {Maiolino}, R. and {Brazzini}, M. and {{\"U}bler}, H. and {Ji}, X. and {Perna}, M. and {Sun}, F. and {Brocchi}, G. and {Carniani}, S. and {Cresci}, G. and {Ivey}, L.~R. and {Juod{\v{z}}balis}, I. and {Marconi}, A. and {Mazzolari}, G. and {Risaliti}, G. and {Trefoloni}, B.},
        title = "{Little Red and Blue Dots: simply stratified Broad Line Regions}",
      journal = {arXiv e-prints},
         year = 2026,
        month = mar,
          eid = {arXiv:2603.22277},
        pages = {arXiv:2603.22277},
          doi = {10.48550/arXiv.2603.22277},
archivePrefix = {arXiv},
       eprint = {2603.22277},
 primaryClass = {astro-ph.GA},
       adsurl = {https://ui.adsabs.harvard.edu/abs/2026arXiv260322277S}
}

@ARTICLE{Juodzbalis2026,
       author = {{Juod{\v{z}}balis}, Ignas and {Maiolino}, Roberto and {Baker}, William M. and {Lake}, Emma Curtis and {Scholtz}, Jan and {D'Eugenio}, Francesco and {Trefoloni}, Bartolomeo and {Isobe}, Yuki and {Tacchella}, Sandro and {Bunker}, Andrew J. and {Carniani}, Stefano and {Charlot}, St{\'e}phane and {Jones}, Gareth C. and {Parlanti}, Eleonora and {Perna}, Michele and {Rinaldi}, Pierluigi and {Robertson}, Brant and {{\"U}bler}, Hannah and {Venturi}, Giacomo and {Willott}, Chris},
        title = "{JADES: comprehensive census of broad-line AGN from reionization to cosmic noon revealed by JWST}",
      journal = {\mnras},
         year = 2026,
        month = mar,
       volume = {546},
       number = {3},
          eid = {stag086},
        pages = {stag086},
          doi = {10.1093/mnras/stag086},
archivePrefix = {arXiv},
       eprint = {2504.03551},
 primaryClass = {astro-ph.GA},
       adsurl = {https://ui.adsabs.harvard.edu/abs/2026MNRAS.546ag086J}
}

@ARTICLE{Inayoshi2025c,
       author = {{Inayoshi}, Kohei and {Ho}, Luis C.},
        title = "{A Critical Evaluation of the Physical Nature of the Little Red Dots}",
      journal = {arXiv e-prints},
         year = 2025,
        month = dec,
          eid = {arXiv:2512.03130},
        pages = {arXiv:2512.03130},
          doi = {10.48550/arXiv.2512.03130},
archivePrefix = {arXiv},
       eprint = {2512.03130},
 primaryClass = {astro-ph.GA},
       adsurl = {https://ui.adsabs.harvard.edu/abs/2025arXiv251203130I}
}

@ARTICLE{Chon2018,
       author = {{Chon}, Sunmyon and {Hosokawa}, Takashi and {Yoshida}, Naoki},
        title = "{Radiation hydrodynamics simulations of the formation of direct-collapse supermassive stellar systems}",
      journal = {\mnras},
         year = 2018,
        month = apr,
       volume = {475},
       number = {3},
        pages = {4104-4121},
          doi = {10.1093/mnras/sty086},
archivePrefix = {arXiv},
       eprint = {1711.05262},
 primaryClass = {astro-ph.GA},
       adsurl = {https://ui.adsabs.harvard.edu/abs/2018MNRAS.475.4104C}
}

@ARTICLE{Shi2024,
       author = {{Shi}, Yanlong and {Kremer}, Kyle and {Hopkins}, Philip F.},
        title = "{From Seeds to Supermassive Black Holes: Capture, Growth, Migration, and Pairing in Dense Protobulge Environments}",
      journal = {\apjl},
         year = 2024,
        month = jul,
       volume = {969},
       number = {2},
          eid = {L31},
        pages = {L31},
          doi = {10.3847/2041-8213/ad5a95},
archivePrefix = {arXiv},
       eprint = {2405.17338},
 primaryClass = {astro-ph.GA},
       adsurl = {https://ui.adsabs.harvard.edu/abs/2024ApJ...969L..31S}
}

@ARTICLE{Loeb1997,
       author = {{Loeb}, Abraham and {Ulmer}, Andrew},
        title = "{Optical Appearance of the Debris of a Star Disrupted by a Massive Black Hole}",
      journal = {\apj},
         year = 1997,
        month = nov,
       volume = {489},
       number = {2},
        pages = {573-578},
          doi = {10.1086/304814},
archivePrefix = {arXiv},
       eprint = {astro-ph/9703079},
 primaryClass = {astro-ph},
       adsurl = {https://ui.adsabs.harvard.edu/abs/1997ApJ...489..573L}
}

@ARTICLE{deGraaff2026,
       author = {{de Graaff}, Anna and {Hviding}, Raphael E. and {Naidu}, Rohan P. and {Greene}, Jenny E. and {Miller}, Tim B. and {Leja}, Joel and {Matthee}, Jorryt and {Brammer}, Gabriel and {Katz}, Harley and {Bezanson}, Rachel and {Boogaard}, Leindert A. and {Bose}, Sownak and {Chisholm}, John and {Cleri}, Nikko J. and {Dayal}, Pratika and {Feldmann}, Robert and {Fudamoto}, Yoshinobu and {Fujimoto}, Seiji and {Furtak}, Lukas J. and {Glazebrook}, Karl and {Gottumukkala}, Rashmi and {Heintz}, Kasper E. and {Kokorev}, Vasily and {Labbe}, Ivo and {Maseda}, Michael V. and {McConachie}, Ian and {Nanayakkara}, Themiya and {Nelson}, Erica and {Nowaczyk}, Przemys{\l}aw and {Oesch}, Pascal A. and {Rix}, Hans-Walter and {Setton}, David J. and {Torralba}, Alberto and {Walter}, Fabian and {Wang}, Bingjie and {Weibel}, Andrea and {van der Wel}, Arjen},
        title = "{Little Red Dots host Black Hole Stars: A unified family of gas-reddened AGN revealed by JWST/NIRSpec spectroscopy}",
      journal = {arXiv e-prints},
         year = 2025,
        month = nov,
          eid = {arXiv:2511.21820},
        pages = {arXiv:2511.21820},
          doi = {10.48550/arXiv.2511.21820},
archivePrefix = {arXiv},
       eprint = {2511.21820},
 primaryClass = {astro-ph.GA},
       adsurl = {https://ui.adsabs.harvard.edu/abs/2025arXiv251121820D}
}

@ARTICLE{Kokorev2026,
       author = {{Kokorev}, Vasily and {Chisholm}, John and {Naidu}, Rohan P. and {Fujimoto}, Seiji and {Atek}, Hakim and {Brammer}, Gabriel and {Finkelstein}, Steven L. and {Akins}, Hollis B. and {Berg}, Danielle A. and {Furtak}, Lukas J. and {Fei}, Qinyue and {Hsiao}, Tiger Yu-Yang and {Labb{\'e}}, Ivo and {Matthee}, Jorryt and {Mu{\~n}oz}, Julian B. and {Oesch}, Pascal A. and {Pan}, Richard and {Rinaldi}, Pierluigi and {Saldana-Lopez}, Alberto and {Schaerer}, Daniel and {Volonteri}, Marta and {Zitrin}, Adi},
        title = "{The Deepest GLIMPSE of a Dense Gas Cocoon Enshrouding a Little Red Dot}",
      journal = {\apj},
         year = 2026,
        month = jun,
       volume = {1004},
       number = {2},
          eid = {153},
        pages = {153},
          doi = {10.3847/1538-4357/ae4ed7},
archivePrefix = {arXiv},
       eprint = {2511.07515},
 primaryClass = {astro-ph.GA},
       adsurl = {https://ui.adsabs.harvard.edu/abs/2026ApJ..1004..153K}
}

@ARTICLE{Sneppen2026,
       author = {{Sneppen}, A. and {Watson}, D. and {Matthews}, J.~H. and {Nikopoulos}, G. and {Allen}, N. and {Brammer}, G. and {Damgaard}, R. and {Heintz}, K.~E. and {Knigge}, C. and {Long}, K.~S. and {Rusakov}, V. and {Sim}, S.~A. and {Witstok}, J.},
        title = "{Inside the cocoon: a comprehensive explanation of the spectra of Little Red Dots}",
      journal = {arXiv e-prints},
         year = 2026,
        month = jan,
          eid = {arXiv:2601.18864},
        pages = {arXiv:2601.18864},
          doi = {10.48550/arXiv.2601.18864},
archivePrefix = {arXiv},
       eprint = {2601.18864},
 primaryClass = {astro-ph.GA},
       adsurl = {https://ui.adsabs.harvard.edu/abs/2026arXiv260118864S}
}

@ARTICLE{Asada2026,
       author = {{Asada}, Yoshihisa and {Inayoshi}, Kohei and {Fei}, Qinyue and {Fujimoto}, Seiji and {Willott}, Chris J.},
        title = "{Origins of the UV Continuum and Balmer Emission Lines in Little Red Dots: Observational Validation of Dense Gas Envelope Models Enshrouding the Active Galactic Nucleus}",
      journal = {\apjl},
         year = 2026,
        month = jul,
       volume = {1005},
       number = {1},
          eid = {L22},
        pages = {L22},
          doi = {10.3847/2041-8213/ae7d0a},
       adsurl = {https://ui.adsabs.harvard.edu/abs/2026ApJ..1005L..22A}
}

@ARTICLE{Volonteri2005a,
       author = {{Volonteri}, Marta and {Rees}, Martin J.},
        title = "{Rapid Growth of High-Redshift Black Holes}",
      journal = {\apj},
         year = 2005,
        month = nov,
       volume = {633},
       number = {2},
        pages = {624-629},
          doi = {10.1086/466521},
archivePrefix = {arXiv},
       eprint = {astro-ph/0506040},
 primaryClass = {astro-ph},
       adsurl = {https://ui.adsabs.harvard.edu/abs/2005ApJ...633..624V}
}

@misc{Steinberg2023,
      title={Stream-Disk Shocks as the Origins of Peak Light in Tidal Disruption Events}, 
      author={Elad Steinberg and Nicholas C. Stone},
      year={2023},
      eprint={2206.10641},
      archivePrefix={arXiv},
      primaryClass={astro-ph.HE},
      url={https://arxiv.org/abs/2206.10641}, 
}

@article{Strubbe2009,
   title={Optical flares from the tidal disruption of stars by massive black holes: Optical signatures of tidally disrupted stars},
   volume={400},
   ISSN={1365-2966},
   url={http://dx.doi.org/10.1111/j.1365-2966.2009.15599.x},
   DOI={10.1111/j.1365-2966.2009.15599.x},
   number={4},
   journal={Monthly Notices of the Royal Astronomical Society},
   publisher={Oxford University Press (OUP)},
   author={Strubbe, Linda E. and Quataert, Eliot},
   year={2009},
   month=Dec, pages={2070–2084} }

@article{Metzger2016,
   title={A bright year for tidal disruptions},
   volume={461},
   ISSN={1365-2966},
   url={http://dx.doi.org/10.1093/mnras/stw1394},
   DOI={10.1093/mnras/stw1394},
   number={1},
   journal={Monthly Notices of the Royal Astronomical Society},
   publisher={Oxford University Press (OUP)},
   author={Metzger, Brian D. and Stone, Nicholas C.},
   year={2016},
   month=June, pages={948–966} }

@article{Ryu2021,
   title={The Impact of Shocks on the Vertical Structure of Eccentric Disks},
   volume={920},
   ISSN={1538-4357},
   url={http://dx.doi.org/10.3847/1538-4357/ac185a},
   DOI={10.3847/1538-4357/ac185a},
   number={2},
   journal={The Astrophysical Journal},
   publisher={American Astronomical Society},
   author={Ryu, Taeho and Krolik, Julian and Piran, Tsvi},
   year={2021},
   month=Oct, pages={130} }

@ARTICLE{Roth2016,
       author = {{Roth}, Nathaniel and {Kasen}, Daniel and {Guillochon}, James and {Ramirez-Ruiz}, Enrico},
        title = "{The X-Ray through Optical Fluxes and Line Strengths of Tidal Disruption Events}",
      journal = {\apj},
         year = 2016,
        month = aug,
       volume = {827},
       number = {1},
          eid = {3},
        pages = {3},
          doi = {10.3847/0004-637X/827/1/3},
archivePrefix = {arXiv},
       eprint = {1510.08454},
 primaryClass = {astro-ph.HE},
       adsurl = {https://ui.adsabs.harvard.edu/abs/2016ApJ...827....3R}
}

@ARTICLE{Abramowicz2002,
       author = {{Abramowicz}, Marek A. and {Igumenshchev}, Igor V. and {Quataert}, Eliot and {Narayan}, Ramesh},
        title = "{On the Radial Structure of Radiatively Inefficient Accretion Flows with Convection}",
      journal = {\apj},
         year = 2002,
        month = feb,
       volume = {565},
       number = {2},
        pages = {1101-1106},
          doi = {10.1086/324717},
archivePrefix = {arXiv},
       eprint = {astro-ph/0110371},
 primaryClass = {astro-ph},
       adsurl = {https://ui.adsabs.harvard.edu/abs/2002ApJ...565.1101A}
}

@ARTICLE{Abramowicz2001,
       author = {{Abramowicz}, Marek A. and {Igumenshchev}, Igor V.},
        title = "{How Dim Could Accreting Black Holes Be?}",
      journal = {\apjl},
         year = 2001,
        month = jun,
       volume = {554},
       number = {1},
        pages = {L53-L54},
          doi = {10.1086/320933},
       adsurl = {https://ui.adsabs.harvard.edu/abs/2001ApJ...554L..53A}
}

@ARTICLE{Watarai2001,
       author = {{Watarai}, Ken-Ya and {Mineshige}, Shin},
        title = "{Slim Disk: Viscosity Prescriptions and Observational Implications}",
      journal = {\pasj},
         year = 2001,
        month = oct,
       volume = {53},
       number = {5},
        pages = {915-921},
          doi = {10.1093/pasj/53.5.915},
archivePrefix = {arXiv},
       eprint = {astro-ph/0109019},
 primaryClass = {astro-ph},
       adsurl = {https://ui.adsabs.harvard.edu/abs/2001PASJ...53..915W}
}

@ARTICLE{Ohsuga2005,
       author = {{Ohsuga}, Ken and {Mori}, Masao and {Nakamoto}, Taishi and {Mineshige}, Shin},
        title = "{Supercritical Accretion Flows around Black Holes: Two-dimensional, Radiation Pressure-dominated Disks with Photon Trapping}",
      journal = {\apj},
         year = 2005,
        month = jul,
       volume = {628},
       number = {1},
        pages = {368-381},
          doi = {10.1086/430728},
archivePrefix = {arXiv},
       eprint = {astro-ph/0504168},
 primaryClass = {astro-ph},
       adsurl = {https://ui.adsabs.harvard.edu/abs/2005ApJ...628..368O}
}

@ARTICLE{Jiang2019,
       author = {{Jiang}, Yan-Fei and {Stone}, James M. and {Davis}, Shane W.},
        title = "{Super-Eddington Accretion Disks around Supermassive Black Holes}",
      journal = {\apj},
         year = 2019,
        month = aug,
       volume = {880},
       number = {2},
          eid = {67},
        pages = {67},
          doi = {10.3847/1538-4357/ab29ff},
archivePrefix = {arXiv},
       eprint = {1709.02845},
 primaryClass = {astro-ph.HE},
       adsurl = {https://ui.adsabs.harvard.edu/abs/2019ApJ...880...67J}
}

@ARTICLE{Yuan2014,
       author = {{Yuan}, Feng and {Narayan}, Ramesh},
        title = "{Hot Accretion Flows Around Black Holes}",
      journal = {\araa},
         year = 2014,
        month = aug,
       volume = {52},
        pages = {529-588},
          doi = {10.1146/annurev-astro-082812-141003},
archivePrefix = {arXiv},
       eprint = {1401.0586},
 primaryClass = {astro-ph.HE},
       adsurl = {https://ui.adsabs.harvard.edu/abs/2014ARA&A..52..529Y}
}

@ARTICLE{Liu2026,
       author = {{Liu}, Hanpu and {Jiang}, Yan-Fei and {Quataert}, Eliot and {Greene}, Jenny E. and {Ma}, Yilun and {Lin}, Xiaojing},
        title = "{Synthetic Spectral Library of Optically Thick Atmospheres for Little Red Dots}",
      journal = {arXiv e-prints},
         year = 2026,
        month = mar,
          eid = {arXiv:2603.02317},
        pages = {arXiv:2603.02317},
          doi = {10.48550/arXiv.2603.02317},
archivePrefix = {arXiv},
       eprint = {2603.02317},
 primaryClass = {astro-ph.GA},
       adsurl = {https://ui.adsabs.harvard.edu/abs/2026arXiv260302317L}
}

@ARTICLE{Sun2026,
       author = {{Sun}, Wendy Q. and {Naidu}, Rohan P. and {Matthee}, Jorryt and {de Graaff}, Anna and {Chisholm}, John and {Greene}, Jenny E. and {Oesch}, Pascal A. and {Torralba}, Alberto and {Hviding}, Raphael E. and {Brammer}, Gabriel and {Simcoe}, Robert A. and {Bose}, Sownak and {Bouwens}, Rychard and {Dayal}, Pratika and {Eilers}, Anna-Christina and {Fei}, Qinyue and {Furtak}, Lukas J. and {Gottumukkala}, Rashmi and {Goulding}, Andy and {Heintz}, Kasper E. and {Hirschmann}, Michaela and {Kokorev}, Vasily and {Leja}, Joel and {Liu}, Zhaoran and {Natarajan}, Priyamvada and {Santarelli}, Andrew D. and {Setton}, David J. and {Smith}, Aaron and {Tacchella}, Sandro and {Volonteri}, Marta and {Walter}, Fabian and {Weibel}, Andrea and {Williams}, Christina C.},
        title = "{Little Red Dot {\ensuremath{-}} Host Galaxy = Black Hole Star: A Gas-Enshrouded Heart at the Center of Every Little Red Dot}",
      journal = {The Open Journal of Astrophysics},
         year = 2026,
        month = may,
       volume = {9},
        pages = {62505},
          doi = {10.33232/001c.162505},
archivePrefix = {arXiv},
       eprint = {2601.20929},
 primaryClass = {astro-ph.GA},
       adsurl = {https://ui.adsabs.harvard.edu/abs/2026OJAp....962505S}
}

@ARTICLE{Salpeter1955,
       author = {{Salpeter}, Edwin E.},
        title = "{The Luminosity Function and Stellar Evolution.}",
      journal = {\apj},
         year = 1955,
        month = jan,
       volume = {121},
        pages = {161},
          doi = {10.1086/145971},
       adsurl = {https://ui.adsabs.harvard.edu/abs/1955ApJ...121..161S}
}

@ARTICLE{Inayoshi2026a,
       author = {{Inayoshi}, Kohei and {Shangguan}, Jinyi and {Chen}, Xian and {Ho}, Luis C. and {Haiman}, Zolt{\'a}n},
        title = "{The Emergence of Little Red Dots from Binary Massive Black Holes}",
      journal = {\apj},
         year = 2026,
        month = may,
       volume = {1002},
       number = {1},
          eid = {25},
        pages = {25},
          doi = {10.3847/1538-4357/ae548d},
archivePrefix = {arXiv},
       eprint = {2505.05322},
 primaryClass = {astro-ph.HE},
       adsurl = {https://ui.adsabs.harvard.edu/abs/2026ApJ..1002...25I}
}

@ARTICLE{Gordon2003,
       author = {{Gordon}, Karl D. and {Clayton}, Geoffrey C. and {Misselt}, K.~A. and {Landolt}, Arlo U. and {Wolff}, Michael J.},
        title = "{A Quantitative Comparison of the Small Magellanic Cloud, Large Magellanic Cloud, and Milky Way Ultraviolet to Near-Infrared Extinction Curves}",
      journal = {\apj},
         year = 2003,
        month = sep,
       volume = {594},
       number = {1},
        pages = {279-293},
          doi = {10.1086/376774},
archivePrefix = {arXiv},
       eprint = {astro-ph/0305257},
 primaryClass = {astro-ph},
       adsurl = {https://ui.adsabs.harvard.edu/abs/2003ApJ...594..279G}
}

@ARTICLE{Zhong2026,
       author = {{Zhong}, Yuxing and {Chen}, Xiaoyang and {Ichikawa}, Kohei and {Kong}, Youwen and {Aoki}, Kentaro and {Yamada}, Satoshi and {Nagao}, Tohru and {Kido}, Daisaburo and {Kawaguchi}, Toshihiro and {Matsuoka}, Yoshiki and {Misawa}, Toru and {Mizukoshi}, Shoichiro and {Onoue}, Masafusa and {Takahashi}, Ayumi and {Toba}, Yoshiki},
        title = "{Blackbody Quasar and Radio Source (BBQSORS): A Candidate of Transitional Little Red Dots with a T {\ensuremath{\sim}} {}10$^{4}$ K Blackbody Spectrum}",
      journal = {\apjl},
         year = 2026,
        month = jun,
       volume = {1004},
       number = {1},
          eid = {L2},
        pages = {L2},
          doi = {10.3847/2041-8213/ae633b},
archivePrefix = {arXiv},
       eprint = {2603.13736},
 primaryClass = {astro-ph.GA},
       adsurl = {https://ui.adsabs.harvard.edu/abs/2026ApJ..1004L...2Z}
}

@ARTICLE{Ubler2023,
       author = {{{\"U}bler}, Hannah and {Maiolino}, Roberto and {Curtis-Lake}, Emma and {P{\'e}rez-Gonz{\'a}lez}, Pablo G. and {Curti}, Mirko and {Perna}, Michele and {Arribas}, Santiago and {Charlot}, St{\'e}phane and {Marshall}, Madeline A. and {D'Eugenio}, Francesco and {Scholtz}, Jan and {Bunker}, Andrew and {Carniani}, Stefano and {Ferruit}, Pierre and {Jakobsen}, Peter and {Rix}, Hans-Walter and {Rodr{\'\i}guez Del Pino}, Bruno and {Willott}, Chris J. and {Boeker}, Torsten and {Cresci}, Giovanni and {Jones}, Gareth C. and {Kumari}, Nimisha and {Rawle}, Tim},
        title = "{GA-NIFS: A massive black hole in a low-metallicity AGN at z {\ensuremath{\sim}} 5.55 revealed by JWST/NIRSpec IFS}",
      journal = {\aap},
         year = 2023,
        month = sep,
       volume = {677},
          eid = {A145},
        pages = {A145},
          doi = {10.1051/0004-6361/202346137},
archivePrefix = {arXiv},
       eprint = {2302.06647},
 primaryClass = {astro-ph.GA},
       adsurl = {https://ui.adsabs.harvard.edu/abs/2023A&A...677A.145U}
}

@ARTICLE{Shen2026,
       author = {{Shen}, Xuejian and {Kannan}, Rahul and {Puchwein}, Ewald and {Smith}, Aaron and {Vogelsberger}, Mark and {Borrow}, Josh and {Garaldi}, Enrico and {Keating}, Laura and {Zier}, Oliver and {McClymont}, William and {Tacchella}, Sandro and {Wang}, Zihao and {Hernquist}, Lars},
        title = "{The THESAN-ZOOM project: star formation efficiencies in high-redshift galaxies}",
      journal = {\mnras},
         year = 2026,
        month = feb,
       volume = {545},
       number = {4},
          eid = {staf2119},
        pages = {staf2119},
          doi = {10.1093/mnras/staf2119},
archivePrefix = {arXiv},
       eprint = {2503.01949},
 primaryClass = {astro-ph.GA},
       adsurl = {https://ui.adsabs.harvard.edu/abs/2026MNRAS.545f2119S}
}

@ARTICLE{Wang2025-SFE,
       author = {{Wang}, Zihao and {Shen}, Xuejian and {Vogelsberger}, Mark and {Li}, Hui and {Kannan}, Rahul and {Puchwein}, Ewald and {Smith}, Aaron and {Borrow}, Josh and {Garaldi}, Enrico and {Keating}, Laura and {Zier}, Oliver and {McClymont}, William and {Tacchella}, Sandro and {Ni}, Yang and {Hernquist}, Lars},
        title = "{The THESAN-ZOOM project: star formation efficiency from giant molecular clouds to galactic scale in high-redshift starbursts}",
      journal = {\mnras},
         year = 2025,
        month = dec,
       volume = {544},
       number = {3},
        pages = {2675-2697},
          doi = {10.1093/mnras/staf1677},
archivePrefix = {arXiv},
       eprint = {2505.05554},
 primaryClass = {astro-ph.GA},
       adsurl = {https://ui.adsabs.harvard.edu/abs/2025MNRAS.544.2675W}
}

@ARTICLE{Umeda2026,
       author = {{Umeda}, Hiroya and {Inayoshi}, Kohei and {Harikane}, Yuichi and {Murase}, Kohta},
        title = "{A Black Hole Envelope Interpretation for Cosmological Demographics of Little Red Dots}",
      journal = {\apj},
         year = 2026,
        month = mar,
       volume = {999},
       number = {2},
          eid = {183},
        pages = {183},
          doi = {10.3847/1538-4357/ae4101},
archivePrefix = {arXiv},
       eprint = {2512.04208},
 primaryClass = {astro-ph.GA},
       adsurl = {https://ui.adsabs.harvard.edu/abs/2026ApJ...999..183U}
}

@ARTICLE{Madau2026b,
       author = {{Madau}, Piero and {Maiolino}, Roberto},
        title = "{Little red dots as obscured little blue dots: relative abundances, luminosities, and black-hole masses}",
      journal = {arXiv e-prints},
         year = 2026,
        month = may,
          eid = {arXiv:2605.05074},
        pages = {arXiv:2605.05074},
          doi = {10.48550/arXiv.2605.05074},
archivePrefix = {arXiv},
       eprint = {2605.05074},
 primaryClass = {astro-ph.GA},
       adsurl = {https://ui.adsabs.harvard.edu/abs/2026arXiv260505074M}
}

@ARTICLE{Sok2026,
       author = {{Sok}, Visal and {Nelson}, Erica J. and {Begelman}, Mitchell C. and {Dexter}, Jason and {D'Eugenio}, Francesco and {Greene}, Jenny E. and {Leja}, Joel and {Whitaker}, Katherine E. and {Bunker}, Andrew J. and {P{\'e}rez-Gonz{\'a}lez}, Pablo G. and {Rinaldi}, Pierluigi and {Torralba}, Alberto and {{\"U}bler}, Hannah},
        title = "{Constraints on the Gas Geometry Surrounding Little Red Dots through Narrow-Line Diagnostics}",
      journal = {arXiv e-prints},
         year = 2026,
        month = jun,
          eid = {arXiv:2606.23778},
        pages = {arXiv:2606.23778},
          doi = {10.48550/arXiv.2606.23778},
archivePrefix = {arXiv},
       eprint = {2606.23778},
 primaryClass = {astro-ph.GA},
       adsurl = {https://ui.adsabs.harvard.edu/abs/2026arXiv260623778S}
}

@misc{Ando2026,
      title={The UV Side of Little Red Dots: Red, Compact, and Iron-Enhanced Rest-UV Emission with a Strong Downturn around Ly$\alpha$}, 
      author={Makoto Ando and Yuichi Harikane and Harley Katz and Kohei Inayoshi and Takumi S. Tanaka},
      year={2026},
      eprint={2606.03522},
      archivePrefix={arXiv},
      primaryClass={astro-ph.GA},
      url={https://arxiv.org/abs/2606.03522}, 
}

@ARTICLE{VandenBerk2001,
       author = {{Vanden Berk}, Daniel E. and {Richards}, Gordon T. and {Bauer}, Amanda and {Strauss}, Michael A. and {Schneider}, Donald P. and {Heckman}, Timothy M. and {York}, Donald G. and {Hall}, Patrick B. and {Fan}, Xiaohui and {Knapp}, G.~R. and {Anderson}, Scott F. and {Annis}, James and {Bahcall}, Neta A. and {Bernardi}, Mariangela and {Briggs}, John W. and {Brinkmann}, J. and {Brunner}, Robert and {Burles}, Scott and {Carey}, Larry and {Castander}, Francisco J. and {Connolly}, A.~J. and {Crocker}, J.~H. and {Csabai}, Istv{\'a}n and {Doi}, Mamoru and {Finkbeiner}, Douglas and {Friedman}, Scott and {Frieman}, Joshua A. and {Fukugita}, Masataka and {Gunn}, James E. and {Hennessy}, G.~S. and {Ivezi{\'c}}, {\v{Z}}eljko and {Kent}, Stephen and {Kunszt}, Peter Z. and {Lamb}, D.~Q. and {Leger}, R. French and {Long}, Daniel C. and {Loveday}, Jon and {Lupton}, Robert H. and {Meiksin}, Avery and {Merelli}, Aronne and {Munn}, Jeffrey A. and {Newberg}, Heidi Jo and {Newcomb}, Matt and {Nichol}, R.~C. and {Owen}, Russell and {Pier}, Jeffrey R. and {Pope}, Adrian and {Rockosi}, Constance M. and {Schlegel}, David J. and {Siegmund}, Walter A. and {Smee}, Stephen and {Snir}, Yehuda and {Stoughton}, Chris and {Stubbs}, Christopher and {SubbaRao}, Mark and {Szalay}, Alexander S. and {Szokoly}, Gyula P. and {Tremonti}, Christy and {Uomoto}, Alan and {Waddell}, Patrick and {Yanny}, Brian and {Zheng}, Wei},
        title = "{Composite Quasar Spectra from the Sloan Digital Sky Survey}",
      journal = {\aj},
         year = 2001,
        month = aug,
       volume = {122},
       number = {2},
        pages = {549-564},
          doi = {10.1086/321167},
archivePrefix = {arXiv},
       eprint = {astro-ph/0105231},
 primaryClass = {astro-ph},
       adsurl = {https://ui.adsabs.harvard.edu/abs/2001AJ....122..549V}
}

@ARTICLE{Davis2007,
       author = {{Davis}, Shane W. and {Woo}, Jong-Hak and {Blaes}, Omer M.},
        title = "{The UV Continuum of Quasars: Models and SDSS Spectral Slopes}",
      journal = {\apj},
         year = 2007,
        month = oct,
       volume = {668},
       number = {2},
        pages = {682-698},
          doi = {10.1086/521393},
archivePrefix = {arXiv},
       eprint = {0707.1456},
 primaryClass = {astro-ph},
       adsurl = {https://ui.adsabs.harvard.edu/abs/2007ApJ...668..682D}
}

@ARTICLE{Lusso2015,
       author = {{Lusso}, E. and {Worseck}, G. and {Hennawi}, J.~F. and {Prochaska}, J.~X. and {Vignali}, C. and {Stern}, J. and {O'Meara}, J.~M.},
        title = "{The first ultraviolet quasar-stacked spectrum at z ≃ 2.4 from WFC3}",
      journal = {\mnras},
         year = 2015,
        month = jun,
       volume = {449},
       number = {4},
        pages = {4204-4220},
          doi = {10.1093/mnras/stv516},
archivePrefix = {arXiv},
       eprint = {1503.02075},
 primaryClass = {astro-ph.GA},
       adsurl = {https://ui.adsabs.harvard.edu/abs/2015MNRAS.449.4204L}
}

@article{Sadowski2016,
   title={Energy flows in thick accretion discs and their consequences for black hole feedback},
   volume={456},
   ISSN={1365-2966},
   url={http://dx.doi.org/10.1093/mnras/stv2854},
   DOI={10.1093/mnras/stv2854},
   number={4},
   journal={Monthly Notices of the Royal Astronomical Society},
   publisher={Oxford University Press (OUP)},
   author={Sadowski, Aleksander and Lasota, Jean-Pierre and Abramowicz, Marek A. and Narayan, Ramesh},
   year={2016},
   month=Jan, pages={3915–3928} 
}

@ARTICLE{Tuna2026,
       author = {{Tuna}, Semih and {Metzger}, Brian D. and {Jiang}, Yan-fei and {Antoni}, Andrea},
        title = "{Super-Eddington Chimneys: On the Cooling Evolution of Tidal Disruption Event Envelopes}",
      journal = {\apj},
         year = 2026,
        month = jun,
       volume = {1004},
       number = {1},
          eid = {7},
        pages = {7},
          doi = {10.3847/1538-4357/ae5f68},
archivePrefix = {arXiv},
       eprint = {2512.14810},
 primaryClass = {astro-ph.HE},
       adsurl = {https://ui.adsabs.harvard.edu/abs/2026ApJ..1004....7T}
}

@ARTICLE{Netzer1990,
       author = {{Netzer}, H. and {Maoz}, D. and {Laor}, A. and {Mendelson}, H. and {Brosch}, N. and {Leibowitz}, E. and {Almoznino}, E. and {Beck}, S. and {Mazeh}, T.},
        title = "{High-Rate Spectroscopic Active Galactic Nucleus Monitoring at the Wise Observatory. II. NGC 5548}",
      journal = {\apj},
         year = 1990,
        month = apr,
       volume = {353},
        pages = {108},
          doi = {10.1086/168594},
       adsurl = {https://ui.adsabs.harvard.edu/abs/1990ApJ...353..108N}
}

@ARTICLE{Geris2026,
       author = {{Geris}, Sophia and {Maiolino}, Roberto and {Ji}, Xihan and {Risaliti}, Guido and {Lanzuisi}, Giorgio and {D'Eugenio}, Francesco and {Isobe}, Yuki and {Jones}, Gareth and {Harshan}, Anishya and {Brazzini}, Matilde and {Juod{\v{z}}balis}, Ignas and {Scholtz}, Jan and {Rinaldi}, Pierluigi and {{\"U}bler}, Hannah and {Baker}, William and {Bunker}, Andrew J. and {Brusa}, Marcella and {Carniani}, Stefano and {Charlot}, Stephane and {Curti}, Mirko and {Comastri}, Andrea and {Lake}, Emma Curtis and {Gilli}, Roberto and {Hainline}, Kevin and {Madau}, Piero and {Marchesi}, Stefano and {Mazzolari}, Giovanni and {Napolitano}, Lorenzo and {Parlanti}, Eleonora and {Pentericci}, Laura and {Ramos Almeida}, Cristina and {Robertson}, Brant and {Silcock}, Maddie S. and {Tripodi}, Roberta and {Venturi}, Giacomo and {Vignali}, Cristian and {Vito}, Fabio and {Zhu}, Yongda},
        title = "{Little Red and Blue Dots: AGN-excited narrow lines, Lyman-$α$ emission, and resemblance to standard quasars}",
      journal = {arXiv e-prints},
         year = 2026,
        month = jun,
          eid = {arXiv:2606.21614},
        pages = {arXiv:2606.21614},
          doi = {10.48550/arXiv.2606.21614},
archivePrefix = {arXiv},
       eprint = {2606.21614},
 primaryClass = {astro-ph.GA},
       adsurl = {https://ui.adsabs.harvard.edu/abs/2026arXiv260621614G}
}

@ARTICLE{Shibata2002,
       author = {{Shibata}, Masaru and {Shapiro}, Stuart L.},
        title = "{Collapse of a Rotating Supermassive Star to a Supermassive Black Hole: Fully Relativistic Simulations}",
      journal = {\apjl},
         year = 2002,
        month = jun,
       volume = {572},
       number = {1},
        pages = {L39-L43},
          doi = {10.1086/341516},
archivePrefix = {arXiv},
       eprint = {astro-ph/0205091},
 primaryClass = {astro-ph},
       adsurl = {https://ui.adsabs.harvard.edu/abs/2002ApJ...572L..39S}
}

@ARTICLE{Shibata2016,
       author = {{Shibata}, Masaru and {Sekiguchi}, Yuichiro and {Uchida}, Haruki and {Umeda}, Hideyuki},
        title = "{Gravitational waves from supermassive stars collapsing to a supermassive black hole}",
      journal = {\prd},
         year = 2016,
        month = jul,
       volume = {94},
       number = {2},
          eid = {021501},
        pages = {021501},
          doi = {10.1103/PhysRevD.94.021501},
archivePrefix = {arXiv},
       eprint = {1606.07147},
 primaryClass = {astro-ph.HE},
       adsurl = {https://ui.adsabs.harvard.edu/abs/2016PhRvD..94b1501S}
}

@article{Merida2026,
   title={The rise and fall of little red dots could be driven by the environment},
   volume={707},
   ISSN={1432-0746},
   url={http://dx.doi.org/10.1051/0004-6361/202557594},
   DOI={10.1051/0004-6361/202557594},
   journal={Astronomy \&amp; Astrophysics},
   publisher={EDP Sciences},
   author={Mérida, Rosa M. and Gaspar, Gaia and Asada, Yoshihisa and Sawicki, Marcin and Christopher Omori, Kiyoaki and Willott, Chris J. and Martis, Nicholas S. and Muzzin, Adam and Noirot, Gaël and Rihtaršič, Gregor and Sarrouh, Ghassan T. E. and Tripodi, Roberta},
   year={2026},
   month=Mar, pages={A212} }

@misc{PG2026,
      title={Little Red Dots: One Photometric Tag Concealing Diverse Spectroscopic Flavors of Massive Star Formation and Black Hole Activity}, 
      author={Pablo G. Pérez-González and Guillermo Barro and Stefano Carniani and Francesco D'Eugenio and George H. Rieke and Roberta Tripodi and Andrew J. Bunker and Xihan Ji and Rui Marques-Chaves and Daniel Schaerer and Giacomo Venturi and Flor Arévalo-González and Santiago Arribas and Pierluigi Rinaldi and Bruno Rodríguez Del Pino and Joris Witstok and Rachana Bhatawdekar and Leindert A. Boogaard and Stephane Charlot and Jacopo Chevallard and Luca Costantin and Mirko Curti and Emma Curtis-Lake and Emanuele Daddi and Kelcey Davis and Mark Dickinson and Callum T. Donnan and Fergus R. Donnan and James S. Dunlop and Daniel J. Eisenstein and Henry C. Ferguson and Román Fernández Aranda and Steven L. Finkelstein and Seiji Fujimoto and Giovanni Gandolfi and Mauro Giavalisco and Norman A. Grogin and Mahmoud Hamed and Michaela Hirschmann and Jeyhan S. Kartaltepe and Dale D. Kocevski and Gene C. K. Leung and Cristina M. Lofaro and Ray A. Lucas and Derek J. McLeod and Jens Melinder and Goran Östlin and Casey Papovich and Laura Pentericci and Borja Pérez-Díaz and Marcia Rieke and Jan Scholtz and Rachel S. Somerville and Thomas M. Stanton and Struan D. Stevenson and Irene Shivaei and Sandro Tacchella and Jonathan R. Trump and Hannah Übler and Xin Wang and Christina C. Williams and Christopher N. A. Willmer and L. Y. Aaron Yung and Yongda Zhu},
      year={2026},
      eprint={2602.20247},
      archivePrefix={arXiv},
      primaryClass={astro-ph.GA},
      url={https://arxiv.org/abs/2602.20247}, 
}

@misc{Merida2026b,
      title={UV-to-optical insights into the BH$^*$ model in little red dots}, 
      author={Rosa M. Mérida and Marcin Sawicki and Chris J. Willott and Gaia Gaspar and Kartheik G. Iyer},
      year={2026},
      eprint={2606.12355},
      archivePrefix={arXiv},
      primaryClass={astro-ph.GA},
      url={https://arxiv.org/abs/2606.12355}, 
}

@misc{Merida2026c,
      title={Testing the BH$^*$ Model: a UV-to-Optical Spectral Fitting of The Cliff}, 
      author={Rosa M. Mérida and Marcin Sawicki and Gaia Gaspar and Chris J. Willott and Kartheik G. Iyer},
      year={2026},
      eprint={2605.07976},
      archivePrefix={arXiv},
      primaryClass={astro-ph.GA},
      url={https://arxiv.org/abs/2605.07976}, 
}

@ARTICLE{Dotan2011,
       author = {{Dotan}, Calanit and {Shaviv}, Nir J.},
        title = "{Super-Eddington slim accretion discs with winds}",
      journal = {\mnras},
         year = 2011,
        month = may,
       volume = {413},
       number = {3},
        pages = {1623-1632},
          doi = {10.1111/j.1365-2966.2011.18235.x},
archivePrefix = {arXiv},
       eprint = {1004.1797},
 primaryClass = {astro-ph.HE},
       adsurl = {https://ui.adsabs.harvard.edu/abs/2011MNRAS.413.1623D}
}

@ARTICLE{Fiacconi2016,
       author = {{Fiacconi}, Davide and {Rossi}, Elena M.},
        title = "{Bright vigorous winds as signposts of supermassive black hole birth}",
      journal = {\mnras},
         year = 2016,
        month = jan,
       volume = {455},
       number = {1},
        pages = {2-16},
          doi = {10.1093/mnras/stv2237},
archivePrefix = {arXiv},
       eprint = {1509.07511},
 primaryClass = {astro-ph.HE},
       adsurl = {https://ui.adsabs.harvard.edu/abs/2016MNRAS.455....2F}
}

@ARTICLE{Zhang2025a,
       author = {{Zhang}, Zijian and {Jiang}, Linhua and {Liu}, Weiyang and {Ho}, Luis C.},
        title = "{Analysis of Multi-epoch JWST Images of {\ensuremath{\sim}}300 Little Red Dots: Tentative Detection of Variability in a Minority of Sources}",
      journal = {\apj},
         year = 2025,
        month = may,
       volume = {985},
       number = {1},
          eid = {119},
        pages = {119},
          doi = {10.3847/1538-4357/adcb3e},
archivePrefix = {arXiv},
       eprint = {2411.02729},
 primaryClass = {astro-ph.GA},
       adsurl = {https://ui.adsabs.harvard.edu/abs/2025ApJ...985..119Z}
}

@ARTICLE{Kokubo2025,
       author = {{Kokubo}, Mitsuru and {Harikane}, Yuichi},
        title = "{Challenging the Active Galactic Nucleus Scenario for JWST/NIRSpec Little Red Dot and Non─Little Red Dot Broad H{\ensuremath{\alpha}} Emitters in Light of Nondetection of NIRCam Photometric Variability and X-Ray}",
      journal = {\apj},
         year = 2025,
        month = dec,
       volume = {995},
       number = {1},
          eid = {24},
        pages = {24},
          doi = {10.3847/1538-4357/ae119e},
archivePrefix = {arXiv},
       eprint = {2407.04777},
 primaryClass = {astro-ph.GA},
       adsurl = {https://ui.adsabs.harvard.edu/abs/2025ApJ...995...24K}
}

@ARTICLE{Setton2025,
       author = {{Setton}, David J. and {Greene}, Jenny E. and {de Graaff}, Anna and {Ma}, Yilun and {Leja}, Joel and {Matthee}, Jorryt and {Bezanson}, Rachel and {Boogaard}, Leindert A. and {Cleri}, Nikko J. and {Katz}, Harley and {Labbe}, Ivo and {Maseda}, Michael V. and {McConachie}, Ian and {Miller}, Tim B. and {Price}, Sedona H. and {Suess}, Katherine A. and {van Dokkum}, Pieter and {Wang}, Bingjie and {Weibel}, Andrea and {Whitaker}, Katherine E. and {Williams}, Christina C.},
        title = "{Little Red Dots at an Inflection Point: Ubiquitous V-shaped Turnover Consistently Occurs at the Balmer Limit}",
      journal = {\apj},
         year = 2025,
        month = dec,
       volume = {995},
       number = {1},
          eid = {118},
        pages = {118},
          doi = {10.3847/1538-4357/ae1500},
archivePrefix = {arXiv},
       eprint = {2411.03424},
 primaryClass = {astro-ph.GA},
       adsurl = {https://ui.adsabs.harvard.edu/abs/2025ApJ...995..118S}
}

@misc{Deugenio2026,
      title={Misaligned or chaotic? A strong break of axial symmetry in the local LRD J1025 revealed with VLT/FORS2 spectropolarimetry}, 
      author={Francesco D'Eugenio and Gabriele Pezzulli and Roberto Maiolino and Alessandro Marconi and Xihan Ji and Cristina Ramos Almeida and Andrea Ferrara and Piero Madau and Xiaojing Lin and José A. Acosta-Pulido and Fuyan Bian and Matilde Brazzini and Zheng Cai and Stefano Carniani and Xiaohui Fan and Ignas Juodžbalis and Robert G. Pascalau and Jan Scholtz and Charlotte Simmonds and Fengwu Sun and Sandro Tacchella},
      year={2026},
      eprint={2607.28715},
      archivePrefix={arXiv},
      primaryClass={astro-ph.GA},
      url={https://arxiv.org/abs/2607.28715}, 
}

@misc{Naidu2026,
      title={Little Red Dots as Intermediate Mass, Super-Eddington Engines: Insights from Type IIn Supernovae and The 1837-1856 Great Eruption of $\eta$ Carinae}, 
      author={Rohan P. Naidu and Jorryt Matthee and Anna de Graaff and Alberto Torralba and Chris Ashall and Harley Katz and John Chisholm and Gabriel Brammer and Luc Dessart and Anna-Christina Eilers and Raphael E. Hviding and David O. Jones and Vasily Kokorev and Joel Leja and Hanpu Liu and Zhaoran Liu and Devesh Nandal and Pascal A. Oesch and Conor L. Ransome and Robert A. Simcoe and Wendy Q. Sun and Andrea Weibel and Mengyuan Xiao},
      year={2026},
      eprint={2606.30711},
      archivePrefix={arXiv},
      primaryClass={astro-ph.GA},
      url={https://arxiv.org/abs/2606.30711}, 
}

@ARTICLE{Gu2026,
       author = {{Gu}, Hua-Peng and {Jiang}, Fangzhou and {Chen}, Xian and {Li}, Ran},
        title = "{Nonequilibrium relativistic core collapse of self-interacting dark matter halos: Limits on the seed black hole mass}",
      journal = {\prd},
         year = 2026,
        month = may,
       volume = {113},
       number = {10},
          eid = {103038},
        pages = {103038},
          doi = {10.1103/6sl3-kjzf},
archivePrefix = {arXiv},
       eprint = {2601.17117},
 primaryClass = {astro-ph.CO},
       adsurl = {https://ui.adsabs.harvard.edu/abs/2026PhRvD.113j3038G}
}

@article{Juodzbalis2026b,
   title={A direct black-hole mass measurement in a little red dot at high redshift},
   volume={653},
   ISSN={1476-4687},
   url={http://dx.doi.org/10.1038/s41586-026-10579-4},
   DOI={10.1038/s41586-026-10579-4},
   number={8116},
   journal={Nature},
   publisher={Springer Science and Business Media LLC},
   author={Juodžbalis, Ignas and Marconcini, Cosimo and D’Eugenio, Francesco and Maiolino, Roberto and Marconi, Alessandro and Übler, Hannah and Scholtz, Jan and Ji, Xihan and Jones, Gareth C. and Perna, Michele and Arribas, Santiago and Bennett, Jake S. and Bromm, Volker and Bunker, Andrew J. and Carniani, Stefano and Charlot, Stéphane and Cresci, Giovanni and Dayal, Pratika and Egami, Eiichi and Fabian, Andrew and Inayoshi, Kohei and Isobe, Yuki and Ivey, Lucy R. and Koudmani, Sophie and Laporte, Nicolas and Liu, Boyuan and Lyu, Jianwei and Mazzolari, Giovanni and Monty, Stephanie and Parlanti, Eleonora and Pérez-González, Pablo G. and Robertson, Brant and Schneider, Raffaella and Sijacki, Debora and Tacchella, Sandro and Trinca, Alessandro and Valiante, Rosa and Volonteri, Marta and Witstok, Joris and Zhang, Saiyang},
   year={2026},
   month=May, pages={1017–1021} }

@ARTICLE{Billand2026,
       author = {{Billand}, Jean-Baptiste and {Elbaz}, David and {Gentile}, Fabrizio and {Tarrasse}, Maxime and {Franco}, Maximilien and {Magnelli}, Benjamin and {Daddi}, Emanuele and {Lyu}, Yipeng and {Dekel}, Avishai and {Pacucci}, Fabio and {Sangalli}, Valentina and {Dickinson}, Mark and {Giavalisco}, Mauro and {Holwerda}, Benne W. and {Kocevski}, Dale D. and {Koekemoer}, Anton M. and {Kokorev}, Vasily and {Lucas}, Ray A. and {P{\'e}rez-Gonz{\'a}lez}, Pablo G.},
        title = "{Investigating the growth of little red dot descendants at z < 4 with the JWST}",
      journal = {\aap},
         year = 2026,
        month = jan,
       volume = {706},
          eid = {A29},
        pages = {A29},
          doi = {10.1051/0004-6361/202556303},
archivePrefix = {arXiv},
       eprint = {2507.04011},
 primaryClass = {astro-ph.GA},
       adsurl = {https://ui.adsabs.harvard.edu/abs/2026A&A...706A..29B}
}

@misc{TangMT2026,
      title={SPURS: Evidence for Clumpy Neutral Envelopes and Ionized IGM Surrounding Little Red Dots in Abell 2744 from Ultra-Deep Rest-UV Spectroscopy}, 
      author={Mengtao Tang and Daniel P. Stark and Charlotte A. Mason and Zuyi Chen and Harley Katz and Max Gronke and Lukas J. Furtak and Seok-Jun Chang and Jorryt Matthee and Lily Whitler and Adi Zitrin and Ryan Endsley and Viola Gelli and Tamojeet Roychowdhury and Peter Senchyna and Michael W. Topping and Meng Zhang},
      year={2026},
      eprint={2604.03563},
      archivePrefix={arXiv},
      primaryClass={astro-ph.GA},
      url={https://arxiv.org/abs/2604.03563}, 
}
\bibliographystyle{aasjournalv7}



\end{document}